\documentclass[12pt]{iopart}

\usepackage{iopams}
\expandafter\let\csname equation*\endcsname\relax
\expandafter\let\csname endequation*\endcsname\relax
\usepackage{amsmath,amssymb}
\usepackage{amsfonts,amsbsy,color}
\usepackage{mathtools}
\usepackage{graphicx}
\usepackage[usenames,dvipsnames]{xcolor}
\usepackage{bm}
\usepackage{tikz,tikz-cd}
\usepackage{caption}
\usepackage{booktabs}

\usepackage[hidelinks]{hyperref}
\hypersetup{
  colorlinks   = true, 
  urlcolor     = blue, 
  linkcolor    = blue, 
  citecolor   = blue 
}

\newcommand{\bbR}{\mathbb{R}}

\newcommand{\bmh}{\bm{h}}
\newcommand{\bmi}{\bm{i}}
\newcommand{\bmj}{\bm{j}}
\newcommand{\bmk}{\bm{k}}

\newcommand{\bmq}{\bm{q}}
\newcommand{\bmr}{\bm{r}}

\newcommand{\bmv}{\bm{v}}

\newcommand{\bmG}{\bm{G}}

\newcommand{\calA}{\mathcal{A}}

\newcommand{\calP}{\mathcal{P}}

\newcommand{\calS}{\mathcal{S}}

\newcommand{\rma}{\mathrm{a}}

\newcommand{\rmn}{\mathrm{n}}

\newcommand{\rms}{\mathrm{s}}

\newcommand{\rmO}{\mathrm{O}}

\newcommand{\rmR}{\mathrm{R}}

\newcommand{\sfI}{\mathsf{I}}

\newcommand{\sfT}{\mathsf{T}}

\newcommand{\ep}{\epsilon}
\newcommand{\rev}{\mathrm{rev}}
\newcommand{\irr}{\mathrm{irr}}
\newcommand{\intq}{\int_{\Lambda^{\prime} \leq |\bmq| \leq \Lambda} \frac{d^d \bmq}{(2\pi)^d}}

\newcommand{\intqo}{\int_{\bmq, \Omega}^{\prime}}
\newcommand{\into}{\int_{-\infty}^{\infty} \frac{d\Omega}{2\pi}}
\newcommand{\tile}{\tilde{e}}

\newcommand{\tilt}{\tilde{\theta}}
\newcommand{\tiln}{\tilde{n}}
\newcommand{\Od}{\tilde{\Omega}_d}
\newcommand{\barc}{\bar{c}_{\mathrm{s}}}
\newcommand{\bard}{\bar{d}_{\mathrm{s}}}
\newcommand{\bart}{\bar{\lambda}_{\theta}}
\newcommand{\barn}{\bar{\lambda}_n}
\newcommand{\bare}{\bar{\lambda}_e}
\newcommand{\barnt}{\bar{\nu}_{\theta}}
\newcommand{\barne}{\bar{\nu}_e}
\newcommand{\barlam}{\bar{\lambda}}
\newcommand{\dbeq}{\overset{\mathrm{D.B.}}{=}}

\begin{document}
\title{Anomalous energy transport in the Berezinskii-Kosterlitz-Thouless phase}

\author{Ken Hiura}
\address{Universal Biology Institute, The University of Tokyo, Tokyo 113-0033, Japan}
\ead{ken.hiura@ubi.s.u-tokyo.ac.jp}

\begin{abstract}
    We study nonlinear fluctuating hydrodynamic theories with charge and energy conservation in and above two dimensions that describe the large-scale behavior of the Hamiltonian XY model in the disordered and ordered phases. Using renormalization group analysis at one-loop order, we show that while Fourier's law holds in the ordered phase above two dimensions and in the disordered phase in any dimension, the energy diffusivity in the ordered phase exactly in two dimensions, the Berezinskii-Kosterlitz-Thouless phase, exhibits a logarithmic divergence in the thermodynamic limit. This divergence arises from elastic energy transport induced by spin-wave excitations.
    \vspace*{1ex}
\end{abstract}

\section{Introduction}

Heat conduction is one of the fundamental transport phenomena in nonequilibrium physics. In heat-conducting systems, the heat conductivity is of primary interest because it characterizes the linear response relation between the energy current density and the spatial temperature gradient. In typical three-dimensional many-particle systems, heat conductivity is asymptotically independent of the system size. This behavior is known as Fourier's law, which is considered as a fundamental law of nonequilibrium thermodynamics \cite{deGrootMazur19962}. However, diverging conductivity with increasing system size is ubiquitously observed in low-dimensional materials \cite{Chang2008,Xu2014,Lee2017}. This anomalous energy transport, indicating a breakdown of Fourier's law, has been the subject of extensive study \cite{LepriLiviPoliti2003,Liu2012,Dhar2008,Livi2023}.

In low-dimensional fluids, convective energy transport plays a crucial role in anomalous heat conduction as a form of energy transfer distinct from simple diffusion \cite{NarayanRamaswamy2002}. It would be natural to ask whether anomalous energy transport can be observed in a lattice system with short-ranged interactions, where convective energy transport associated with translational motion is absent. One might speculate that anomalous heat conduction could occur if a soft mode emerges due to the spontaneous breaking of a continuous symmetry at low temperatures, leading to elastic energy transport. However, large thermal fluctuations in low-dimensional systems destroy long-range order, and according to the Hohenberg-Mermin-Wagner theorem \cite{Hohenberg1967,MerminWagner1966,Mermin1968}, spontaneous breaking of continuous symmetries is prohibited in and below two dimensions. At the same time, breakdowns of conventional hydrodynamics are often observed due to anomalous fluctuations in the conserved current in low-dimensional systems \cite{AlderWainwright1970,PomeauResibois1975,ForsterNelsonStephen1977}. Thus, while continuous symmetry breaking can occur in higher dimensions, anomalous transport tends to emerge in lower dimensions. This raises the nontrivial question of whether anomalous heat conduction can occur in lattice systems at low temperatures.

In this context, we consider the ferromagnetic Hamiltonian XY model on a square lattice \cite{LeonciniVergaRuffo1998} as a potential candidate for exhibiting such anomalous behavior. This model conserves the charge associated with the $\rmO(2)$ symmetry and the energy. Numerical studies of the two-dimensional Hamiltonian XY model \cite{DellagoPosch1997,DelfiniLepriLivi2005} indicate that, while the disordered phase at high temperatures follows Fourier's law, the heat conductivity at low temperatures diverges logarithmically in the thermodynamic limit, suggesting anomalous heat conduction in the Berezinskii-Kosterlitz-Thouless (BKT) phase \cite{Berezinskii1971,Berezinskii1972,KosterlitzThouless1973,Kosterlitz1974}. To understand the universal aspects of such energy transport in lattice spin systems, it may be necessary to develop a hydrodynamic theory independent of details of specific microscopic models.

In this paper, we study energy transport in nonlinear fluctuating hydrodynamic models with charge and energy conservation that we expect to describe the large-distance and long-time behavior of the Hamiltonian XY model in and above two dimensions. We find that energy transport is purely diffusive in the disordered phase, while an elastic energy current emerges in the ordered phase due to the presence of spin-wave excitations. Using a dynamic renormalization group analysis at one-loop order, we show that the size-dependent energy diffusivity converges to a finite value in the thermodynamic limit in the ordered phase above two dimensions, as well as in the disordered phase in any dimension. In contrast, in the ordered phase in exactly two dimensions, the energy diffusivity diverges logarithmically with the system size. These theoretical results suggest anomalous energy transport in the BKT phase and are consistent with the previous numerical studies \cite{DellagoPosch1997,DelfiniLepriLivi2005}. To the best of my knowledge, this work is the first to apply renormalization group to anomalous energy transport in lattice spin systems far from criticality.

This paper is organized as follows. We introduce the Hamiltonian XY model as a microscopic model and propose hydrodynamic theories that describe the large-scale behavior of this model in Section \ref{sec:hydro}. We compute the renormalized diffusivities using dynamic renormalization group in Section \ref{sec:rg}. Concluding remarks are given in Section \ref{sec:remark}

\section{Hydrodynamics}
\label{sec:hydro}

\subsection{Microscopic model}
\label{subsec:micro}

The hydrodynamic theory presented in this paper is independent of the microscopic details of specific models. Nevertheless, to clarify the systems of interest, we introduce a microscopic model that we expect to be described by the hydrodynamic theory. As such a model, we consider the ferromagnetic Hamiltonian XY model on the $d$-dimensional square lattice $\Lambda$ with a lattice constant $a$. Let $\theta_{\bmi}$ and $p_{\bmi}$ denote the angular variable and momentum of the XY spin at site $\bmi \in \Lambda$, respectively. The Hamiltonian of the system is given by
\begin{align*}
    H = \sum_{\bmi \in \Lambda} \frac{p_{\bmi}^2}{2 I} + \frac{1}{2} \sum_{\bmi \in \Lambda} \sum_{\bmj \in B(\bmi)} V(\theta_{\bmi} - \theta_{\bmj}),
\end{align*}
where $I$ is the moment of inertia, $B(\bmi)$ is the set of the nearest-neighbor sites of site $\bmi$, and $V(\theta) = J(1 - \cos \theta)$, with $J$ being the positive coupling constant. The Hamiltonian possesses a continuous $\rmO(2)$ symmetry, meaning it is invariant under a global continuous shift of the angular variables: $\theta_{\bmi} \to \theta_{\bmi} + \alpha$ for a constant $\alpha$. The time evolution of the system is described by the Hamilton equation
\begin{align}
    \dot{\theta}_{\bmi} &= \frac{\partial H}{\partial p_{\bmi}} = \frac{p_{\bmi}}{I},
    \\ \label{eq:mcc}
    \dot{p}_{\bmi} &= - \frac{\partial H}{\partial \theta_{\bmi}} = - \sum_{\bmj \in B(\bmi)} V^{\prime}(\theta_{\bmi} - \theta_{\bmj}).
\end{align}
The local energy $e_{\bmi}$ at site $\bmi$ is defined as
\begin{align*}
    e_{\bmi} \coloneqq \frac{p_{\bmi}^2}{2 I} + \frac{1}{2} \sum_{\bmj \in B(\bmi)} V(\theta_{\bmi} - \theta_{\bmj}).
\end{align*}
The conserved quantities in this system are $N \coloneqq \sum_{\bmi \in \Lambda} p_{\bmi}$ and the total energy $H$. The quantity $N$ is the conserved charge associated with the $O(2)$ symmetry. The local conservation laws for these two quantities are expressed as Eq.~\eqref{eq:mcc} and
\begin{align} \label{eq:mec}
    \dot{e}_{\bmi} = \sum_{\bmj \in B(\bmi)} j^e(\bmj,\bmi),
\end{align}
respectively. Here, $j^e(\bmi,\bmj)$ is the energy current from site $\bmi$ to $\bmj$ defined by
\begin{align*}
    j^e(\bmi,\bmj) \coloneqq \frac{1}{2} V^{\prime}(\theta_{\bmi} - \theta_{\bmj}) (p_{\bmi} + p_{\bmj}) = -j^e(\bmj, \bmi).
\end{align*}
A previous numerical study \cite{LeonciniVergaRuffo1998} demonstrates that the steady state of the two-dimensional Hamiltonian XY model undergoes a BKT transition \cite{Berezinskii1972,KosterlitzThouless1973,Kosterlitz1974} from the BKT phase to the disordered phase. The aim of this paper is to develop a hydrodynamic theory for energy transport in the disordered and the BKT phases of this model. If we accept the universality of anomalous transport phenomena in these phases, the results presented in the following sections are expected to apply to other spin models, such as modified versions of the XY model \cite{DomanySchickSwendsen1984,DellagoPosch1997} that exhibit a first-order phase transition between the disordered and BKT phase, rather than the BKT transition.

\subsection{Derivation of hydrodynamics}
\label{subsec:hydro}

In this subsection, we derive two fluctuating hydrodynamic equations that describe the Hamiltonian XY model in the disordered and ordered phases, respectively. In this paper, we provide a phenomenological derivation based on the second law of thermodynamics and the detailed balance condition. A major part of the derivation is not new and follows Refs. \cite{ChaikinLubensky1995}. While the nonlinear fluctuating hydrodynamic equation in the ordered phase could be derived directly from the microscopic Hamiltonian XY model using the projection operator method \cite{Hiura2023}, we focus here on the phenomenological approach.

The hydrodynamic description in the disordered phase proposed in this paper applies to systems with energy conservation and charge conservation associated with $\mathrm{O}(2)$ symmetry. In addition to these properties, the hydrodynamics in the ordered phase requires the (quasi-)long-range order of the order parameter. As noted in Section \ref{subsec:micro}, it is expected that the standard Hamiltonian XY model \cite{LeonciniVergaRuffo1998} and its modified version \cite{DomanySchickSwendsen1984,DellagoPosch1997} belong to this class.

\subsubsection{Disordered phase}

To construct a hydrodynamic theory of the Hamiltonian XY model, we begin by choosing a complete set of slow variables. In the disordered phase, we assume that only the long-wavelength components of the two conserved fields, the charge density and energy density fields, constitute this set. Let $n$ and $e$ denote the charge density and the energy density fields, respectively. These fields obey the local conservation laws,
\begin{align} \label{eq:cc}
    \partial_t n &= - \nabla \cdot \bmj^n,
    \\ \label{eq:ec}
    \partial_t e &= - \nabla \cdot \bmj^e,
\end{align}
where $\bmj^n$ and $\bmj^e$ are the conserved charge and energy current densities, respectively. These equations are the continuous space analog of the local conservation laws in lattice systems, Eqs.~\eqref{eq:mcc} and \eqref{eq:mec}. Our goal is to determine the constitutive relations that specify the form of the conserved current densities as functions of $n$, $e$, and their derivatives. To achieve this, we utilize the fundamental thermodynamic relation and the second law of thermodynamics. The thermodynamic relation in the disordered phase is expressed as
\begin{align} \label{eq:dotr}
    d s = \beta de - \beta \mu dn,
\end{align}
where $s = s(e,n)$ is the thermodynamic entropy density, $\beta$ is the inverse temperature, and $\mu$ is the chemical potential. We define the entropy functional $\calS$ as the spatial integration of the entropy density,
\begin{align}
    \calS [e,n] \coloneqq \int_{\bmr} s(e(\bmr),n(\bmr)).
\end{align}
Here, $\int_{\bmr} \coloneqq \int d^d \bmr$ is the spatial integration over the entire space. Using the continuity equations \eqref{eq:cc} and \eqref{eq:ec} with the thermodynamic relation \eqref{eq:dotr}, we find that the total entropy production rate is expressed as
\begin{align}
    \frac{d}{dt} \calS = \int_{\bmr} \left[ \nabla \beta \cdot (\bmj^e - \mu \bmj^n) - \beta \bmj^n \cdot \nabla \mu \right].
\end{align}
To derive this, we have assumed that all the fields decay sufficiently rapidly at infinity in space or satisfy the periodic boundary condition in a bounded region. 

We decompose the conserved current densities into three parts: reversible, irreversible and noise components. Specifically, we write $\bmj^n = \bmj^{n,\rev} + \bmj^{n,\irr} + \bmj^{n, \rmn}$ and $\bmj^e = \bmj^{e,\rev} + \bmj^{e,\irr} + \bmj^{e, \rmn}$. The reversible parts are characterized by the condition that they do not contribute entropy production, whereas the irreversible parts are defined by their dissipation-producing nature. First, we require that the entropy production rate vanishes when $\bmj = \bmj^{\rev}$. This leads to the reversible parts being determined as
\begin{align}
    \bmj^{n,\rev} = 0, \quad \bmj^{e,\rev} = \mu \bmj^{n,\rev} = 0
\end{align}
at the zeroth order of the derivative expansion. Second, we impose that the entropy production rate must be positive when $\bmj = \bmj^{\rev} + \bmj^{\irr}$. The simplest choice satisfying this condition is
\begin{align} \label{eq:doirr}
    \bmj^{n,\irr} = - \gamma_n \nabla \mu, \quad \bmj^{e,\irr} = \mu \bmj^{n,\irr} + \gamma_e \nabla \beta = - \gamma_n \mu \nabla \mu + \gamma_e \nabla \beta,
\end{align}
where $\gamma_n$ and $\gamma_e$ denote the transport coefficients associated with charge and energy, respectively. The positivity of the entropy production rate implies that $\gamma_n > 0$ and $\gamma_e > 0$. For later convenience, we express Eq.~\eqref{eq:doirr} in terms of the Onsager coefficients and the thermodynamic forces. The thermodynamic forces are defined as the derivatives of the entropy functional with respect to slow variables. Using the thermodynamic relation \eqref{eq:dotr}, we obtain the thermodynamic forces as
\begin{align}
    \frac{\delta \calS}{\delta n (\bmr)} = - \beta \mu, \quad \frac{\delta \calS}{\delta e (\bmr)} = \beta. 
\end{align}
Eq.~\eqref{eq:doirr} is then expressed as
\begin{align}
    - \nabla \cdot \bmj^{n,\irr} (\bmr) &= \int_{\bmr^{\prime}} \left[ L^{nn}(\bmr, \bmr^{\prime}) \frac{\delta \calS}{\delta n (\bmr^{\prime})} + L^{ne}(\bmr, \bmr^{\prime}) \frac{\delta \calS}{\delta e (\bmr^{\prime})} \right],
    \\
    - \nabla \cdot \bmj^{e,\irr} (\bmr) &= \int_{\bmr^{\prime}} \left[ L^{en}(\bmr, \bmr^{\prime}) \frac{\delta \calS}{\delta n (\bmr^{\prime})} + L^{ee}(\bmr, \bmr^{\prime}) \frac{\delta \calS}{\delta e (\bmr^{\prime})} \right],
\end{align}
where
\begin{align*}
    L^{nn}(\bmr, \bmr^{\prime}) &= - \nabla \cdot \gamma_n \beta^{-1} \nabla \delta (\bmr - \bmr^{\prime}),
    \\
    L^{ne}(\bmr, \bmr^{\prime}) &= - \nabla \cdot \gamma_n \beta^{-1} \mu \nabla \delta (\bmr - \bmr^{\prime}),
    \\
    L^{en}(\bmr, \bmr^{\prime}) &= - \nabla \cdot \gamma_n \beta^{-1} \mu \nabla \delta (\bmr - \bmr^{\prime}),
    \\
    L^{ee}(\bmr, \bmr^{\prime}) &= - \nabla \cdot \left( \gamma_e + \gamma_n \beta^{-1} \mu^2 \right) \nabla \delta (\bmr - \bmr^{\prime})
\end{align*}
are the Onsager coefficients. Note that the Onsager matrix composed of these Onsager coefficients is positive-definite and it satisfies the reciprocity relations. Finally, we determine the noise parts using the detailed balance condition. Since the noise terms physically represent random deviations of the current densities from the deterministic terms, $\bmj^{\rev} + \bmj^{\irr}$, it is reasonable to assume that the noise terms are expressed as white Gaussian noises in both space and time. Furthermore, in the long-wavelength limit, the equilibrium distribution $p_{\mathrm{eq}}$ of $e$ and $n$ can be approximated as
\begin{align}
    \ln p_{\mathrm{eq}}[e,n] \simeq \calS [e,n] + \mathrm{const.}
\end{align}
at the leading order of the derivative expansion. Under these assumptions, the following choice satisfies the detailed balance condition \cite{Green1952,GrahamHaken1971}:
\begin{align*}
    \bmj^{n,\rmn} = \sqrt{2 \gamma_n \beta^{-1}} \bxi^n, \quad \bmj^{e,\rmn} = \sqrt{2 \gamma_e} \bxi^e + \mu \sqrt{2 \gamma_n \beta^{-1}} \bxi^n,
\end{align*}
where $\bxi^n$ and $\bxi^e$ are independent white Gaussian noises with zero means and covariances,
\begin{align} \label{eq:noisecovll}
    \langle \bxi^n (\bmr,t) [\bxi^n (\bmr^{\prime},t^{\prime})]^{\sfT} \rangle &= \delta^d (\bmr - \bmr^{\prime}) \delta (t-t^{\prime}),
    \\ \label{eq:noisecovee}
    \langle \bxi^e (\bmr,t) [\bxi^e (\bmr^{\prime},t^{\prime})]^{\sfT} \rangle &= \delta^d (\bmr - \bmr^{\prime}) \delta (t-t^{\prime}),
    \\ \label{eq:noisecovel}
    \langle \bxi^e (\bmr,t) [\bxi^n (\bmr^{\prime},t^{\prime})]^{\sfT} \rangle &= 0.
\end{align}
Consequently, the fluctuating hydrodynamic equations in the disordered phase are given by
\begin{align} \label{eq:hdl}
    \partial_t n &= - \nabla \cdot \left( - \gamma_n \nabla \mu + \sqrt{2 \gamma_n \beta^{-1}} \bxi^n \right),
    \\ \label{eq:hde}
    \partial_t e &= - \nabla \cdot \left( \mu \left( - \gamma_n \nabla \mu + \sqrt{2 \gamma_n \beta^{-1}} \bxi^n \right) + \gamma_e \nabla \beta + \sqrt{2 \gamma_e} \bxi^e \right).
\end{align}
The dynamics of charge and energy fluctuations in the disordered phase is purely diffusive. Away from the critical point, there is no reason to expect anomalous transport in the disordered phase. This conclusion is supported by the power-counting argument presented in Section \ref{subsec:doRG}.

\subsubsection{Ordered phase}

In the ordered phase, the gradient of the phase field may be included as a slow variable in addition to the locally conserved fields \cite{ChaikinLubensky1995}. The phase field can be defined microscopically as follows. Consider the Hamiltonian XY model introduced in Section \ref{subsec:micro}. For a coarse-grained length $a_0 \gg a$, we define the two-component magnetization $\vec{\varphi} = (\varphi_1, \varphi_2)$ at $\bmr \in \bbR^d$ as
\begin{align} \label{eq:mag}
    \varphi_1 (\bmr) = \sum_{i \in B(\bmr;a_0)} \cos \theta_i, \quad \varphi_2 (\bmr) = \sum_{i \in B(\bmr ; a_0)} \sin \theta_i,
\end{align}
where $B(\bmr ; a_0) = \{ \bmi \in \Lambda : | \bmi - \bmr | \leq a_0 \}$ is the coarse-grained region around $\bmr$ with a radius of $a_0$. The coarse-grained phase field is then defined as $\theta \coloneqq \tan^{-1}(\varphi_2/\varphi_1)$.

In this paper, we assume that the local amplitude field $m \coloneqq \sqrt{\varphi_1^2 + \varphi_2^2}$ remains finite and that amplitude fluctuations around the finite equilibrium value decay within a finite time, even in the long-wavelength limit. Above two dimensions and at sufficiently low temperatures where the $\rmO(2)$ symmetry is spontaneously broken, this assumption is justified. In exactly two dimensions, however, this assumption becomes subtle, as the total magnetization vanishes in the thermodynamic limit at finite temperatures. Nevertheless, since the decay of the total magnetization is algebraic with respect to the linear dimension of the averaged region, and the spatial average in Eq.~\eqref{eq:mag} is taken over a mesoscopic length $a_0$, which is much smaller than the system size, this assumption may still be reasonable.

Let us derive the hydrodynamic equation in the ordered phase. The approach is similar to that used in the disordered phase. The thermodynamic relation in the ordered phase is given by
\begin{align} \label{eq:otr}
    ds = \beta de - \beta \mu dn - \beta \bmh_{\theta} \cdot d \bmv_{\theta},
\end{align}
where $\bmv_{\theta} \coloneqq \nabla \theta$ is the gradient of the phase and $\bmh_{\theta}$ its conjugate field \cite{ChaikinLubensky1995}.

In this paper, we focus on the low-temperature region where vortex excitations can be neglected, allowing the phase field $\theta$ to be treated as a scalar field. In this regime, the transverse component of $\bmv_{\theta}$ induced by vortex excitations can be ignored, and the equation of motion for $\theta$ can be expressed as
\begin{align} \label{eq:josephson}
    \partial_t \theta = \mu - j^{\theta},
\end{align}
where the scalar field $j^{\theta}$ will be determined later. Here, the appearance of the chemical potential $\mu$ can be understood as follows \cite{ChaikinLubensky1995,HalperinHohenberg1969}. Applying an external chemical potential $\mu_{\mathrm{ex}}$ in the Hamiltonian $H \to H - \mu_{\mathrm{ex}} N$ induces an additional angular frequency $- \mu_{\mathrm{ex}}$ in the equation of motion for the phase $\theta$. To ensure $\partial_t \theta = 0$ in equilibrium, the equation of motion for $\theta$ should takes the form $\partial_t \theta = \mu - \mu_{\mathrm{ex}}$, if we neglect the gradient terms, which vanish in equilibrium. In Eq.~\eqref{eq:josephson}, we expand the equation of motion for $\theta$ around this form, setting the external chemical potential to $\mu_{\mathrm{ex}} = 0$. Consequently, $\bmv_{\theta}$ acquires only a longitudinal component during its time evolution, as described by
\begin{align} \label{eq:peom}
    \partial_t \bmv_{\theta} = \nabla (\mu - j^{\theta}),
\end{align}
as desired.

The entropy functional $\calS$ is defined as
\begin{align}
    \calS [e,n, \bmv_{\theta}] = \int_{\bmr} s(e (\bmr),n (\bmr), \bmv_{\theta} (\bmr)).
\end{align}
Using the equations of motion \eqref{eq:cc}, \eqref{eq:ec}, and \eqref{eq:peom} with the thermodynamic relation \eqref{eq:otr}, we find that the entropy production rate is given by
\begin{align}
    \frac{d}{dt} \calS = \int_{\bmr} \left[ \nabla \beta \cdot \left( \bmj^e - \mu \bmj^n - \bmh_{\theta} j^{\theta} \right) - \beta (\bmj^n + \bmh_{\theta}) \cdot \nabla \mu - \beta j^{\theta} \nabla \cdot \bmh_{\theta} \right].
\end{align}
We decompose $j^{\theta}$ as $j^{\theta} = j^{\theta,\rev} + j^{\theta, \irr} + j^{\theta, \rmn}$. By identifying
\begin{align} \label{eq:orev}
    j^{\theta, \rev} = 0, \quad \bmj^{n,\rev} = - \bmh_{\theta}, \quad \bmj^{e,\rev} = \mu \bmj^{n, \rev} + \bmh_{\theta} j^{\theta ,\rev} = - \mu \bmh_{\theta}.
\end{align}
we find that the condition for vanishing entropy production rate in the reversible parts is satisfied. This choice is physically reasonable from a microscopic point of view. Indeed, at low temperatures in two or higher dimensions, phase fluctuations are small, and the interaction potential $V$ in Section \ref{subsec:micro} can be approximated by a harmonic form, $V(\theta_{\bmi} - \theta_{\bmj}) \approx \frac{J}{2} (\theta_{\bmi} - \theta_{\bmj})^2$. This approximation leads to elastic contributions,
\begin{align}
    V^{\prime}(\theta_{\bmi} - \theta_{\bmj}) \approx J(\theta_{\bmi} - \theta_{\bmj})
\end{align}
in the charge current and
\begin{align}
    j^e(\bmi, \bmj) \approx J(\theta_{\bmi} - \theta_{\bmj}) \frac{p_{\bmi} + p_{\bmj}}{2}
\end{align}
in the energy current. By identifying $\bmh_{\theta} \sim J(\theta_{\bmj} - \theta_{\bmi})$ and $\mu \sim p_{\bmi}$, Eqs.~\eqref{eq:josephson} and \eqref{eq:orev} are consistent with the above approximations.

To ensure the positivity of the entropy production rate, we define
\begin{align}
    j^{\theta, \irr} &= - \gamma_{\theta} \nabla \cdot \bmh_{\theta},
    \\
    \bmj^{n,\irr} &= - \gamma_n \nabla \mu,
    \\
    \bmj^{e,\irr} &= \gamma_e \nabla \beta - \gamma_n \mu \nabla \mu - \gamma_{\theta} \bmh_{\theta} (\nabla \cdot \bmh_{\theta}),
\end{align}
where $\gamma$, $\gamma_n$, and $\gamma_e$ are the transport coefficients, with $\gamma, \gamma_n, \gamma_e > 0$. These constitutive relations can be expressed in terms of the Onsager coefficients and thermodynamic forces,
\begin{align}
    - \nabla j^{\theta,\irr} (\bmr) = \int_{\bmr^{\prime}} \left[ L^{\bmv_{\theta} \bmv_{\theta}}(\bmr, \bmr^{\prime}) \frac{\delta \calS}{\delta \bmv_{\theta}(\bmr^{\prime})} + L^{\bmv_{\theta} n} (\bmr, \bmr^{\prime}) \frac{\delta \calS}{\delta n(\bmr^{\prime})} + L^{\bmv_{\theta} e} (\bmr, \bmr^{\prime}) \frac{\delta \calS}{\delta e(\bmr^{\prime})} \right],
    \\
    - \nabla \cdot \bmj^{n,\irr} (\bmr) = \int_{\bmr^{\prime}} \left[ L^{n \bmv_{\theta}}(\bmr, \bmr^{\prime}) \frac{\delta \calS}{\delta \bmv_{\theta}(\bmr^{\prime})} + L^{nn} (\bmr, \bmr^{\prime}) \frac{\delta \calS}{\delta n(\bmr^{\prime})} + L^{n e} (\bmr, \bmr^{\prime}) \frac{\delta \calS}{\delta e(\bmr^{\prime})} \right],
    \\
    - \nabla \cdot \bmj^{e,\irr} (\bmr) = \int_{\bmr^{\prime}} \left[ L^{e \bmv_{\theta}}(\bmr, \bmr^{\prime}) \frac{\delta \calS}{\delta \bmv_{\theta}(\bmr^{\prime})} + L^{e n} (\bmr, \bmr^{\prime}) \frac{\delta \calS}{\delta n(\bmr^{\prime})} + L^{e e} (\bmr, \bmr^{\prime}) \frac{\delta \calS}{\delta e(\bmr^{\prime})} \right],
\end{align}
where
\begin{align}
    \frac{\delta \calS}{\delta \bmv_{\theta} (\bmr)} = - \beta \bmh_{\theta} (\bmr)
\end{align}
is the thermodynamic force associated with the gradient of the phase and
\begin{align*}
    L^{\bmv_{\theta} \bmv_{\theta}} (\bmr, \bmr^{\prime}) &= - \nabla \gamma_{\theta} \beta^{-1} \nabla \cdot \delta (\bmr - \bmr^{\prime}),
    \\
    L^{\bmv_{\theta} n} (\bmr, \bmr^{\prime}) &= 0,
    \\
    L^{\bmv_{\theta} e} (\bmr, \bmr^{\prime}) &= - \nabla \gamma_{\theta} \beta^{-1} \bmh_{\theta} \cdot \nabla \delta (\bmr - \bmr^{\prime}),
    \\
    L^{n \bmv_{\theta}} (\bmr, \bmr^{\prime}) &= 0,
    \\
    L^{nn} (\bmr, \bmr^{\prime}) &= - \nabla \cdot \gamma_n \beta^{-1} \nabla \delta (\bmr - \bmr^{\prime}),
    \\
    L^{n e} (\bmr, \bmr^{\prime}) &= - \nabla \cdot \gamma_n \beta^{-1} \mu \nabla \delta (\bmr - \bmr^{\prime}),
    \\
    L^{e \bmv_{\theta}} (\bmr, \bmr^{\prime}) &= - \nabla \cdot \gamma_{\theta} \beta^{-1} \bmh_{\theta} \nabla \cdot \delta (\bmr - \bmr^{\prime}),
    \\
    L^{e n} (\bmr, \bmr^{\prime}) &= - \nabla \cdot \gamma_n \beta^{-1} \mu \nabla \delta (\bmr - \bmr^{\prime}),
    \\
    L^{ee} (\bmr, \bmr^{\prime}) &= - \nabla \cdot \left[ \left( \gamma_e + \gamma_n \beta^{-1} \mu^2 \right) \sfI + \gamma_{\theta} \beta^{-1} \bmh_{\theta} \bmh_{\theta}^{\sfT} \right] \nabla \delta (\bmr - \bmr^{\prime}),
\end{align*}
are the Onsager coefficients. Here, $\sfI$ is the $d \times d$ identity matrix. The Onsager matrix composed of these Onsager coefficients is positive-definite and it satisfies the reciprocity relation. The forms of the irreversible parts suggest that the noise parts obeying the detailed balance condition are given by
\begin{align}
    j^{\theta, \rmn} &= \sqrt{2 \gamma_{\theta} \beta^{-1}} \xi^{\theta},
    \\
    \bmj^{n,\rmn} &= \sqrt{2 \gamma_n \beta^{-1}} \bxi^n,
    \\
    \bmj^{e,\rmn} &= \sqrt{2 \gamma_e} \bxi^e + \mu \sqrt{2 \gamma_n \beta^{-1}} \bxi^n + \bmh_{\theta} \sqrt{2 \gamma_{\theta} \beta^{-1}} \xi^{\theta},
\end{align}
where $\xi^{\theta}$ is a white Gaussian noise with zero mean and unit variance, and statistically independent of $\bxi^n$ and $\bxi^e$.

Consequently, the fluctuating hydrodynamic equations in the ordered phase are given by
\begin{align} \label{eq:oht}
    \partial_t \theta &= \mu + \gamma_{\theta} \nabla \cdot \bmh_{\theta} - \sqrt{2 \gamma_{\theta} \beta^{-1}} \xi^{\theta},
    \\ \label{eq:ohc}
    \partial_t n &= - \nabla \cdot \left( - \bmh_{\theta} - \gamma_n \nabla \mu + \sqrt{2 \gamma_n \beta^{-1}} \bxi^n \right),
    \\
    \partial_t e &= - \nabla \cdot \left( \bmh_{\theta} \left(- \mu - \gamma_{\theta} \nabla \cdot \bmh_{\theta} + \sqrt{2 \gamma_{\theta} \beta^{-1}} \xi^{\theta} \right) \right.
    \notag \\ \label{eq:ohe}
    &\qquad \left. + \mu \left( - \bmh_{\theta} - \gamma_n \nabla \mu + \sqrt{2 \gamma_n \beta^{-1}} \bxi^n \right) + \gamma_e \nabla \beta + \sqrt{2 \gamma_e} \bxi^e \right).
\end{align}
In this paper, we use the equation of motion for $\theta$ instead of that for $\bmv_{\theta}$. In contrast to the disordered phase, there exists an elastic reversible energy current, represented by the term $- \mu \bmh_{\theta}$, in addition to the diffusion current, which was suggested by Ref. \cite{DellagoPosch1997}. This elastic energy current, which transports energy faster than diffusion, can induce anomalous energy diffusion at large scales. To investigate the renormalization effect of this nonlinear reversible current on energy transport at large scales, we employ the renormalization group method in Section \ref{sec:rg}.

\section{Renormalization group}
\label{sec:rg}

In this section, we investigate the large-scale behavior of the nonlinear fluctuating hydrodynamics presented in the previous section using a perturbative renormalization group.

\subsection{Disordered phase}
\label{subsec:doRG}

\subsubsection{Linear theory}

We first examine the disordered phase. We consider a homogeneous equilibrium state characterized by the energy density $e_0$ and the charge density $n = 0$. The equation of motion for dynamical fluctuations around this equilibrium state is obtained by expanding the hydrodynamic equations \eqref{eq:hdl} and \eqref{eq:hde} around the equilibrium values $(e(\bmr), n(\bmr)) = (e_0, 0)$. Due to the symmetry $s(e,n) = s(e, -n)$, the entropy density function can be expanded as
\begin{align*}
    s(e,n) = s_0 + \frac{\delta e}{T_0} - \frac{(\delta e)^2}{2 C_0 T_0^2} - \frac{n^2}{2 I_0 T_0} + \cdots,
\end{align*}
where $s_0 = s(e_0, 0)$ denotes the entropy density at the reference equilibrium state, $\delta e = e - e_0$ the deviation of the energy density from the equilibrium value, $T_0$ the equilibrium temperature, $C_0$ the heat capacity, and $I_0$ the charge susceptibility. From the concavity of the entropy function, $C_0 > 0$ and $I_0 > 0$. The temperature and the chemical potential are given by
\begin{align} \label{eq:beta}
    \beta^{-1} = \left( \frac{\partial s}{\partial e} \right)^{-1} = T_0 \left( 1 - \frac{\delta e}{C_0 T_0} + \cdots \right)^{-1}
\end{align}
and
\begin{align} \label{eq:mu}
    \mu = - \beta^{-1} \frac{\partial s}{\partial n} = \frac{n}{I_0} + \cdots,
\end{align}
respectively. The ellipses denote higher-order corrections. Furthermore, the transport coefficients can be expanded around the equilibrium state as $\gamma_n = \gamma_{n0} + \cdots$ and $\gamma_e = \gamma_{e 0} + \cdots$, where $\gamma_{n0}$ and $\gamma_{e0}$ are transport coefficients evaluated at $(e,n) = (e_0,0)$, and the ellipses indicate variations due to fluctuations in energy $\delta e$ and charge $n$. From Eqs.~\eqref{eq:hdl} and \eqref{eq:hde}, we derive the linearized hydrodynamic equations in the disordered phase,
\begin{align} \label{eq:dll}
    \partial_t n &= - \nabla \cdot \left( - \nu_n \nabla n + \sqrt{2 D_n} \bm{\xi}^n \right),
    \\ \label{eq:dle}
    \partial_t e &= - \nabla \cdot \left( - \nu_e \nabla e + \sqrt{2 D_e} \bm{\xi}^e \right).
\end{align}
Here, $\nu_{n}$ and $\nu_e$ are the charge and energy diffusivities defined by
\begin{align} \label{eq:nune}
    \nu_n \coloneqq \frac{\gamma_{n0}}{I_0}, \quad \nu_e \coloneqq \frac{\gamma_{e 0}}{C_0 T_0^2}.
\end{align}
The noise strengths, $D_n$ and $D_e$, are defined by
\begin{align} \label{eq:Dne}
    D_n \coloneqq \gamma_{n0} T_0, \quad D_e \coloneqq \gamma_{e0}.
\end{align}
There are two diffusion modes in Eqs.~\eqref{eq:dll} and \eqref{eq:dle} characterized by the dispersion relations, $\omega_n (\bmk) = - i \nu_n \bmk^2$ and $\omega_e (\bmk) = - i \nu_e \bmk^2$.

\subsubsection{Nonlinear theory}

We introduce the scaling dimensions of parameters to study nonlinear corrections to the linear theory systematically. Let $[A] = \chi_A$ be the scaling dimension of $A$. We use $[\bmr] = -1$ and $[t] = -z$. From Eqs.~\eqref{eq:noisecovll} and \eqref{eq:noisecovee}, $[\bm{\xi}^l] = [\bm{\xi}^e] = \frac{d+z}{2}$. From Eqs.~\eqref{eq:dll} and \eqref{eq:dle},
\begin{align} \label{eq:sdle}
    [\nu_n] = z-2, \quad [D_n] = -d + 2 \chi_n + z - 2, \quad [\nu_e] = z-2, \quad [D_e] = -d + 2 \chi_e + z -2.
\end{align}
From these results, we find that the linear theory, Eqs. \eqref{eq:dll} and \eqref{eq:dle}, is scale invariant for any set of parameters with the scaling dimensions,
\begin{align} \label{eq:dsd}
    z = 2, \quad \chi_n = \chi_e = \frac{d}{2},
\end{align}
at which the scaling dimensions of the parameters in Eq. \eqref{eq:sdle} vanish.

The higher-order fluctuations in the expansion of the entropy density function and the transport coefficients introduce nonlinear correction terms to the linear theory \eqref{eq:dll} and \eqref{eq:dle}. However, the scaling dimensions of the coupling constants for these nonlinear terms are all negative under the diffusive scaling characterized by Eq.~\eqref{eq:dsd}. As a result, the nonlinear corrections do not qualitatively affect the large-scale behavior of the linear theory. Consequently, in the disordered phase, charge and energy transport at large scales are well described by the linear theory, with no anomalous behavior, even in two dimensions. This conclusion is consistent with the numerical simulations reported in Refs. \cite{DellagoPosch1997,DelfiniLepriLivi2005}, which shows that the heat conductivity in the disordered phase of the two-dimensional Hamiltonian XY model converges in the thermodynamic limit.

\subsection{Ordered phase}

\subsubsection{Linear theory}

We next study the ordered phase. We suppose that the entropy density function in the ordered phase can be expanded as
\begin{align}
    s (e,n, \bmv_{\theta}) = s_0 + \frac{\delta e}{T_0} - \frac{(\delta e)^2}{2 C_0 T_0} - \frac{n^2}{2 I_0 T_0} - \frac{\rho_0}{2 T_0} \bmv_{\theta}^2 + \cdots,
\end{align}
where we have used the symmetry $s(e,n,\bmv_{\theta}) = s(e,-n,\bmv_{\theta}) = s(e,n,- \bmv_{\theta})$. The coefficient $\rho_0$ is the spin-wave stiffness, which is positive from the concavity of $s$. The conjugate thermodynamic variables are given by Eqs.~\eqref{eq:beta}, \eqref{eq:mu}, and
\begin{align}
    \bmh_{\theta} = - \beta^{-1} \frac{\partial s}{\partial \bmv_{\theta}} = \rho_0 \bmv_{\theta} + \cdots.
\end{align}
From Eqs.~\eqref{eq:oht}, \eqref{eq:ohc}, and \eqref{eq:ohe}, the linearized fluctuating hydrodynamic equations in the ordered phase are found to be
\begin{align} \label{eq:olt}
    \partial_t \theta &= c_{\rms} n + \nu_{\theta} \triangle \theta - \sqrt{2 D_{\theta}} \xi^{\theta},
    \\ \label{eq:oln}
    \partial_t n &= - \nabla \cdot \left( - d_{\rms} \nabla \theta - \nu_n \nabla n + \sqrt{2 D_n} \bm{\xi}^l \right),
    \\ \label{eq:ole}
    \partial_t e &= - \nabla \cdot \left( - \nu_e \nabla e + \sqrt{2 D_e} \bm{\xi}^e \right).
\end{align}
The phase diffusivity $\nu_{\theta}$, the noise strength $D_{\theta}$, and the reversible linear coupling constants, $c_{\rms}$ and $d_{\rms}$, are defined by
\begin{align} \label{eq:nut}
    \nu_{\theta} \coloneqq \rho_0 \gamma_{\theta 0}, \quad D_{\theta} \coloneqq \gamma_{\theta 0} T_0, \quad c_{\rms} \coloneqq \frac{1}{I_0}, \quad d_{\rms} \coloneqq \rho_0.
\end{align}
In contrast to the linear theory in the disordered phase, there are two damped sound modes and one diffusion mode in Eqs.~\eqref{eq:olt}, \eqref{eq:oln}, and \eqref{eq:ole}, characterized by the dispersion relations,
\begin{align} \label{eq:omegapm}
    \omega_{\pm}(\bmk) = - i \frac{\nu_{\theta} + \nu_n}{2} \bmk^2 \pm \sqrt{c_{\rms} d_{\rms}} |\bmk| \sqrt{1 - \frac{(\nu_{\theta} - \nu_n)^2}{4 c_{\rms} d_{\rms}} \bmk^2},
\end{align}
and $\omega_0 (\bmk) = - i \nu_e \bmk^2$, respectively. The speed of sound reads as $c = \sqrt{c_{\rms} d_{\rms}} \coloneqq \sqrt{ \rho_0 / I_0}$.

The scaling dimensions of the above parameters are
\begin{align*}
    [\nu_{\theta}] = z-2, \quad [D_{\theta}] = -d + 2 \chi_{\theta} + z,
    \\
    [c_{\rms}] = z + \chi_{\theta} - \chi_n, \quad [d_{\rms}] = z -2 + \chi_n - \chi_{\theta}.
\end{align*}
If $c_{\rms} =0$ and $d_{\rms} = 0$, the linear theory is purely diffusive and scale-invariant. This theory is characterized by the scaling dimensions
\begin{align} \label{eq:gsd}
    z = 2, \quad \chi_{\theta} = \frac{d}{2} -1, \quad \chi_n = \chi_e  = \frac{d}{2}.
\end{align}
Note that $c_{\rms}$ and $d_{\rms}$ acquire positive scaling dimensions, $[c_{\rms}] = [d_{\rms}] = 1$, under this diffusive scaling.

\subsubsection{Nonlinear theory}
\label{subsubsec:powercounting}

Hereafter, we use simply $e$ to denote the deviation $\delta e$. There are three marginal second-order nonlinear terms that can be derived from the expansion of Eqs.~\eqref{eq:oht}, \eqref{eq:ohc} and \eqref{eq:ohe}: $\lambda_{\theta} e n$ in the equation of motion for $\theta$, $- \nabla \cdot (- \lambda_n e \nabla \theta)$ in the equation of motion for $n$, and $- \nabla \cdot (- \lambda_e n \nabla \theta)$ in the equation of motion for $e$, where $\lambda_{\theta}$, $\lambda_n$, $\lambda_e$ are coupling constants. Note that these nonlinear terms arise from the second-order expansion of the reversible parts. The first two terms are regarded as fluctuations in the speed of sound caused by temperature fluctuations, and the third term represents an elastic, reversible energy current induced by spin-wave excitations.

These nonlinear terms can also be derived based on symmetry considerations. The reversible terms in the equation of motion for $\theta$ must form a scalar field and be odd under time-reversal transformation. The possible form satisfying these constraints and expressible in terms of $\bmv_{\theta}$, $n$ and $e$ is given by $c_{\rms} n + \lambda_{\theta} e n + \cdots$, where the ellipses denote higher-order corrections. Similarly, the reversible charge current density must be a vector field and even under time-reversal transformation, leading to $\bmj^{n, \rev} = - d_{\rms} \bmv_{\theta} - \lambda_n e \bmv_{\theta} + \cdots$. The reversible energy current density, on the other hand, must be a vector field and odd under time-reversal transformation. Since no linear term satisfies these conditions, the leading-order expression is $\bmj^{e,\rev} = - \lambda_e n \bmv_{\theta} + \cdots$. 

The scaling dimensions of these coupling constants are
\begin{align*}
    [\lambda_{\theta}] = z + \chi_{\theta} - \chi_n - \chi_e, \quad [\lambda_n] = z-2 - \chi_{\theta} + \chi_n - \chi_e, \quad [\lambda_e] = z-2 - \chi_{\theta} - \chi_n + \chi_e.
\end{align*}
In the diffusive scaling given by Eq. \eqref{eq:gsd}, $[\lambda_{\theta}] = [\lambda_n] = [\lambda_e] = 1-d/2$, which takes $0$ in $d=2$ and negative in $d>2$. The dimensions of other nonlinear terms are negative under this scaling in and above two dimensions $d \geq 2$. Therefore, we study the correction to the linear theory caused by these nonlinear terms. As a result, the nonlinear fluctuating hydrodynamic equations we study are given by
\begin{align} \label{eq:oht1}
    \partial_t \theta &= (c_{\rms} + \lambda_{\theta} e)n + \nu_{\theta} \triangle \theta - \sqrt{2 D_{\theta}} \xi^{\theta},
    \\ \label{eq:ohc1}
    \partial_t n &= - \nabla \cdot \left( - (d_{\rms} + \lambda_n e) \nabla \theta - \nu_n \nabla n + \sqrt{2 D_n} \bm{\xi}^n \right),
    \\ \label{eq:ohe1}
    \partial_t e &= - \nabla \cdot \left( - \lambda_e n \nabla \theta - \nu_e \nabla e + \sqrt{2 D_e} \bm{\xi}^e \right).
\end{align}
The upper critical dimension is determined by setting the scaling dimensions of the nonlinear couplings to zero, yielding $d=2$. For $d > 2$, the nonlinear terms are irrelevant, so the large-scale behavior will be described by the linear theory. In contrast, for $d \leq 2$, we expect non-trivial behavior, such as anomalous transport.

\subsubsection{Dynamic renormalization group}
\label{subsec:drg}

We analyze the hydrodynamic model described by Eqs. \eqref{eq:oht1}, \eqref{eq:ohc1}, and \eqref{eq:ohe1} using perturbative renormalization group. Let $\calA [\Phi ; \Lambda, \calP]$ denotes the Martin-Siggia-Rose-Jansen-De Dominicis action \cite{MartinSiggiaRose1973,Janssen1976,DeDominicis1976} for Eqs.~\eqref{eq:oht1}, \eqref{eq:ohc1}, \eqref{eq:ohe1}. Here, $\Phi(k) = [\theta(k), \tilde{\theta}(k), n(k), \tilde{n}(k), e(k), \tilde{e}(k)]^{\mathsf{T}}$ is the set of the Fourier components of fields of slow variables and their response fields, characterized by the wavenumber vector and the frequency, $k = (\bmk, \omega)$. An ultraviolet cutoff $\Lambda$ is introduced, and $\calP = (\nu_{\theta}, D_{\theta}, \nu_n, D_n, \nu_e, D_e, c_{\rms}, d_{\rms}, \lambda_{\theta}, \lambda_n, \lambda_e)$ refers to the set of parameters defined at the scale $\Lambda$. The action $\calA$ is decomposed into the free part $\calA_0$ and the interaction part $\calA_1$. The free part $\calA_0$ is given by
\begin{align*}
    \calA_0 = \frac{1}{2} \int_k \sum_{\alpha, \alpha^{\prime}} \Phi_{\alpha}(-k) [\bmG^{-1}(k ; \Lambda, \calP)]_{\alpha \alpha^{\prime}} \Phi_{\alpha^{\prime}}(k).
\end{align*}
Here, $\int_k \coloneqq \int_{|\bmk| \leq \Lambda} \frac{d^d\bmk}{(2\pi)^d} \int \frac{d\omega}{2 \pi}$ and the summation on $\alpha$ runs over the set of field labels $\{ \theta, \tilt, n, \tiln, e, \tile \}$. The inverse of the propagator $\bmG(k ; \Lambda, \calP)$ is
\begin{align*}
        \bmG^{-1}(k ; \Lambda, \calP) \coloneqq 
        \begin{bmatrix}
            0 & i \omega + \nu_{\theta} \bmk^2 & 0 & d_{\rms} \bmk^2 & 0 & 0 \\
            - i \omega + \nu_{\theta} \bmk^2 & -2 D_{\theta} & - c_{\rms} & 0 & 0 & 0 \\
            0 & - c_{\rms} & 0 & i \omega + \nu_n \bmk^2 & 0 & 0 \\
            d_{\rms} \bmk^2 & 0 & - i \omega + \nu_n \bmk^2 & - 2 D_n \bmk^2 & 0 & 0 \\
            0 & 0 & 0 & 0 & 0 & i \omega + \nu_e \bmk^2 \\
            0 & 0 & 0 & 0 & - i \omega + \nu_e \bmk^2 & -2 D_e \bmk^2
        \end{bmatrix}.
\end{align*}
The components of the propagator are presented in \ref{app:bare}. The interaction part $\calA_1$ has three vertices corresponding to the nonlinear terms in Eqs.~\eqref{eq:oht1}, \eqref{eq:ohc1}, \eqref{eq:ohe1}. The explicit form is
\begin{align*}
    \calA_1 &= \int_{\bmr,t} \left[ - \lambda_{\theta} \tilt e n - \lambda_n \tiln \nabla \cdot (e \nabla \theta) - \lambda_e \tile \nabla \cdot (l \nabla \theta) \right]
    \\
    &= \int_{k_1, k_2, k_3} \hat{\delta} (k_1 + k_2 + k_3) \left[ - \lambda_{\theta} \tilt (k_1) e(k_2) n(k_3) \right.
    \\
    &\qquad \left. - \lambda_n (\bmk_1 \cdot \bmk_2) \tiln (k_1) \theta (k_2) e(k_3) - \lambda_e (\bmk_1 \cdot \bmk_2) \tile (k_1) \theta (k_2) n(k_3)  \right],
\end{align*}
where $\hat{\delta}(k) = (2 \pi)^{d+1} \delta^d (\bmk) \delta (\omega)$.

The dynamic renormalization group is constructed in the following way. Let $\Lambda^{\prime} = \Lambda e^{- \delta l}$ be a new cutoff scale with an infinitesimal RG time increment $\delta l$. We define
\begin{align*}
    A^{<} (k) = \Theta (\Lambda^{\prime} - |\bmk|) A(k), \quad A^{>}(k) = \Theta (|\bmk| - \Lambda^{\prime})
\end{align*}
for any random variable $A$. The effective action at the scale $\Lambda^{\prime}$ is defined through 
\begin{align} \label{eq:defRG}
    e^{- \calA [\Phi^{<} ; \Lambda^{\prime}, \calP^{\prime}]} = \int d[\Phi^{>}] \ \rme^{- \calA [\Phi^{<} + \Phi^{>} ; \Lambda, \calP]}.
\end{align}
For small $\delta l$, the change in the parameter is expressed as
\begin{align*}
    \calP^{\prime} \simeq \calP (1 + \delta l \cdot \delta \calP),
\end{align*}
where $\delta \calP = \delta \calP (\calP)$ is a function of $\calP$. This infinitesimal transformation of the parameter defines the renormalization group transformation from the initial parameter $\calP_0$ to the renormalized parameter at the scale $\Lambda (l) \coloneqq \Lambda \rme^{-l}$,
\begin{align*}
    \calP \mapsto \calP (l)
\end{align*}
with an initial data $\calP (0) = \calP$. The RG equation is expressed as
\begin{align*}
    \frac{d \calP (l)}{dl} = \delta \calP (\calP(l)) \calP(l).
\end{align*}
Within the perturbation theory of momentum-shell RG, the correction to the parameter is calculated using only one-loop diagrams, as contributions from $n$-loop diagrams are of order $O((\delta l)^n)$.

\subsubsection{Detailed balance condition}

We derived the Langevin equations, Eqs.~\eqref{eq:oht}, \eqref{eq:ohc}, and \eqref{eq:ohe}, by imposing the detailed balance condition. Meanwhile, the model described by Eqs.~\eqref{eq:oht1}, \eqref{eq:ohc1}, and \eqref{eq:ohe1}, generally breaks the detailed balance condition for two reasons. First, expanding the equations up to a finite order in deviations from the equilibrium values does not necessarily preserve the detailed balance condition, even if the original equations satisfy the detailed balance condition. Second, in Eqs. \eqref{eq:oht1}, \eqref{eq:ohc1}, and \eqref{eq:ohe1}, the parameters $D_{\theta}$, $\nu_{\theta}$, $D_n$, $\nu_n$, $c_{\rms}$, $d_{\rms}$, $\lambda_{\theta}$, $\lambda_n$, and $\lambda_e$ are treated as independent, even though these parameters are determined by the derivatives of $s$ and the transport coefficients, making these parameters interdependent rather than independent. Note that treating these parameters as independent is convenient for defining the perturbative corrections $\delta \calP$ concisely, as discussed in \ref{subsubsec:potential}.

In order to identify the detailed balanced hydrodynamic equation including only the marginal and relevant terms, we proceed in this paper as follows. First, we retain only the relevant and marginal terms in the RG sense among the terms derived from the expansion and treat all the coupling constants as independent. This procedure is explained in Section \ref{subsubsec:powercounting}. Then, we impose the necessary constraints on these parameters to restore the detailed balance condition.

Let us provide the constraint on the parameters that makes Eqs.~\eqref{eq:oht1}, \eqref{eq:ohc1}, and \eqref{eq:ohe1} detailed balanced. We can prove that Eqs.~\eqref{eq:oht1}, \eqref{eq:ohc1}, and \eqref{eq:ohe1} obey the detailed balance condition with respect to the Gaussian stationary distribution,
\begin{align}
    p_{\mathrm{eq}} [e,n,\theta] \propto \exp \left[ - \int_{\bmr} \left(  \frac{1}{2} \frac{\nu_e}{D_e} e^2 + \frac{1}{2} \frac{\nu_n}{D_n} n^2 + \frac{1}{2} \frac{\nu_{\theta}}{D_{\theta}} (\nabla \theta)^2 \right) \right],
\end{align}
if and only if
\begin{align} \label{eq:db}
    c_{\rms} \frac{\nu_{\theta}}{D_{\theta}} = d_{\rms} \frac{\nu_n}{D_n}, \quad \lambda_{\theta} \frac{\nu_{\theta}}{D_{\theta}} = \lambda_n \frac{\nu_n}{D_n} = \lambda_e \frac{\nu_e}{D_e}.
\end{align}
Indeed, according to Ref. \cite{GrahamHaken1971}, the detailed balance condition for Eqs.~\eqref{eq:oht1}, \eqref{eq:ohc1}, and \eqref{eq:ohe1} is equivalent to
\begin{align*}
    \nu_{\theta} \triangle \theta &= D_{\theta} \frac{\delta}{\delta \theta} (- \ln p_{\mathrm{eq}}),
    \\
    \nabla \cdot (\nu_n \nabla n) &= (- D_n \triangle) \frac{\delta}{\delta n} (- \ln p_{\mathrm{eq}}),
    \\
    \nabla \cdot (\nu_e \nabla e) &= (- D_e \triangle ) \frac{\delta}{\delta e} (- \ln p_{\mathrm{eq}})
\end{align*}
for the irreversible parts and
\begin{align*}
    \int_{\bmr} \left[ (c + \lambda_{\theta} e) n \frac{\delta}{\delta \theta (\bmr)} + \nabla \cdot \left[ (d_{\rms} + \lambda_n e) \nabla \theta \right] \frac{\delta }{\delta n (\bmr)} + \nabla \cdot \left[ \lambda_e n \nabla \theta \right] \frac{\delta}{\delta e (\bmr)} \right] (- \ln p_{\mathrm{eq}}) = 0
\end{align*}
for the reversible parts. The first condition for the irreversible parts is verified without any constraints on parameters. The second condition for the reversible parts is written as
\begin{align*}
    &\int_{\bmr} \left[ \left( c_{\rms} \frac{\nu_{\theta}}{D_{\theta}} - d_{\rms} \frac{\nu_n}{D_n} \right) \nabla n \cdot \nabla \theta + \left( \lambda_{\theta} \frac{\nu_{\theta}}{D_{\theta}} - \lambda_n \frac{\nu_n}{D_n} \right) e \nabla n \cdot \nabla \theta \right.
    \\
    &\qquad \qquad \left. + \left( \lambda_{\theta} \frac{\nu_{\theta}}{D_{\theta}} - \lambda_e \frac{\nu_e}{D_e} \right) n \nabla e \cdot \nabla \theta \right] = 0,
\end{align*}
which is equivalent to Eq.~\eqref{eq:db}. Hereafter, we call the constraint \eqref{eq:db} the detailed balance condition.

In this paper, we assume that the detailed balance condition ~\eqref{eq:db} holds at the initial scale $\Lambda$. By explicitly calculating the one-loop corrections, we show that if Eq.~\eqref{eq:db} is satisfied at the initial scale $\Lambda$,  it remains valid at the new scale $\Lambda^{\prime}$. Thus, the detailed balance condition is preserved along the RG flow within the perturbation theory. In \ref{app:db}, we further confirm this property using the time-reversal symmetry and the Ward-Takahashi identities.

\subsubsection{Dimensionless couplings}
\label{subsubsec:nondim}

There are $11$ parameters in Eqs.~\eqref{eq:oht1}, \eqref{eq:ohc1}, and \eqref{eq:ohe1}. We introduce $7$ independent dimensionless parameters,
\begin{align*}
    &\barnt  = \frac{\nu_{\theta}}{\nu_n}, \quad \barne = \frac{\nu_e}{\nu_n},
    \\
    &\bar{c}_s = \frac{c_{\rms}}{\nu_n \Lambda (l)} \sqrt{\frac{\nu_{\theta}}{D_{\theta}}} \sqrt{\frac{D_n}{\nu_n}}, \quad \bar{d}_s = \frac{d_{\rms}}{\nu_n \Lambda (l)} \sqrt{\frac{D_{\theta}}{\nu_{\theta}}} \sqrt{\frac{\nu_n}{D_n}},
    \\
    &\bar{\lambda}_{\theta} = \frac{\lambda_{\theta} \Lambda(l)^{\frac{d-2}{2}}}{\nu_n} \sqrt{\frac{\nu_{\theta}}{D_{\theta}} \frac{D_n}{\nu_n} \frac{D_e}{\nu_e}} \Od^{1/2},
    \\
    &\bar{\lambda}_l = \frac{\lambda_n \Lambda(l)^{\frac{d-2}{2}}}{\nu_n} \sqrt{\frac{D_{\theta}}{\nu_{\theta}} \frac{\nu_n}{D_n} \frac{D_e}{\nu_e}} \Od^{1/2},
    \\
    &\bar{\lambda}_e = \frac{\lambda_e \Lambda(l)^{\frac{d-2}{2}}}{\nu_n} \sqrt{\frac{D_{\theta}}{\nu_{\theta}} \frac{D_n}{\nu_n} \frac{\nu_e}{D_e}} \Od^{1/2},
\end{align*}
where $\Od$ is the surface area of the $(d-1)$-dimensional unit sphere divided by $(2\pi)^d$. In the above expressions, all the parameters are evaluated at the scale $\Lambda(l) = \Lambda e^{-l}$. The reasoning behind the selection of these dimensionless parameters is discussed in \ref{app:nondim}. If the system is detailed balanced at the scale $\Lambda (l)$, $\barc = \bard$ and $\bart = \barn = \bare$, reducing the number of independent dimensionless parameters to $4$ in detailed balanced system.

The corrections to the dimensionless parameters are given by
\begin{align*}
    \delta \barnt &= \delta \nu_{\theta} - \delta \nu_n,
    \\
    \delta \barne &= \delta \nu_e - \delta \nu_n,
    \\
    \delta \barc &= 1 + \delta c_{\rms} - \delta \nu_n + \frac{1}{2} ( \delta \nu_{\theta} - \delta D_{\theta} - \delta \nu_n + \delta D_n ),
    \\
    \delta \bard &= 1 + \delta d_{\rms} - \delta \nu_n + \frac{1}{2} \left( - \delta \nu_{\theta} + \delta D_{\theta} + \delta \nu_n - \delta D_n \right),
    \\
    \delta \bart &= \frac{2-d}{2} + \delta \lambda_{\theta} - \delta \nu_n + \frac{1}{2} \left( \delta \nu_{\theta} - \delta D_{\theta} - \delta \nu_n + \delta D_n - \delta \nu_e + \delta D_e \right),
    \\
    \delta \barn &= \frac{2-d}{2} + \delta \lambda_n - \delta \nu_n + \frac{1}{2} \left( - \delta \nu_{\theta} + \delta D_{\theta} + \delta \nu_n - \delta D_n - \delta \nu_e + \delta D_e \right),
    \\
    \delta \bare &= \frac{2-d}{2} + \delta \lambda_e - \delta \nu_n + \frac{1}{2} \left( - \delta \nu_{\theta} + \delta D_{\theta} - \delta \nu_n + \delta D_n + \delta \nu_e - \delta D_e \right).
\end{align*}
In \ref{app:db}, we show that the time-reversal symmetry leads to the fluctuation-dissipation relations,
\begin{align} \label{eq:fdt}
    \delta \nu_{\theta} = \delta D_{\theta}, \quad \delta \nu_n = \delta D_n, \quad \delta \nu_e = \delta D_e.
\end{align}
The detailed balance condition \eqref{eq:db}, together with Eq.~\eqref{eq:fdt}, also yields
\begin{align} \label{eq:fdt2}
    \delta c_{\rms} = \delta d_{\rms}, \quad \delta \lambda_{\theta} = \delta \lambda_n = \delta \lambda_e.
\end{align}
We further confirm these properties by explicitly calculating the one-loop corrections. Hence, the corrections of the dimensionless parameters are significantly simplified to
\begin{align*}
    \delta \barnt &= \delta \nu_{\theta} - \delta \nu_n,
    \\
    \delta \barne &= \delta \nu_e - \delta \nu_n,
    \\
    \delta \barc &= \delta \bard = 1 + \delta c_{\rms} - \delta \nu_n,
    \\
    \delta \bart &= \delta \barn = \delta \bare = \frac{2-d}{2} + \delta \lambda_{\theta} - \delta \nu_n.
\end{align*}

\subsection{RG equations}
\label{subsec:RGeq}

The one-loop corrections to the parameters are given by
\begin{align}
    \delta \nu_{\theta} &= \delta D_{\theta} = \bart^2 \frac{\barne + \barnt}{\barnt ((\barnt + \barne)(1 + \barne) + \barc^2)},
    \\
    \delta \nu_n &= \delta D_n = \barn^2 \frac{1}{d} \frac{\barne + 1}{(\barnt + \barne)(1+ \barne) + \barc^2},
    \\ \label{eq:correctnue}
    \delta \nu_e &= \delta D_e = \bare^2 \frac{1}{d} \frac{1}{\barne (\barnt + 1)},
    \\
    \delta c_{\rms} &= \delta d_{\rms} = 0,
    \\
    \delta \lambda_{\theta} &= \delta \lambda_n = \delta \lambda_e = 0.
\end{align}
The details of calculations are presented in \ref{app:loop}. The detailed balance conditions \eqref{eq:fdt} and \eqref{eq:fdt2} are actually satisfied. The RG equations become
\begin{align} \label{eq:RGt}
    \frac{d \barnt}{dl} &= \left( \frac{\barne + \barnt}{(\barnt + \barne)(1 + \barne) + \barc^2} - \frac{1}{d} \frac{\barnt (1 + \barne)}{(\barnt + \barne)(1 + \barne) + \barc^2} \right) \barlam^2,
    \\ \label{eq:RGe}
    \frac{d \barne}{dl} &= \left( \frac{1}{d} \frac{1}{1 + \barnt} - \frac{1}{d} \frac{\barne (1 + \barne)}{(\barnt + \barne)(1+ \barne) + \barc^2} \right) \barlam^2,
    \\ \label{eq:RGc}
    \frac{d \barc}{dl} &= \left( 1 - \frac{\barlam^2}{d} \frac{\barne + 1}{(\barnt + \barne)(1 + \barne) + \barc^2} \right) \barc,
    \\ \label{eq:RGlam}
    \frac{d \barlam}{dl} &= \frac{2-d}{2} \barlam - \frac{1}{d} \frac{1 + \barne}{(\barnt + \barne)(1 + \barne) + \barc^2} \barlam^3,
\end{align}
where $\barlam \coloneqq \bart = \barn = \bare$. 

Before analyzing the RG equations, we recall the assumptions on the parameters. First, the diffusivities $\nu$ and noise strengths $D$ are assumed to be positive, ensuring the positivity of the entropy production rate and the concavity of the entropy density function $s$. Similarly, $c_{\rms}$ and $d_{\rms}$ are also positive, following from the concavity of $s$. As a result, $\barnt$, $\barne$, $\barc$ are positive. The nonlinear coupling constants $\lambda_{\theta}$, $\lambda_n$, and $\lambda_e$ are assumed to be small to justify the perturbative calculations. However, their signs are not physically significant. This can be seen from the RG equations ~\eqref{eq:RGt}, \eqref{eq:RGe}, \eqref{eq:RGc} and \eqref{eq:RGlam}. If the sign of $\barlam$ is inverted, or the signs of $\lambda_{\theta}$, $\lambda_n$ and $\lambda_e$ are simultaneously inverted, the RG equations remain unchanged. Thus, the physical properties do not depend on the sign of $\barlam$. Hereafter, we assume $\barlam > 0$ at $l=0$ without loss of generality.

\subsection{Properties of the RG solution with finite speed of sound}
\label{subsec:finitec}

The behavior of the RG flow differs depending on whether sound waves are present or absent. In the Hamiltonian XY model introduced in Section \ref{subsec:micro}, the speed of sound at temperatures close to zero is given by $c \approx a \sqrt{J / I} > 0$, which follows from the harmonic expansion of $V$. Therefore, the systems with finite speed of sound are of particular interest. We begin by analyzing the case where the speed of sound is finite at the initial scale.

If the initial value of the dimensionless speed of sound $\barc (0)$ is positive, $\barc (l)$ grows exponentially in $l$. To handle this, it is useful to introduce
\begin{align}
    R(l) \coloneqq \frac{1}{\barc} = \frac{\nu_n (l) \Lambda(l)}{c}.
\end{align}
Note that the speed of sound $c = \sqrt{c_s d_s}$ is not renormalized, as $\delta c_{\rms} = \delta d_{\rms} = 0$. The parameter $R$ acts as a length scale, determining whether phase and charge fluctuations are underdamped or not. When $R \gg 1$, many modes with the wavenumber $\bmk$ satisfy $c |\bmk| < \nu_n \bmk^2$, indicating they are overdamped. Conversely, when $R \ll 1$, all overdamped modes have already been eliminated from the dynamical degrees of freedom, leaving only underdamped fluctuations.

The RG equation for $R$ is
\begin{align} \label{eq:RGR}
    \frac{d R}{dl} = \left( -1 + \frac{\barlam^2}{d} \frac{(1 + \barne ) R^2}{1 + (\barnt + \barne)(1 + \barne) R^2} \right) R.
\end{align}
The remaining RG equations are expressed in terms of $R$ instead of $\barc$ as
\begin{align} \label{eq:RGt1}
    \frac{d \barnt}{dl} &= \left( \frac{(\barne + \barnt) R^2}{1 + (\barnt + \barne)(1 + \barne) R^2} - \frac{1}{d} \frac{\barnt ( \barne + 1) R^2}{1 + (\barnt + \barne)(1+ \barne) R^2} \right) \barlam^2,
    \\ \label{eq:RGe1}
    \frac{d \barne}{dl} &= \left( \frac{1}{d} \frac{1}{1 + \barnt } - \frac{1}{d} \frac{\barne (1 + \barne ) R^2}{1 + (\barnt + \barne)(1 + \barne) R^2} \right) \barlam^2,
    \\ \label{eq:RGlam1}
    \frac{d \barlam}{dl} &= \frac{2-d}{2} \barlam - \frac{1}{d} \frac{(1 + \barne) R^2}{1 + (\barnt + \barne)(1 + \barne) R^2} \barlam^3.
\end{align}
The RG equations for the renormalized diffusivities are given by
\begin{align} \label{eq:RGt0}
    \frac{d \nu_{\theta}}{dl} &= \frac{(\barnt + \barne) R^2}{\barnt (1 + (\barnt + \barne)(1 + \barne) R^2)} \barlam^2 \nu_{\theta},
    \\ \label{eq:RGn0}
    \frac{d \nu_n}{dl} &= \frac{1}{d} \frac{(1 + \barne) R^2}{1 + (\barnt + \barne)(1 + \barne) R^2} \barlam^2 \nu_n,
    \\ \label{eq:RGe0}
    \frac{d \nu_e}{dl} &= \frac{1}{d} \frac{1}{\barne (1 + \barnt)} \barlam^2 \nu_e.
\end{align}
The scale-dependence of the renormalized diffusivities is thus determined from Eqs.~\eqref{eq:RGt0}, \eqref{eq:RGn0}, and \eqref{eq:RGe0}, together with the solution of the RG equations for the dimensionless parameters.

\begin{figure}[h]
    \centering
    \includegraphics[width=16cm]{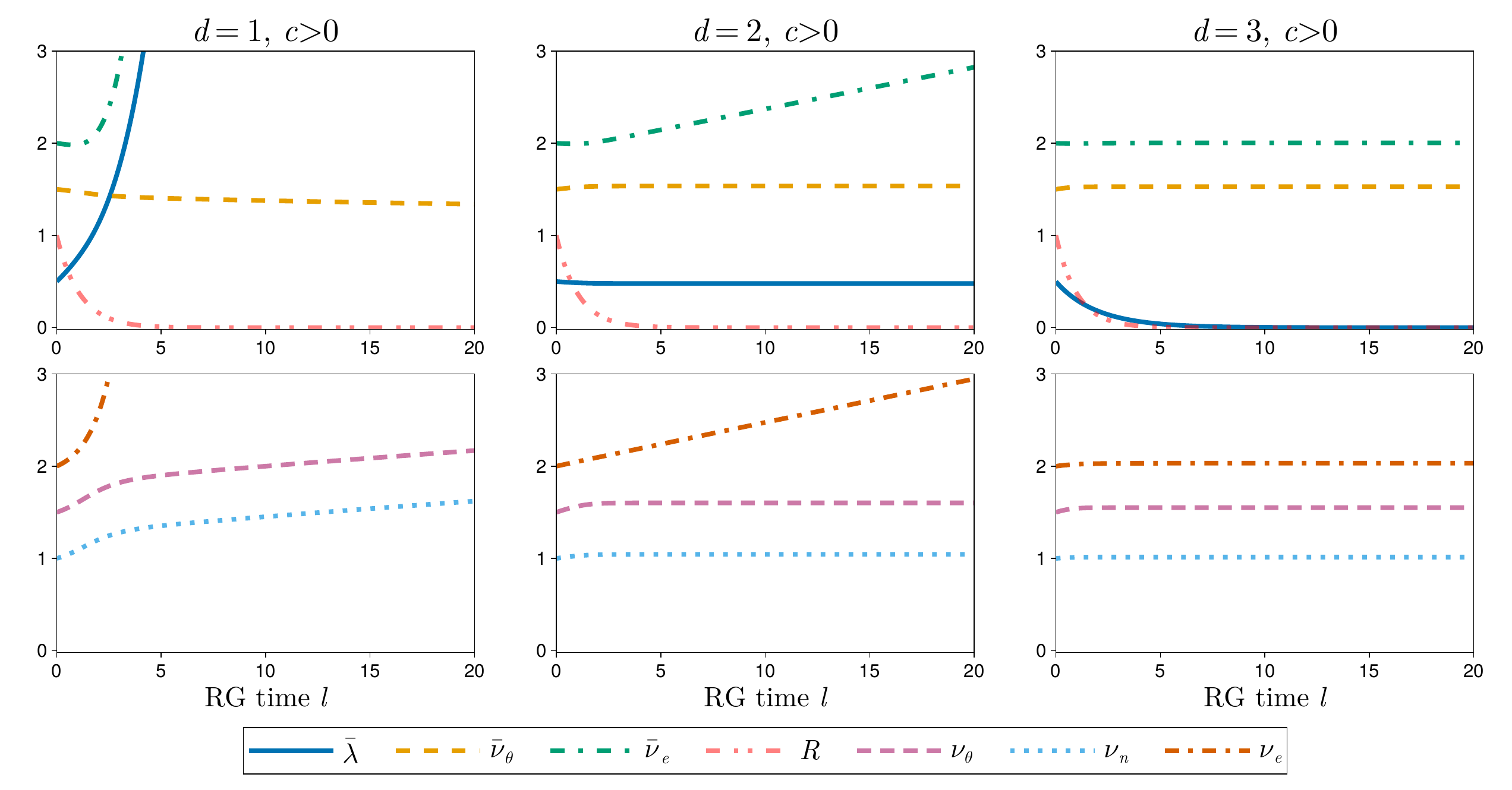}
    \caption{RG flows in the presence of sound waves. The initial values of the parameters are $\barlam = 0.5$, $\barnt = 1.5$, $\barne = 2.0$, $\nu_{\theta} = 1.5$, $\nu_n = 1.0$, $\nu_e = 2.0$, and $R = 1.0$.}
    \label{fig:RGiter}
\end{figure}

For our model, the behavior of the renormalized parameters differs significantly above and below two dimensions. Above two dimensions $d > 2$, both $R$ and $\barlam$ decay exponentially. As the second terms of the right-hand sides of Eqs.~\eqref{eq:RGR} and \eqref{eq:RGlam1} become much smaller than the first terms, the asymptotic behavior of $R$ and $\barlam$ is well described by
\begin{align}
    R(l) \sim \rme^{-l}, \quad \barlam (l) \sim \rme^{- \frac{d-2}{2} l}
\end{align}
for $l \gtrsim 1$. The right-hand sides of Eqs.~\eqref{eq:RGt1} and \eqref{eq:RGe1} also decay rapidly, leading $\barnt$ and $\barne$ to converge to finite values,
\begin{align}
    \barnt (l) \simeq \barnt (\infty), \quad \barne (l) \simeq \barne (\infty).
\end{align}
The asymptotic values, $\barnt (\infty)$ and $\barne (\infty)$, are determined by solving the RG equations. The trajectories obtained from numerical integration of the RG equations for $d=3$ are shown in Fig. \ref{fig:RGiter}.

In exactly two dimensions $d=2$, the RG equation for $\barlam$ becomes
\begin{align} \label{eq:RGlamd2}
    \frac{d \barlam}{dl} = - \frac{1}{d} \frac{(1 + \barne) R^2}{1 + (\barnt + \barne)(1 + \barne) R^2} \barlam^3.
\end{align}
Since the right-hand side is negative, $\barlam (l)$ remains small if $\barlam (0)$ is initially small. Therefore, the perturbative RG analysis is applicable to this case as long as the nonlinear couplings are small at the initial scale. To examine the asymptotic behavior in $d=2$, we assume that the growth of $\barne$ and $\barnt$ is much slower than the exponential decay of $R$, $\barnt \barne R^2 \ll 1$ and $\barne^2 R^2 \ll 1$. Under this assumption, the right-hand side of Eq.~\eqref{eq:RGlamd2} decays rapidly, leading $\barlam(l)$ to converges to a finite value $\barlam (\infty)$, which depends on the initial condition. A similar argument shows that $\barnt (l)$ converges quickly to a finite value $\barnt (\infty)$. Thus, the RG equation for $\barne$ can be asymptotically approximated by
\begin{align}
    \frac{d \barne}{dl} \simeq \frac{1}{d} \frac{\barlam(\infty)^2}{1 + \barnt (\infty)}
\end{align}
for $l \gtrsim l_0 \sim 1$, resulting in linear growth for $\barne$:
\begin{align}
    \barne (l) \simeq \barne (l_0) + \frac{1}{d} \frac{\barlam(\infty)^2}{1 + \barnt (\infty)} (l-l_0)
\end{align}
This result is consistent with the assumptions that $\barnt \barne R^2 \ll 1$ and $\barne^2 R^2 \ll 1$. From these findings, we obtain
\begin{align}
    \frac{d \nu_{\theta}}{dl} \simeq 0, \quad \frac{d \nu_n}{dl} \simeq 0, \quad \frac{d \nu_e}{dl} \simeq \frac{1}{d} \frac{\barlam (\infty)^2}{1 + \barnt (\infty)} \frac{1}{\barne (l_0) + \frac{1}{d} \frac{\barlam (\infty)^2}{1 + \barnt (\infty)} (l-l_0)} \nu_e
\end{align}
for $l \gtrsim 1$. The first two equations indicate that $\nu_{\theta}(l) \sim \nu_{\theta}(\infty)$ and $\nu_n (l) \sim \nu_n (\infty)$. The final equation suggests that the energy diffusivity diverges as
\begin{align} \label{eq:2dnue}
    \nu_e (l) \simeq \nu_e (l_0) \left( 1 + \frac{1}{d} \frac{\barlam (\infty)^2}{1 + \barnt (\infty)} \frac{1}{\bar{\nu}_e (l_0)} (l-l_0) \right) \sim l
\end{align}
where the last asymptotics holds when $l \gg 1$. The numerical solution in Fig. \ref{fig:RGiter} clearly shows the linear growth of $\barne (l)$ and $\nu_e (l)$.

We note a key difference from the case of isothermal incompressible fluids. In such fluids, the dimensionless nonlinear coupling that quantifies the strength of the convective term decays to zero as a power law in two dimensions \cite{ForsterNelsonStephen1977}. This much slower decay than exponential decay leads to the divergence of the shear viscosity. In contrast, in our model, the presence of the sound waves causes the nonlinear couplings to remain finite even in two dimensions due to the rapid increase in the dimensionless speed of sound $\barc$. Nevertheless, while the corrections to the phase and charge diffusivities decay rapidly as a result of this increase and thus $\nu_{\theta}(l)$ and $\nu_n(l)$ remain finite as $l \to \infty$, the correction to the energy diffusivity in Eq.~\eqref{eq:correctnue} is not affected by the exponential growth of $\barc$, resulting in the linear growth of $\nu_e (l)$ at large $l$.

Our model was derived under the assumption that the phase field, or its gradient, can be treated as a slow degree of freedom at low temperatures in and above two dimensions. Therefore, strictly speaking, our model cannot be directly applied to one-dimensional systems. Nevertheless, the resulting continuum model remains well-defined in one dimension. With this in mind, we explore the RG equations below the critical dimension $d=2$. Below two dimensions $d < 2$, $\barlam$ will grow exponentially. For $1 < d < 2$, the estimations
\begin{align}
    \barlam \sim \rme^{\frac{2-d}{2} l}, \quad R \sim \rme^{-l}, \quad \barnt \simeq \barnt (\infty), \quad \barne \sim \rme^{(2-d) l}
\end{align}
are consistent with the RG equations. Under these estimations, $(\barnt + \barne)(1 + \barne) R^2 \sim e^{2(1-d) l} \ll 1$, and the ratios of the second terms to the first terms on the right-hand sides of Eqs.~\eqref{eq:RGR} and \eqref{eq:RGlam1} are
\begin{align}
    \barlam^2 \barne R^2 \sim \rme^{2(1-d) l} \ll 1, \quad \barne R^2 \barlam^2 \sim \rme^{(1-d) l} \ll 1,
\end{align}
respectively. Even in the case of $d=1$, numerical observations still show exponential growth of $\barlam$ and $\barne$, as well as exponential decay of $R$, though the growth rates slightly deviate from the earlier estimations and $\barnt (l)$ may decay to zero slowly. In any case, the exponential growth of $\barlam$ signals the breakdown of the perturbative RG calculation. Thus, our perturbative RG analysis fails to capture the large-scale behavior below the critical dimensions $d=2$. 

\subsection{Properties of the RG solution with zero speed of sound}
\label{subsec:zeroc}

In this section, we examine the RG solutions in the absence of sound waves. These solutions are obtained by fine-tuning the speed of sound at the initial scale to $\barc (0) = 0$. As previously mentioned, the large-scale behavior at sufficiently low temperatures is governed by the RG equations with sound waves. Even if the speed of sound $c$ is close to zero at higher temperatures, the limiting behavior as $l \to \infty$ will still be described by the RG equations with sound waves, as $\barc$ grows exponentially with $l$. However, the initial transient behavior can be approximated by the RG flow on the subspace $\barc = 0$. This suggests that finite-size systems with a sufficiently small speed of sound can be effectively described by the RG solution with zero speed of sound. This crossover phenomena is discussed in Section \ref{subsubsec:crossover}.

The RG equations in Eqs.~\eqref{eq:RGt}, \eqref{eq:RGe} and \eqref{eq:RGlam} with $\barc = 0$ reduce to
\begin{align} \label{eq:RGn2}
    \frac{d \barnt}{dl} &= \left( \frac{1}{1 + \barne} - \frac{1}{d} \frac{\barnt}{\barnt + \barne} \right) \barlam^2,
    \\ \label{eq:RGe2}
    \frac{d \barne}{dl} &= \left( \frac{1}{d} \frac{1}{1 + \barnt} - \frac{1}{d} \frac{\barne}{\barnt + \barne} \right) \barlam^2,
    \\ \label{eq:RGlam2}
    \frac{d \barlam}{dl} &= \frac{\ep}{2} \barlam - \frac{1}{d} \frac{1}{\barnt + \barne} \barlam^3,
\end{align}
where $\ep = 2-d$. 

Above two dimensions $d > 2$, $\barlam$ decay exponentially as $\barlam (l) \sim e^{- \frac{d-2}{2} l}$. As the right-hand sides of Eqs.~\eqref{eq:RGn2} and \eqref{eq:RGe2} decay rapidly, $\barnt (l)$ and $\barne (l)$ converge to finite values, $\barnt (l) \simeq \barnt (\infty), \quad \barne (l) \simeq \barne (\infty)$ for $l \gtrsim \frac{1}{d-2}$. Therefore, $\nu_{\theta}$, $\nu_l$, and $\nu_e$ converge to finite values rapidly in $l \to \infty$.

\begin{figure}[h]
    \centering
    \includegraphics[width=16cm]{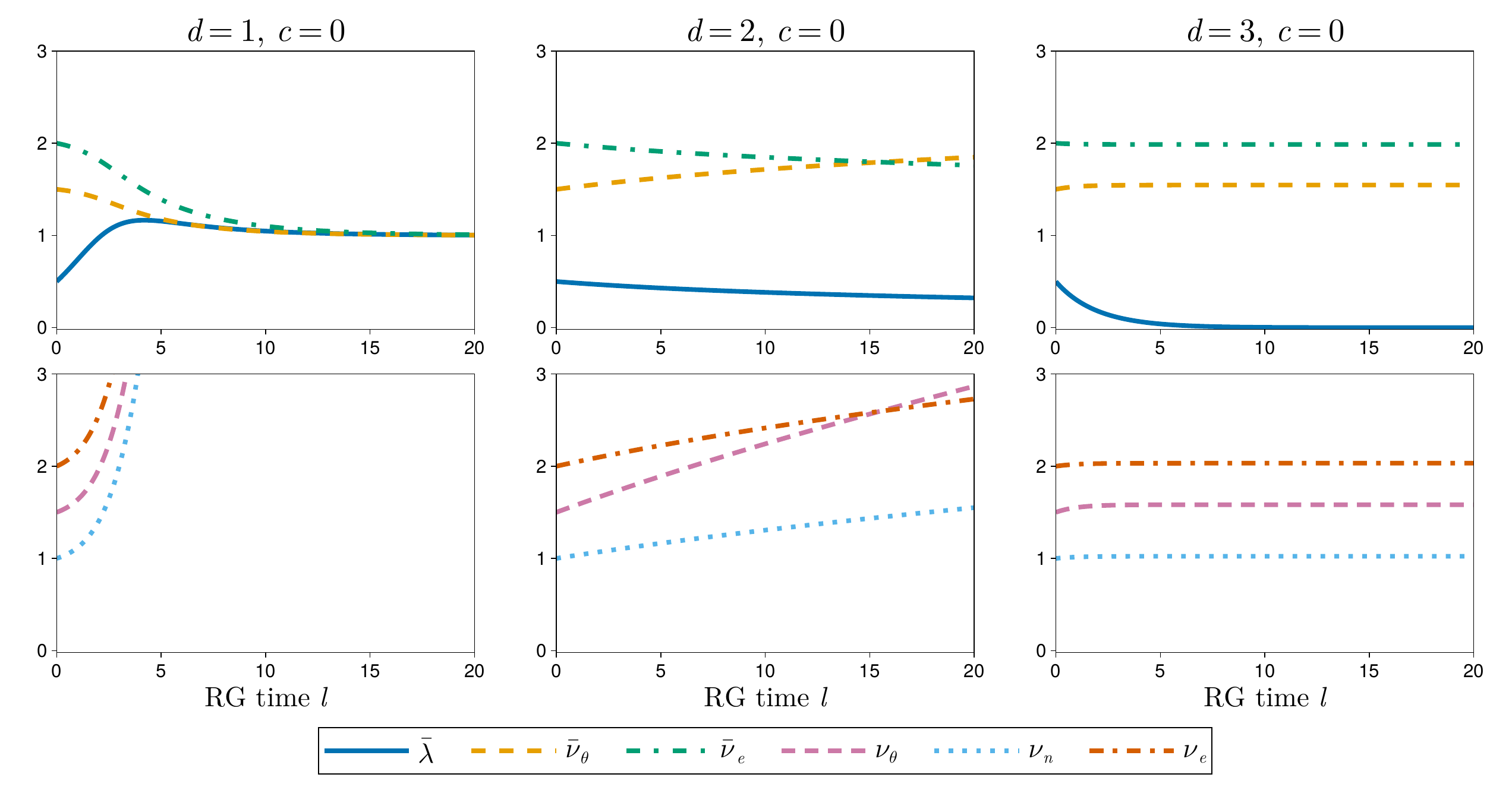}
    \caption{RG flows in the abcense of sound waves. The initial values of the parameters are $\barlam = 0.5$, $\barnt = 1.5$, $\barne = 2.0$, $\nu_{\theta} = 1.5$, $\nu_n = 1.0$, and $\nu_e = 2.0$.}
    \label{fig:RGiterc}
\end{figure}

In exactly two dimensions $d=2$, numerical integration of Eqs.~\eqref{eq:RGn2}, \eqref{eq:RGe2} and \eqref{eq:RGlam2} suggests that $\barlam$ decreases monotonically with $l$, while $\barnt$ grows in $l$ for the range of initial values. An example of the RG flow for an initial condition is presented in Fig. \ref{fig:RGiterc}.

To analyze this behavior, we introduce $u = - \ln \barlam$ and rewrite Eqs.~\eqref{eq:RGn2}, \eqref{eq:RGe2} and \eqref{eq:RGlam2} as
\begin{align} \label{eq:d2c0nt}
    \frac{d \barnt}{d u} &= 2 \frac{\barnt + \barne}{1 + \barne} - \barnt,
    \\ \label{eq:d2c0ne}
    \frac{d \barne}{d u} &= \frac{\barnt}{1 + \barnt} (1 - \barne).
\end{align}
Note that $u \to \infty$ as $\barlam \to 0$. Since $\barnt > 0$, Eq.~\eqref{eq:d2c0ne} implies $\barne \to 1$ in the limit $u \to \infty$. In this asymptotic regime, Eq.~\eqref{eq:d2c0nt} becomes
\begin{align}
    \frac{d \barnt}{d u} \simeq 1,
\end{align}
leading to $\barnt \simeq u$ as $u \to \infty$. Then, Eq.~\eqref{eq:RGlam2} can be written as
\begin{align}
    \frac{d u}{dl} \simeq \frac{\rme^{-2 u}}{2 u}.
\end{align}
The solution is given by
\begin{align}
    u (l) = \frac{1}{2} \left( 1 + W \left( \frac{2 (l+C)}{\rme} \right) \right),
\end{align}
or
\begin{align}
    \barlam (l) = \left[ \frac{1}{2(l+C)} W \left( \frac{2 (l+C)}{\rme} \right) \right]^{\frac{1}{2}},
\end{align}
where $W$ is the Lambert W function and $C$ is an integration constant. In the limit $l \to \infty$, the nonlinear coupling $\barlam$ exhibits a very slow decay as
\begin{align}
    \barlam (l) \simeq \left[ \frac{\ln l}{2 l} \right]^{\frac{1}{2}}.
\end{align}
From these asymptotic properties, we obtain
\begin{align}
    \frac{d \nu_{\theta}}{dl} \simeq \frac{\nu_{\theta}}{2 l}, \quad \frac{d \nu_n}{dl} \simeq \frac{\nu_n}{2 l}, \quad \frac{d \nu_e}{dl} \simeq \frac{\nu_e}{2 l}.
\end{align}
As a result, the diffusivities follow a power-law growth in $l$,
\begin{align} \label{eq:d2c0power}
    \nu_{\theta} \sim l^{\frac{1}{2}}, \quad \nu_{n} \sim l^{\frac{1}{2}}, \quad \nu_{e} \sim l^{\frac{1}{2}}.
\end{align}
However, numerical simulations suggest that convergence to this asymptotic regime is very slow. Therefore, observing this behavior in numerical simulations of microscopic models or experiments may be challenging.

Below two dimensions $d < 2$, we find a stable fixed point by setting the right-hand sides of Eqs.~\eqref{eq:RGn2}, \eqref{eq:RGe2} and \eqref{eq:RGlam2} to zero,
\begin{align} \label{eq:fixedpt}
    (\barnt^{*}, \barne^{*}, \barlam^{*}) = \left( \frac{d}{\ep}, 1 , \sqrt{d} \right).
\end{align}
The asymptotic convergence to this fixed point for $d=1$ is indeed observed in the numerical integration of the RG equations (Fig. \ref{fig:RGiterc}). Note that the dimensionless parameters reach to this fixed point if the dimensionless speed of sound $\barc$ is fine-tuned to zero, whereas it is unstable along the direction of $\barc$ \cite{Sato2020}. In the $\ep$ expansion, the fixed point value of the nonlinear coupling should be of order $O(\ep)$ for the perturbative RG calculation to be applicable just below the critical dimension. However, the fixed point value $\barlam^*$ in Eq.~\eqref{eq:fixedpt} is $\sqrt{d}$, which remains finite even in the limit $\ep \to 0$. As a result, the reliability of this fixed point (Eq.~\eqref{eq:fixedpt}) is uncertain. For now, we leave this issue for future investigation and proceed with analyzing the RG solutions around this fixed point.

At this fixed point,
\begin{align}
    \delta \nu_{\theta}^* = \delta \nu_n^* = \delta \nu_e^* = \frac{2-d}{2}.
\end{align}
Therefore, the diffusivities diverge as
\begin{align}
    \nu_{\theta}(l) \sim e^{\frac{2-d}{2} l}, \quad \nu_{n}(l) \sim e^{\frac{2-d}{2} l}, \quad \nu_{e}(l) \sim e^{\frac{2-d}{2} l}
\end{align}
and their ratios approach to universal values,
\begin{align}
    \frac{\nu_{\theta}(l)}{\nu_n(l)} \to \frac{d}{\ep}, \quad \frac{\nu_e (l)}{\nu_n (l)} \to 1
\end{align}
as $l \to \infty$. The dynamic exponent $z$ and the scaling dimensions of the fields $\chi_{\Phi_{\alpha}}$ at this non-Gaussian fixed point are determined by the equations $z-2 + \delta \nu_{\alpha}^* = 0$, $-d + 2 \chi_{\theta} + z + \delta D_{\theta}^* = 0$, $-d + 2 \chi_n + z - 2 + \delta D_n^* = 0$, and $-d + 2 \chi_e + z-2 + \delta D_e^* = 0$, yielding
\begin{align}
    z = 1 + \frac{d}{2}, \quad \chi_{\theta} = \frac{d-2}{2}, \quad \chi_n = \chi_e = \frac{d}{2}.
\end{align}
This non-trivial dynamic exponent was previously found in the literature on anomalous transport \cite{ForsterNelsonStephen1977,NarayanRamaswamy2002}. Note that these scaling dimensions of the fields are consistent with the Gaussianity of the stationary distribution.

\subsection{Renormalized diffusivity}
\label{subsec:scaling}

Consider a two-point function
\begin{align}
    C_{\alpha \alpha^{\prime}}(\bmk, \omega ; \Lambda, \calP) \hat{\delta}(k+k^{\prime}) \coloneqq \langle \Phi_{\alpha} (\bmk, \omega) \Phi_{\alpha^{\prime}} (\bmk^{\prime}, \omega^{\prime}) \rangle_{\calA[\Lambda, \calP]},
\end{align}
where the bracket $\langle \cdot \rangle_{\calA [\Lambda, \calP]}$ denotes the ensemble average with respect to the action $\calA [\cdot ; \Lambda, \calP]$. From Eqs.~\eqref{eq:defRG}, we obtain
\begin{align}
    C_{\alpha \alpha^{\prime}} (\bmk, \omega ; \Lambda_0, \calP_0) = C_{\alpha \alpha^{\prime}} (\bmk, \omega ; \Lambda (l), \calP (l)).
\end{align}
For $(\alpha, \alpha^{\prime}) = (e,e)$, this becomes
\begin{align} \label{eq:sce}
    C_{ee}(\bmk, \omega ; \Lambda, \calP) = C_{ee} (\bmk, \omega ; \Lambda (l), \calP (l)).
\end{align}
We define the wavenumber- and frequency-dependent energy diffusivity $\nu_{e \rmR}$ and noise strength $D_{e \rmR}$ as
\begin{align} \label{eq:wfd}
    C_{ee} (\bmk, \omega ; \Lambda, \calP) = \frac{2 D_{e \rmR}(\bmk, \omega ; \Lambda, \calP) \bmk^2}{\omega^2 + (\nu_{e \rmR}(\bmk, \omega; \Lambda, \calP) \bmk^2)^2}.
\end{align}
From Eqs.~\eqref{eq:sce} and \eqref{eq:wfd}, we read the relation for the energy diffusivity and the noise strength,
\begin{align} \label{eq:nueR}
    \nu_{e \rmR} (\bmk, \omega ;\Lambda, \calP) &= \nu_{e \rmR}(\bmk, \omega ; \Lambda (l), \calP (l))
    \\ \label{eq:DeR}
    D_{e \rmR} (\bmk, \omega ; \Lambda, \calP) &= D_{e \rmR}(\bmk, \omega ; \Lambda (l), \calP (l)).
\end{align}
Similarly, the wavenumber- and frequency-dependent phase and charge diffusivities and noise strengths can be defined by replacing the initial parameters, $\nu_{\theta}$, $\nu_n$, $D_{\theta}$, and $D_n$, in the propagators in \ref{app:bare} with $\nu_{\theta \rmR}(\bmk, \omega)$, $\nu_{n \rmR}(\bmk, \omega)$, $D_{\theta \rmR}(\bmk, \omega)$, and $D_{n \rmR}(\bmk, \omega)$. Since $c_{\rms}$ and $d_{\rms}$ are not renormalized, we find the same relations for these parameters as in Eqs.~\eqref{eq:nueR} and \eqref{eq:DeR}.

Applying a dimensional analysis, we can write
\begin{align}
    \nu_{e \rmR} (\bmk, \omega ; \Lambda, \calP) &= \nu_e \Psi_{\nu_e} \left( \frac{\bmk}{\Lambda}, \frac{\omega}{\nu_n \Lambda^2} ; \barnt, \barne, R, \barlam \right)
\end{align}
for a dimensionless function $\Psi_{\nu_e}$. Combining this with Eq.~\eqref{eq:nueR}, we obtain
\begin{align} \label{eq:homnue}
    \nu_{e \rmR} (\bmk, \omega; \Lambda, \calP) = \nu_e (l) \Psi_{\nu_e} \left( \frac{\bmk}{\Lambda (l)}, \frac{\omega}{\nu_n(l) \Lambda(l)^2} ; \barnt (l), \barne (l), R (l), \barlam (l) \right).
\end{align}
The large-distance and long-time behavior of the wavenumber- and frequency-dependent diffusivity follows from this relation. Similar relations hold for $\nu_{\theta \rmR}$ and $\nu_{n \rmR}$.

\subsubsection{Above two dimensions $d > 2$ with $c > 0$}
\label{subsubsec:abovec}

A key observation is that if $\barlam$ is sufficiently small and both $\bmk$ and $\omega$ are not close to zero, the wavenumber- and frequency-dependent diffusivity can be approximated by that of the linear theory, with only small corrections from the nonlinear terms. This implies that $\Psi_{\nu_e} = 1 + \rmO(\barlam^2)$ for small $\barlam$, as the correction to the energy diffusivity in the linear theory is of order $\rmO (\barlam^2)$.

We can freely choose the RG time $l$ in Eq.~\eqref{eq:homnue}. For $\bmk = \bm{0}$, a convenient choice is
\begin{align} \label{eq:somega}
    \frac{\omega}{\nu_n (l_*) \Lambda (l_*)^2} = 1.
\end{align}
Note that while the combination $\omega / (\nu_n (l_*) \Lambda (l_*)^2)$ remains fixed at unity, $l_* \to \infty$ as $\omega \to 0$. More explicitly,
\begin{align}
    l_* \simeq \frac{1}{2} \ln \left[ \frac{\nu_n (\infty) \Lambda^2}{\omega} \right]
\end{align}
in the limit $\omega \to 0$. Since $\barlam (l_*) \sim \rme^{- \frac{d-2}{2}l_*}$ and $\nu_e (l_*) \simeq \nu_e (\infty) (1 + \rmO(\barlam (l_*)^2))$ as $l_* \to \infty$ for $d > 2$ with $c > 0$, it follows from Eq.~\eqref{eq:homnue} that
\begin{align}
    \nu_{e \rmR} (\bm{0}, \omega) \simeq \nu_e (\infty) + \rmO \left( \omega^{\frac{d-2}{2}} \right)
\end{align}
in the low-frequency limit $\omega \to 0$, which is consistent with the expected long-time tail behavior. For the phase and charge diffusivities, we observe a faster decay in the low-frequency limit,
\begin{align}
    \nu_{\theta \rmR} (\bm{0}, \omega) \simeq \nu_{\theta}(\infty) + \rmO \left( \omega^{\frac{d}{2}} \right), \quad \nu_{n \rmR} (\bm{0}, \omega) \simeq \nu_n (\infty) + \rmO \left( \omega^{\frac{d}{2}} \right)
\end{align}
in the limit $\omega \to 0$, as the corrections to the phase and charge diffusivities are of order $\rmO (R^2 \barlam^2)$.

\subsubsection{In exactly two dimensions $d =2$ with $c > 0$}

As previously mentioned, $\barlam (l)$ remains small as long as $\barlam (0)$ is small. Consequently, we find
\begin{align}
    \nu_{e \rmR}(\bm{0}, \omega) \simeq \nu_e (l_*) \sim \ln \left( \frac{1}{\omega} \right)
\end{align}
as $\omega \to 0$, where $l_*$ is defined in Eq.~\eqref{eq:somega}. The asymptotic behavior of the size-dependent energy diffusivity $\nu_e (L)$ is determined by the frequency-dependent energy diffusivity $\nu_{e \rmR} (\bm{0}, \omega)$ evaluated at the characteristic frequency $\omega = c / L$,
\begin{align} \label{eq:logdiv}
    \nu_e (L) \sim \ln L.
\end{align}
This procedure corresponds to the conventional prescription where the temporal integration in the Green-Kubo formula is truncated at the characteristic time scale $L/c$ to obtain size-dependent transport coefficients \cite{Dhar2008}.

The previous numerical study \cite{DellagoPosch1997} of the modified XY model reports that the autocorrelation function of the energy current decays very slowly in the BKT phase. Later, a more detailed analysis \cite{DelfiniLepriLivi2005} for the XY model in the BKT phase quantitatively confirms that the heat conductivity exhibits a logarithmic divergence. The logarithmic divergence of the size-dependent heat conductivity found in this paper, Eq. \eqref{eq:logdiv}, is consistent with these previous numerical studies. Rigorously speaking, distinguishing between different types of divergence, such as $\ln L$ and $(\ln L)^{\alpha}$ ($\alpha > 0$), based on the existing data may be challenging. Notably, a divergence of the form $(\ln L)^{\frac{1}{2}}$ actually appears in the size-dependent shear viscosity of a two-dimensional isothermal fluid \cite{ForsterNelsonStephen1977}. The present work provides additional theoretical support for the logarithmic divergence of heat conductivity.

The phase and charge diffusivities converge to finite values in the low-frequency limit, indicating normal transport behavior for the phase and charge density fields even in two dimensions. However, unlike in the case of $d > 2$, where simple power-law corrections are observed, we find logarithmic corrections,
\begin{align}
    \nu_{\theta}(\bm{0}, \omega) \simeq \nu_{\theta} (\infty) + \rmO \left( \omega \ln \left( \frac{1}{\omega} \right) \right), \quad \nu_{n}(\bm{0}, \omega) \simeq \nu_{n} (\infty) + \rmO \left( \omega \ln \left( \frac{1}{\omega} \right) \right).
\end{align}

\subsubsection{Zero speed of sound $c=0$ cases}

Above two dimensions with $c = 0$, the results are similar to those in Section \ref{subsubsec:abovec}. The only difference is that the corrections to the phase and charge diffusivities in the low-frequency limit are $\omega^{\frac{d-2}{2}}$ instead of $\omega^{\frac{d}{2}}$, due to the absence of sound waves.

In exactly two dimensions with $c=0$, the power-law divergence of $\nu_{\theta}(l)$, $\nu_n(l)$, and $\nu_e(l)$ in the asymptotic regime discussed in Section \ref{subsec:zeroc} imply a logarithmic divergence of $\nu_{\theta \rmR}(\bm{0}, \omega)$, $\nu_{n \rmR}(\bm{0}, \omega)$ and $\nu_{e \rmR} (\bm{0}, \omega)$ in the low-frequency limit,
\begin{align}
    \nu_{\theta}(\bm{0}, \omega) \sim \left[ \ln \left( \frac{1}{\omega} \right) \right]^{\frac{1}{2}}, \quad \nu_{n}(\bm{0}, \omega) \sim \left[ \ln \left( \frac{1}{\omega} \right) \right]^{\frac{1}{2}}, \quad \nu_{e}(\bm{0}, \omega) \sim \left[ \ln \left( \frac{1}{\omega} \right) \right]^{\frac{1}{2}}.
\end{align}

Below two dimensions with $c=0$, the fixed-point value of the dimensionless nonlinear coupling is of order unity, meaning that perturbative analysis from the linear theory will fails. Nevertheless, we can obtain
\begin{align} \label{eq:nued1}
    \nu_{e \rmR} (\bm{0}, \omega) \simeq \nu_e (l) \Psi_{\nu_e} \left( \bm{0}, \frac{\omega}{\nu_n (l) \Lambda(l)^2} ; \barnt^*, \barne^*, \barlam^* \right)
\end{align}
in the limit $l \to \infty$, where $(\barnt^*, \barne^*, \barlam^*)$ are the fixed-point values of the dimensionless parameters in Eq.~\eqref{eq:fixedpt}. If we use the matching condition \eqref{eq:somega} in Eq.~\eqref{eq:nued1}, the right-hand side depends on $l$ only through $\nu_e (l)$. Eq.~\eqref{eq:somega} and $\nu_n (l) \sim e^{\frac{2-d}{2}l}$ as $l \to \infty$ imply
\begin{align}
    l_* \simeq \frac{2}{2+d} \ln \left( \frac{1}{\omega} \right),
\end{align}
which ensure $l_* \to \infty$ as $\omega \to 0$. Similar results apply for $\nu_{\theta \rmR}$ and $\nu_{n \rmR}$. Consequently, all diffusivities exhibit power-law divergences in the low-frequency limit as
\begin{align}
    \nu_{\theta \rmR} (\bm{0}, \omega) \sim \omega^{- \frac{2-d}{2+d}}, \quad \nu_{n \rmR} (\bm{0}, \omega) \sim \omega^{- \frac{2-d}{2+d}}, \quad \nu_{e \rmR} (\bm{0}, \omega) \sim \omega^{- \frac{2-d}{2+d}}.
\end{align}
Similar findings have been reported in the literature on anomalous transport \cite{NarayanRamaswamy2002,ForsterNelsonStephen1977}.

\subsubsection{Crossover}
\label{subsubsec:crossover}

\begin{figure}[h]
    \centering
    \includegraphics[width=16cm]{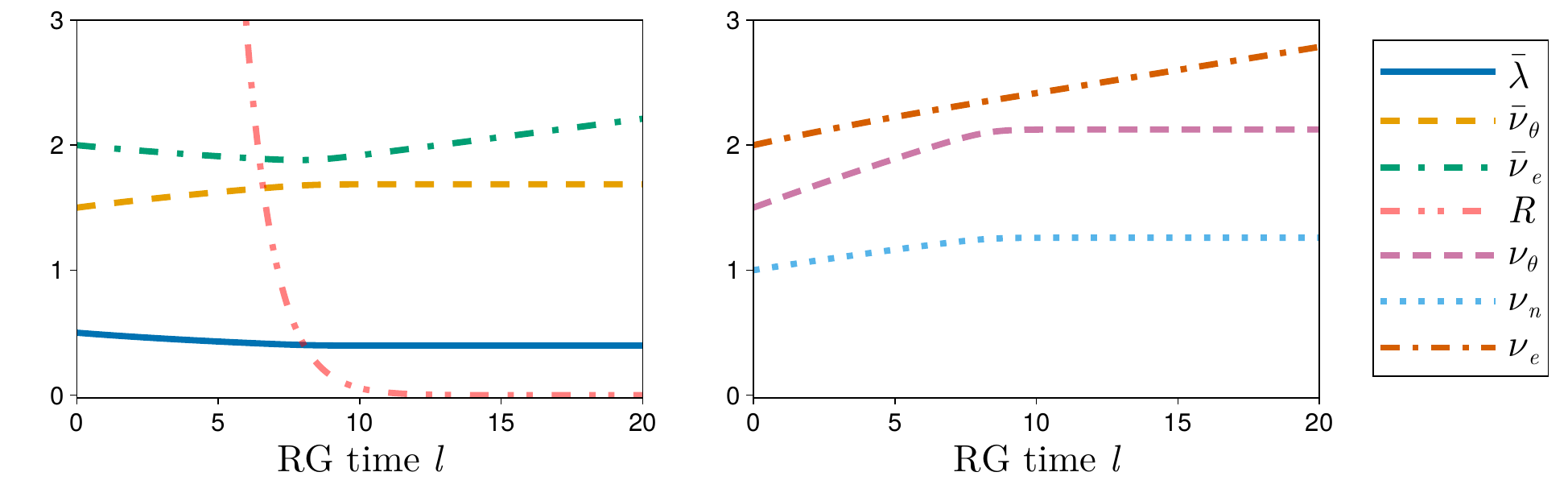}
    \caption{RG flows for $d=2$ with a small speed of sound. The initial values of the parameters are $\barlam = 0.5$, $\barnt = 1.5$, $\barne = 2.0$, $\nu_{\theta} = 1.5$, $\nu_n = 1.0$, $\nu_e = 2.0$, and $R = 10^3$.}
    \label{fig:crossover}
\end{figure}

We conclude by discussing the crossover phenomena from zero to finite speed of sound. In dimensions $d>2$, the behavior of the RG flow remains the same regardless of the presence of sound waves. The RG equations cannot be applied to the case of $d < 2$ when $c > 0$. Therefore, in this section, we focus on the case of $d = 2$. From the RG equations \eqref{eq:RGt}, \eqref{eq:RGe}, and \eqref{eq:RGlam}, the crossover RG time $l_{\mathrm{cr}}$ is determined by the condition
\begin{align}
    \barc^2 \sim (\barnt + \barne) (1 + \barne).
\end{align}
The trajectories obtained from the numerical integration of the RG equations for $d=2$ with a small initial $\barc$ are shown in Fig. \ref{fig:crossover}. The initial values are $\barnt (0) = 1.5$, $\barne (0) = 2.0$, and $\barc (0) = 10^{-3}$, or equivalently, $R (0) = 10^3$. The crossover time is roughly estimated as $l_{\mathrm{cr}} \sim 8.06$, using these initial values for $\barnt$, $\barne$, and the estimation $\barc (l) \sim \barc (0) \rme^{l}$. Fig. \ref{fig:crossover} shows a crossover occurring around this estimated time.

The energy diffusivity increases with $l$ over the entire range of $l$. However, the phase and charge diffusivities tend to diverge weakly for $l \ll l_{\mathrm{cr}}$, while converging to finite values for $l \gg l_{\mathrm{cr}}$. This behavior suggests that the size-dependent phase and charge diffusivities in two dimensions exhibit a logarithmic diverging tendency up to the crossover length scale $L_{\mathrm{cr}} \sim e^{l_{\mathrm{cr}}}$, but saturate to finite values for $L \gg L_{\mathrm{cr}}$. Further numerical simulations of microscopic models are necessary to validate these findings.

\section{Concluding remarks}
\label{sec:remark}

We have proposed two fluctuating hydrodynamic equations with charge and energy conservation that describe the large-scale behavior of the Hamiltonian XY model in the disordered and ordered phases. In the disordered phase, the dynamics is purely diffusive, while in the ordered phase, we find elastic, nonlinear energy transport in addition to diffusion. Our analysis shows that the large-scale behavior in the disordered phase is well described by the linear diffusion equation, and thus the diffusivities are finite in the thermodynamic limit. Using the dynamic renormalization group, we demonstrated that the energy diffusivity in exactly two dimensions diverges as $\nu_e (L) \sim \ln L$ in the thermodynamic limit $L \to \infty$. These predictions are consistent with previous numerical simulations of the Hamiltonian XY model. In concluding, we offer brief comments on future directions for this study.

\subsection{Effect of vortices}
\label{subsec:vortex}

We conjecture that as long as the effect of vortices is neglected, the anomalous energy transport in exactly two dimensions predicted in this paper is robust and persists without involving any crossover, in contrast to the one-dimensional system discussed in Section \ref{subsec:remark1d}. This raises an important question: how do vortices influence energy transport in two dimensions? A numerical study \cite{DelfiniLepriLivi2005} suggests that the logarithmic divergence of the heat conductivity of the two-dimensional Hamiltonian XY model at low temperatures weakens as the temperature increases, and that the heat conductivity becomes finite above the BKT transition point. A possible explanation for this temperature-dependent behavior is as follows. While the spin-wave hydrodynamics in this paper can describe the universal form of the diverging heat conductivity at low temperatures, it does not account for vortex excitations, which reduce both the spin-wave stiffness and the amplitude, since the spin-wave approximation neglects the periodic nature of the phase field. As our analysis suggests, the finiteness of the spin-wave stiffness and the amplitude is crucial for anomalous transport. As temperature rises, vortex excitations diminish the anomalous part of the heat conductivity, which may eventually vanish at the BKT transition point, where the renormalized stiffness drops discontinuously to zero. This scenario will be investigated in future studies. As a related issue, we note a numerical study that suggests a vanishing phase diffusivity near the BKT transition point \cite{LepriRuffo2001}.

\subsection{One-dimensional system}
\label{subsec:remark1d}

We comment on the connection to recent studies on one-dimensional nonlinear fluctuating hydrodynamics. In one dimension, the nonlinear fluctuating hydrodynamics in Eqs.~\eqref{eq:oht1}, \eqref{eq:ohc1}, and \eqref{eq:ohe1} reduce to
\begin{align}
    &\partial_t v_{\theta} + \partial_x \left( - c_{\rms} n - \lambda_{\theta} n e - \nu_{\theta} \partial_x v_{\theta} + \sqrt{2 D_{\theta}} \xi^{\theta} \right) = 0,
    \\
    & \partial_t n + \partial_x \left( - d_{\rms} v_{\theta} - \lambda_n v_{\theta} e - \nu_n \partial_x n + \sqrt{2 D_n} \xi^n \right) = 0,
    \\
    & \partial_t e + \partial_x \left( - \lambda_e v_{\theta} n - \nu_e \partial_x e + \sqrt{2 D_e} \xi^e \right) = 0,
\end{align}
where we have used $v_{\theta} = \partial_x \theta$ instead of $\theta$. Following \cite{Spohn2014a}, we introduce a normal coordinate
\begin{align}
    \begin{bmatrix}
        \phi_{+} \\
        \phi_{-} \\
        \phi_{0}
    \end{bmatrix}
    = 
    \begin{bmatrix}
        - \frac{\sqrt{d_{\rms}}}{Z_1} & \frac{\sqrt{c_{\rms}}}{Z_1} & 0 \\
        \frac{\sqrt{d_{\rms}}}{Z_1} & \frac{\sqrt{c_{\rms}}}{Z_1} & 0 \\
        0 & 0 & \frac{1}{Z_0}
    \end{bmatrix}
    \begin{bmatrix}
        v_{\theta} \\
        n \\
        e
    \end{bmatrix},
\end{align}
where the normalization factors $Z_1$ and $Z_0$ are given by
\begin{align}
    Z_1 = \sqrt{\frac{D_{\theta}}{\nu_{\theta}} d_{\rms} + \frac{D_n}{\nu_n} c_{\rms}}, \quad Z_0 = \sqrt{\frac{D_r}{\nu_e}}.
\end{align}
Then, Eqs.~\eqref{eq:oht1}, \eqref{eq:ohc1}, and \eqref{eq:ohe1} under the detailed balance condition \eqref{eq:db} can be written as
\begin{align}
    &\partial_t \phi_+ + \partial_x \left( c \phi_{+} + \lambda_{\theta} \sqrt{\frac{\nu_{\theta}}{D_{\theta}} \frac{D_n}{\nu_n} \frac{D_e}{\nu_e}} \phi_{+} \phi_0 - \frac{\nu_{\theta} + \nu_n}{2} \partial_x \phi_+ - \frac{- \nu_{\theta} + \nu_n}{2} \partial_x \phi_{-} \right.
    \notag \\
    &\qquad \qquad \left. + \frac{\sqrt{2 D_{\theta}} + \sqrt{2 D_n}}{2} \xi^{\theta} + \frac{- \sqrt{2 D_{\theta}} + \sqrt{2 D_n}}{2} \xi^n  \right) = 0,
    \\
    &\partial_t \phi_- + \partial_x \left( - c \phi_{+} - \lambda_{n} \sqrt{\frac{D_{\theta}}{\nu_{\theta}} \frac{\nu_n}{D_n} \frac{D_e}{\nu_e}} \phi_{-} \phi_0 - \frac{-\nu_{\theta} + \nu_n}{2} \partial_x \phi_+ - \frac{\nu_{\theta} + \nu_n}{2} \partial_x \phi_{-} \right.
    \notag \\
    &\qquad \qquad \left. + \frac{- \sqrt{2 D_{\theta}} + \sqrt{2 D_n}}{2} \xi^{\theta} + \frac{\sqrt{2 D_{\theta}} + \sqrt{2 D_n}}{2} \xi^n  \right) = 0,
    \\
    &\partial_t \phi_0 + \partial_x \left( \lambda_e \sqrt{\frac{D_{\theta}}{\nu_{\theta}} \frac{D_n}{\nu_n} \frac{\nu_e}{D_e}} (\phi_{+}^2 - \phi_{-}^2) - \nu_e \partial_x \phi_0 + \sqrt{2 D_e} \xi^n  \right) = 0.
\end{align}
Thus, the hydrodynamic equation for $d=1$ proposed in this paper falls within the class of one-dimensional nonlinear fluctuating hydrodynamics developed in Ref. \cite{Spohn2014a}. The universal scaling behavior of time correlation functions is studied using the mode-coupling approximation in Ref. \cite{Spohn2014a}. A related discussion on the fluctuating hydrodynamics of the one-dimensional Hamiltonian XY model, also known as coupled rotors, can be found in Ref. \cite{Spohn2014b}.

In Section \ref{subsec:finitec}, we argued that our perturbative RG analysis breaks down in one dimension. Previous numerical simulations \cite{GendelmanSavin2000,GiardinaLiviPolitiVassalli2000,YangHu2005} suggest finite conductivity in one dimension. The theoretical understanding of normal transport in one dimension is particularly challenging. It is important to note that the one-dimensional Hamiltonian XY model exhibits a disordered phase at any finite temperature. Therefore, even if a hydrodynamic description is applicable to the one-dimensional system, amplitude fluctuations, which are neglected in our present analysis, may be need to be accounted for. Incorporating these amplitude fluctuations into a continuum description could allow for inclusion of spatially localized excitations that act as scatterers. In this case, the domain size, i.e., the mean distance between localized excitations, might serve as an additional infrared cutoff, leading to finite heat conductivity. This approach would require integrating fluctuations with finite amplitudes, complicating the perturbative analysis. A functional renormalization group approach might provide a useful tool for studying the non-perturbative nature of the one-dimensional problem. For instance, a non-perturbative analysis could potentially clarify the crossover from infinite to finite conductivity reported in Refs. \cite{YangHu2005,GendelmanSavin2005}.

\subsection{Related models}

The next step in our research is to extend the current analysis to other systems in order to further explore the connection between heat conductivity and phase transitions. One promising system for investigation is the $\rmO(N)$ vector model. At low temperatures, this model may be effectively described by the $\rmO(N)$ nonlinear $\sigma$ model. In the $\rmO (N)$ nonlinear $\sigma$ model with $N > 2$, more than one angle variable is necessary to specify the orientation of the order parameter, in contrast to the XY or $\rmO (2)$ model, where only a single angle $\theta$ is needed. These angle variables interact strongly with each other. It is widely believed that, even in two dimensions, the spin-wave stiffness is renormalized to zero at long wavelength for any finite temperatures due to this interaction. This implies the absence of quasi-long range order in the two-dimensional $\rmO (N)$ model with $N > 2$ \cite{Migdal1975,Polyakov1975,BrezinZinnJustin1976}. The presence of elastic energy transport is crucial for anomalous energy transport, and its presence is expected under (quasi-)long range order of the order parameter. Therefore, a renormalization group analysis of the $\rmO(N)$ nonlinear $\sigma$ model for $N>2$ and $d=2$, incorporating energy conservation, could potentially predict finite heat conductivity at low temperatures in contrast to our analysis for $N=2$.

\ack
The author would like to thank the participants of the YITP workshop YITP-T-24-04 on ``Advances in Fluctuating Hydrodynamics: Bridging the Micro and Macro Scales'' for their helpful comments. The author also thanks the anonymous referee for their helpful comments on the asymptotic behavior of the RG equation in $d=2$ with zero sound velocity in Section \ref{subsec:zeroc}. This work is supported by JSPS KAKENHI Grant No. JP22J00337 and JST ERATO Grant No. JPMJER2302. 

\clearpage

\begin{table}[h]
    \centering
    \caption{List of the symbols}
    \begin{tabular}{cl}  
        \toprule
        Symbol & Description \\
        \midrule
        $e$   & Energy density \\
        $n$   & Charge density \\
        $\theta$ & Phase \\
        $\bmv_{\theta}$ & Gradient of phase \\
        \midrule
        $\bmj^e$ & Energy current density \\
        $\bmj^n$ & Charge current density \\
        $j^{\theta}$ & Decay rate of phase \\ 
        \midrule
        $s$ & Entropy density \\
        $\beta$ & Inverse temperature \\
        $\mu$ & Chemical potential \\
        $\bmh_{\theta}$ & Field conjugate to phase \\
        $T_0$ & Temperature at the reference equilibrium state \\
        $C_0$ & Heat capacity at the reference equilibrium state \\
        $I_0$ & Charge susceptibility at the reference equilibrium state \\
        $\rho_0$ & Spin-wave stiffness at the reference equilibrium state \\
        \midrule
        $\gamma_e$ & Transport coefficient associated with energy \\
        $\gamma_n$ & Transport coefficient associated with charge \\
        $\gamma_{\theta}$ & Transport coefficient associated with phase \\
        \midrule
        $\nu_e$ & Energy diffusivity \\
        $\nu_n$ & Charge diffusivity \\
        $\nu_{\theta}$ & Phase diffusivity \\
        $D_e$ & Energy noise strength \\
        $D_n$ & Charge noise strength \\
        $D_{\theta}$ & Phase noise strength \\
        $c_{\rms}$, $d_{\rms}$ & Linear coupling constants \\
        $\lambda_{\theta}$, $\lambda_n$, $\lambda_e$ & Nonlinear coupling constants \\
        \bottomrule
    \end{tabular}
\end{table}

\clearpage

\appendix

\section{Details of dynamical action}

\subsection{Propagators and vertex functions}
\label{app:bare}

\begin{figure}[h]
    \centering
    \begin{minipage}[b]{0.3\columnwidth}
        \centering
        \includegraphics[width=4.0cm]{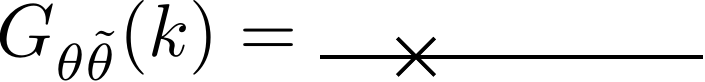}
    \end{minipage}
    \begin{minipage}[b]{0.3\columnwidth}
        \centering
        \includegraphics[width=4.0cm]{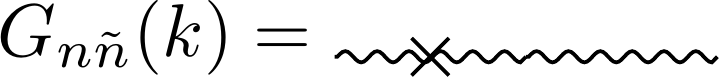}
    \end{minipage}
    \begin{minipage}[b]{0.3\columnwidth}
        \centering
        \includegraphics[width=4.0cm]{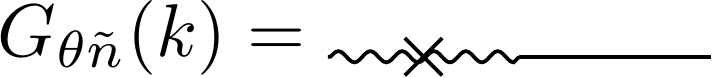}
    \end{minipage}

    \hspace{2mm}

    \begin{minipage}[b]{0.3\columnwidth}
        \centering
        \includegraphics[width=4.0cm]{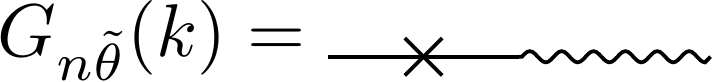}
    \end{minipage}
    \begin{minipage}[b]{0.3\columnwidth}
        \centering
        \includegraphics[width=4.0cm]{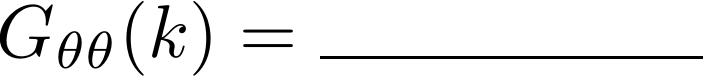}
    \end{minipage}
    \begin{minipage}[b]{0.3\columnwidth}
        \centering
        \includegraphics[width=4.0cm]{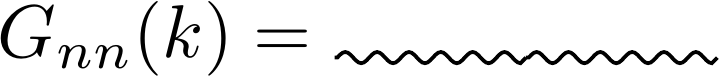}
    \end{minipage}

    \hspace{2mm}

    \begin{minipage}[b]{0.3\columnwidth}
        \centering
        \includegraphics[width=4.0cm]{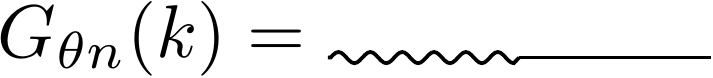}
    \end{minipage}
    \begin{minipage}[b]{0.3\columnwidth}
        \centering
        \includegraphics[width=4.0cm]{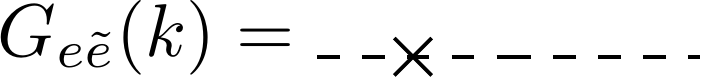}
    \end{minipage}
    \begin{minipage}[b]{0.3\columnwidth}
        \centering
        \includegraphics[width=4.0cm]{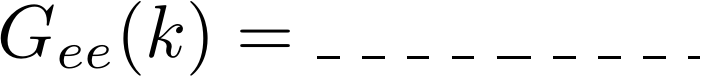}
    \end{minipage}

    \caption{Diagrammatic representations of propagators.}
    \label{fig:propagator}
\end{figure}

We write the $(\alpha, \alpha^{\prime})$-components of the propagator $\bmG (k ; \Lambda, \calP) = \bmG(-k ; \Lambda, \calP)^{\sfT}$ at the initial scale $\Lambda$ simply as $G_{\alpha \alpha^{\prime}}(k)$. The components in the phase-charge sector are given by
\begin{align}
    G_{\theta \tilt} (k) & = \frac{- i \omega + \nu_n \bmk^2}{(i \omega - i \omega_{+})(i \omega - i \omega_{-})},
    \\
    G_{n \tiln} (k) & = \frac{-i \omega + \nu_{\theta} \bmk^2}{(i \omega - i \omega_{+})(i \omega - i \omega_{-})},
    \\
    G_{\theta \tiln} (k) &=  \frac{c_{\rms}}{(i \omega - i \omega_{+})(i \omega - i \omega_{-})},
    \\
    G_{n \tilt} (k) &=  \frac{-d_{\rms} \bmk^2}{(i \omega - i \omega_{+})(i \omega - i \omega_{-})}
    \\
    G_{\theta \theta} (k) & = \frac{2 c_{\rms}^2 D_n \bmk^2 + 2 D_{\theta} (\omega^2 + \nu_n^2 \bmk^4)}{(i \omega - i\omega_{+})(i\omega - i\omega_{-})(i\omega + i\omega_{+})(i\omega + i\omega_{-})},
    \\
    G_{nn} (k) &=  \frac{2 d_{\rms}^2 D_{\theta} \bmk^4 + 2 D_n \bmk^2 (\omega^2 + \nu_{\theta}^2 \bmk^4)}{(i \omega - i\omega_{+})(i\omega - i\omega_{-})(i\omega + i\omega_{+})(i\omega + i\omega_{-})},
    \\
    G_{\theta n} (k) &=  \frac{2i (d_{\rms} D_{\theta} + c_{\rms} D_n)\bmk^2 \omega + 2(c_{\rms} \nu_{\theta} D_n - d_{\rms} D_{\theta} \nu_n) \bmk^4}{(i \omega - i\omega_{+})(i\omega - i\omega_{-})(i\omega + i\omega_{+})(i\omega + i\omega_{-})},
\end{align}
where $\omega_{\pm}$ is given in Eq.~\eqref{eq:omegapm}. The components in the energy sector are given by
\begin{align}
    G_{e \tile} (k) &=  \frac{1}{-i \omega + \nu_e \bmk^2},
    \\
    G_{ee} (k) &= \frac{2 D_e \bmk^2}{(-i \omega + \nu_e \bmk^2)(i \omega + \nu_e \bmk^2)}.
\end{align}
The diagrammatic representations of these propagators are presented in Fig. \ref{fig:propagator}, where the momentum $k$ flows from the second index to the first index.

\begin{figure}[h]
    \centering
    \includegraphics[width=6.0cm]{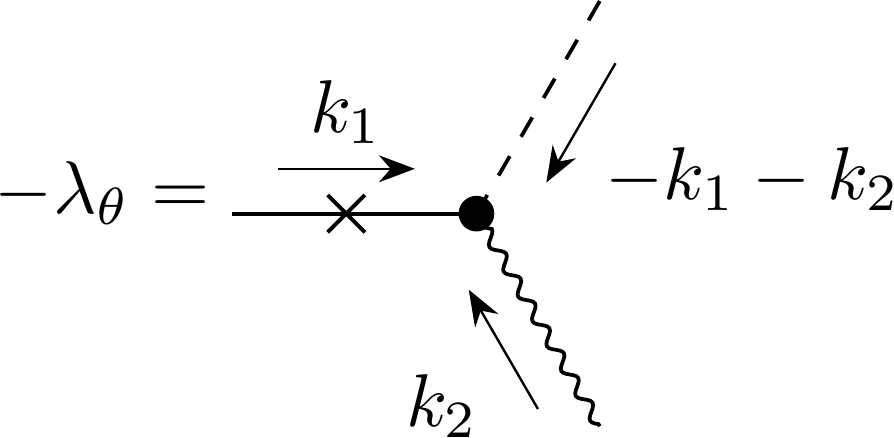}

    \hspace{2mm}

    \includegraphics[width=7.0cm]{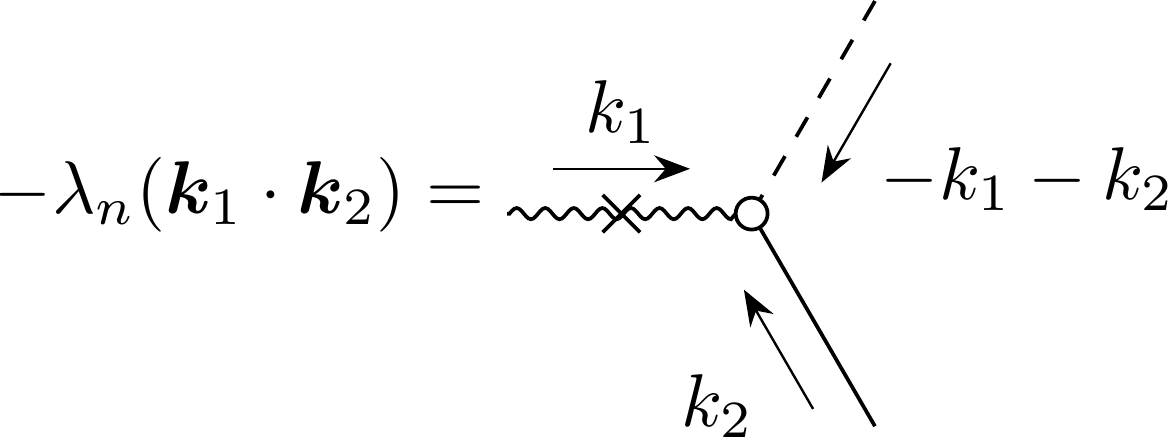}

    \hspace{2mm}

    \includegraphics[width=7.0cm]{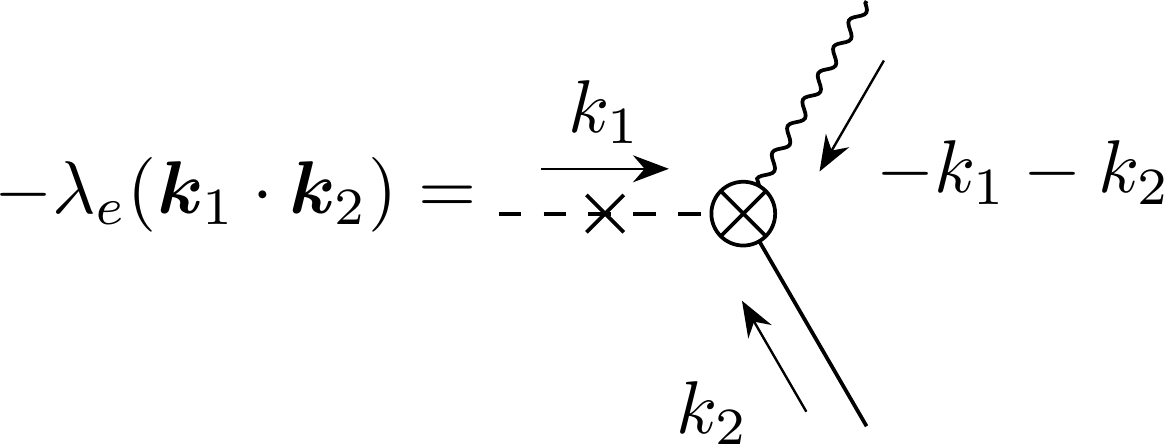}

    \caption{Diagrammatic representations of vertex functions.}
    \label{fig:vertex}
\end{figure}

Under the detailed balance condition \eqref{eq:db}, the correlation functions are expressed in terms of the response functions,
\begin{align} \label{eq:dbtt}
    G_{\theta \theta} (\bmk, \omega) &= \frac{D_{\theta}}{\nu_{\theta} \bmk^2} \left( G_{\theta \tilt} (\bmk, \omega) + G_{\theta \tilt} (\bmk, - \omega) \right),
    \\ \label{eq:dbnn}
    G_{nn} (\bmk, \omega) &= \frac{D_{n}}{\nu_n} \left( G_{n \tiln} (\bmk, \omega) + G_{n \tiln} (\bmk, - \omega) \right),
    \\ \label{eq:dbtn}
    G_{\theta n} (\bmk, \omega) &= \frac{D_{n}}{\nu_n} \left( G_{\theta \tiln} (\bmk, \omega) - G_{\theta \tiln} (\bmk, - \omega) \right)
    \\ \label{eq:dbnt}
    &= - \frac{D_{\theta}}{\nu_{\theta} \bmk^2} \left( G_{n \tilt} (\bmk, \omega) - G_{n \tilt} (\bmk, - \omega) \right),
    \\ \label{eq:dbee}
    G_{ee} (\bmk, \omega) &= \frac{D_{e}}{\nu_{e}} \left( G_{e \tile} (\bmk, \omega) + G_{e \tile} (\bmk, - \omega) \right).
\end{align}
and the response functions, $G_{n \tilt}$ and $G_{\theta \tiln}$, satisfy a reciprocal relation,
\begin{align} \label{eq:dbntres}
    G_{n \tilt} (k) = - \frac{\nu_{\theta}}{D_{\theta}} \frac{D_n}{\nu_n} \bmk^2 G_{\theta \tiln} (k).
\end{align}
The diagrammatic representations of the vertex functions are presented in Fig. \ref{fig:vertex}

\section{Details of RG analysis}
\label{app:rg}

\subsection{Dimensionless couplings}
\label{app:nondim}

We present an argument motivating us to introduce the dimensionless couplings in Section \ref{subsubsec:nondim}. Under the trivial rescaling $t^{\prime} = \alpha_t t$, $\bmr^{\prime} = \alpha_r \bmr$, $\Phi^{\prime} = \alpha_{\Phi} \Phi$, $\sqrt{\alpha_r^d \alpha_t} \xi^{\Phi \prime}(\bmr^{\prime}, t^{\prime}) = \xi^{\Phi}(\bmr,t)$, we obtain
\begin{align}
    \partial_{t^{\prime}} \theta^{\prime} &= c_{\rms} \frac{\alpha_{\theta}}{\alpha_t \alpha_n} n^{\prime} + \lambda_{\theta} \frac{\alpha_{\theta}}{\alpha_t \alpha_n \alpha_e} e^{\prime} n^{\prime} + \nu_{\theta} \frac{\alpha_r^2}{\alpha_t} \triangle^{\prime} \theta^{\prime} + \sqrt{2 D_{\theta} \frac{\alpha_{\theta}^2 \alpha_r^d}{\alpha_t}} \xi^{\theta \prime},
    \\
    \partial_{t^{\prime}} n^{\prime} &= - \nabla^{\prime} \cdot \left( - d_{\rms} \frac{\alpha_r^2 \alpha_n}{\alpha_t \alpha_{\theta}} \nabla^{\prime} \theta^{\prime} - \lambda_n \frac{\alpha_r^2 \alpha_n}{\alpha_t \alpha_{\theta} \alpha_e} e^{\prime} \nabla^{\prime} \theta^{\prime} - \nu_n \frac{\alpha_r^2}{\alpha_t} \nabla^{\prime} n^{\prime} + \sqrt{2 D_n \frac{\alpha_n^2 \alpha_r^{d+2}}{\alpha_t}} \bxi^{n \prime} \right),
    \\
    \partial_{t^{\prime}} e^{\prime} &= - \nabla^{\prime} \cdot \left( - \lambda_e \frac{\alpha_r^2 \alpha_e}{\alpha_t \alpha_n \alpha_{\theta}} n^{\prime} \nabla^{\prime} \theta^{\prime} - \nu_e \frac{\alpha_r^2}{\alpha_t} \nabla^{\prime} e^{\prime} + \sqrt{2 D_e \frac{\alpha_e^2 \alpha_r^{d+2}}{\alpha_t}} \bxi^{e \prime} \right).
\end{align}
We impose the conditions
\begin{align}
    \alpha_r = \Lambda, \quad \nu_n \frac{\alpha_r^2}{\alpha_t} = 1, \quad D_n \frac{\alpha_n^2 \alpha_r^{d+2}}{\alpha_t} = 1
\end{align}
to ensure that the charge diffusivity and noise strength are unity. These requirements determine
\begin{align}
    \alpha_t = \nu_n \Lambda^2, \quad \alpha_n = \sqrt{\frac{\nu_n}{D_n \Lambda^d}}.
\end{align}
Then, the phase and energy diffusivities become
\begin{align}
    \nu_{\theta} \frac{\alpha_r^2}{\alpha_t} = \frac{\nu_{\theta}}{\nu_n} = \barnt, \quad \nu_e \frac{\alpha_r^2}{\alpha_t} = \frac{\nu_e}{\nu_n} = \barne.
\end{align}
Furthermore, we require that the diffusivities match the noise strengths:
\begin{align}
    D_{\theta} \frac{\alpha_{\theta}^2 \alpha_r^d}{\alpha_t} = \frac{\nu_{\theta}}{\nu_n}, \quad D_e \frac{\alpha_e^2 \alpha_r^{d+2}}{\alpha_t} = \frac{\nu_e}{\nu_n},
\end{align}
which determine
\begin{align}
    \alpha_{\theta} = \sqrt{\frac{\nu_{\theta}}{D_{\theta} \Lambda^{d-2}}}, \quad \alpha_e = \sqrt{\frac{\nu_e}{D_e \Lambda^d}}.
\end{align}
With these choices, the reversible couplings in the rescaled equations are given by
\begin{align*}
    &c_{\rms} \frac{\alpha_{\theta}}{\alpha_t \alpha_n} = \barc, \quad d_{\rms} \frac{\alpha_r^2 \alpha_n}{\alpha_t \alpha_{\theta}} = \bard,
    \\
    &\lambda_{\theta} \frac{\alpha_{\theta}}{\alpha_t \alpha_n \alpha_e} = \bart \Od^{-\frac{1}{2}}, \quad \lambda_n \frac{\alpha_r^2 \alpha_n}{\alpha_t \alpha_{\theta} \alpha_e} = \barn \Od^{- \frac{1}{2}}, \quad \lambda_e \frac{\alpha_r^2 \alpha_e}{\alpha_t \alpha_n \alpha_{\theta}} = \bare \Od^{- \frac{1}{2}}.
\end{align*}
The multiplicative factor $\Od^{-\frac{1}{2}}$ makes the expression of the corrections simpler.

\subsection{One-loop corrections}
\label{app:loop}

\subsubsection{Preliminaries}

\begin{figure}[h]
    \centering
    \begin{minipage}[b]{0.40\columnwidth}
        \centering
        \includegraphics[width=4cm]{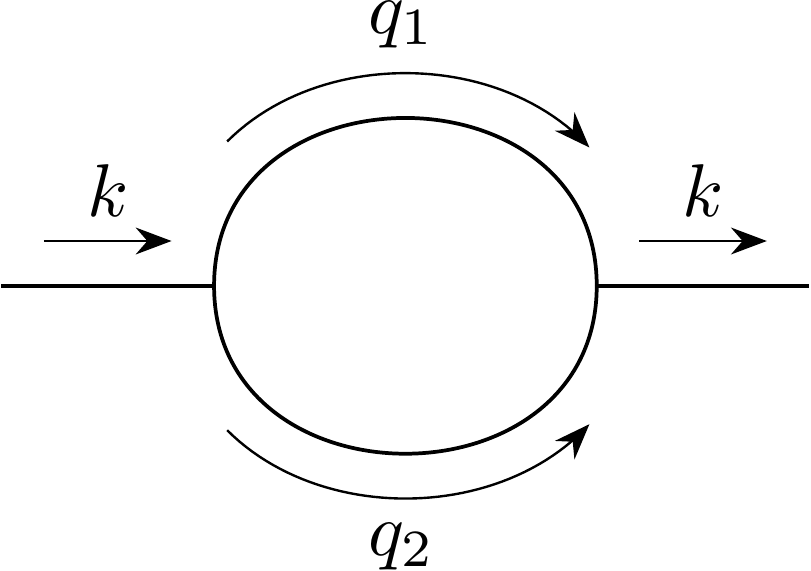}
        \caption*{(a)}
    \end{minipage}
    \begin{minipage}[b]{0.40\columnwidth}
        \centering
        \includegraphics[width=4cm]{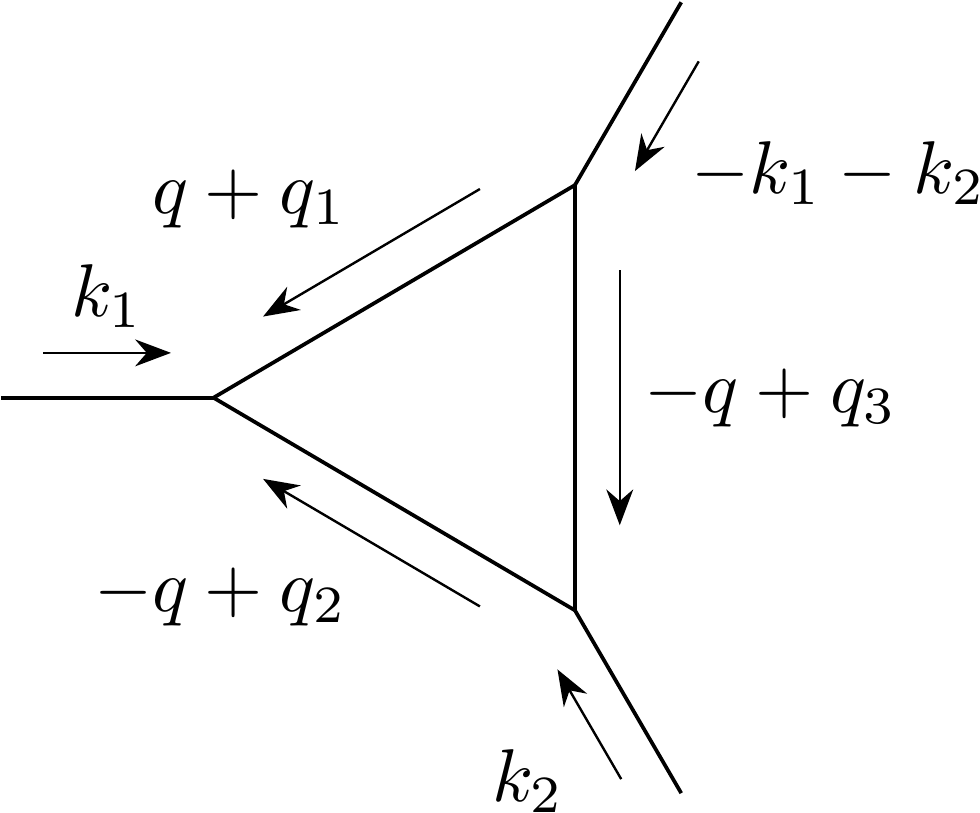}
        \caption*{(b)}
    \end{minipage}
    \caption{(a) One-loop diagram for propagator (b) one-loop diagram for vertex}
    \label{fig:integral}
\end{figure}

When calculating the one-loop diagram contributing to the propagator in Fig. \ref{fig:integral} (a), we encounter the following integral,
\begin{align}
    \bar{F} (\bmk) \coloneqq \int_{\bbR^d} \frac{d^d \bmq_1}{(2\pi)^d} \int_{\bbR^d} \frac{d^d \bmq_2}{(2\pi)^d} F(\bmq_1, \bmq_2) \theta^{+}(|\bmq_1|) \theta^{+}(|\bmq_2|) \hat{\delta} (\bmq_1 + \bmq_2 - \bmk).
\end{align}
where $F$ is a function and
\begin{align*}
    \theta^{+}(|\bmq|) &\coloneqq \theta (\Lambda - |\bmq|) \theta (|\bmq| - \Lambda^{\prime})
    \\
    &=
    \begin{cases}
        1 & \Lambda^{\prime} < |\bmq| < \Lambda,
        \\
        1/2 & |\bmq| = \Lambda^{\prime}, \Lambda,
        \\
        0 & \text{otherwise}.
    \end{cases} 
\end{align*}
In performing this integral, it is important to note that
\begin{align*}
    \theta^{+}(|\bmq + \bmk|) = \theta^{+}(|\bmq|) + (\hat{\bmq} \cdot \bmk) [\theta (\Lambda - |\bmq|) \delta (|\bmq| - \Lambda^{\prime}) - \theta (|\bmq| - \Lambda^{\prime}) \delta (\Lambda - |\bmq|)] + \rmO(\bmk^2),
\end{align*}
and thus, in general,
\begin{align*}
    \bar{F}(\bmk) \neq \intq F(\bmq, \bmk - \bmq).
\end{align*}
To simplify the computation, we use the following identity \cite{BereraYoffe2010}:
\begin{align}
    \theta^{+}(|\bmq_{+}|) \theta^{+}(|\bmq_{-}|) = \theta^{+}(|\bmq|) + \rmO(\bmk^2),
\end{align}
where
\begin{align}
    \bmq_{\pm} = \pm \bmq + \frac{\bmk}{2}.
\end{align}
By shifting the integration variable $\bmq \to \bmq + \frac{\bmk}{2}$ and applying the above identity, we find that
\begin{align} \label{eq:qpm}
    \bar{F}(\bmk) &= \int_{\bbR^d} \frac{d^d \bmq}{(2\pi)^d} F (\bmq_{+}, \bmq_{-}) \theta^{+}(|\bmq_{+}|) \theta^{+}(|\bmq_{-}|) 
    \notag \\
    &= \intq F(\bmq_{+}, \bmq_{-}) (1 + \rmO(\bmk^2)).
\end{align}
In our calculations, the leading contributions come from the first terms on the right-hand side, while the correction terms become of higher-order in $\bmk$ in the long-wavelength limit. Therefore, we can safely neglect these terms. Accordingly, we apply Eq.~\eqref{eq:qpm} to compute one-loop corrections to the propagators. When we compute the one-loop correction to the vertex in Fig. \ref{fig:integral} (b), we also use the following identity,
\begin{align*}
    \theta^{+}(|\bmq + \bmq_1|) \theta^{+}(|- \bmq + \bmq_2|) \theta^{+}(|- \bmq + \bmq_3|) = \theta^{+}(|\bmq|) + O(\bmk^2),
\end{align*}
and
\begin{align*}
    \bmq_1 + \bmq_2 + \bmk_1 = 0, \quad \bmq_1 + \bmq_3 + \bmk_1 + \bmk_2 = 0, \quad -\bmq_2 + \bmq_3 + \bmk_2 = 0
\end{align*}
for
\begin{align*}
    \bmq_1 = - \frac{2}{3} \bmk_1 - \frac{1}{3} \bmk_2, \quad \bmq_2 = - \frac{1}{3} \bmk_1 + \frac{1}{3} \bmk_2, \quad \bmq_3 = - \frac{1}{3} \bmk_1 - \frac{2}{3} \bmk_2.
\end{align*}
Here, $\rmO(\bmk^2)$ denotes second-order correction with respect to the external momenta, $\bmk_1$ and $\bmk_2$.

\subsubsection{One-loop corrections to propagators and vertex functions}
\label{subsubsec:potential}

For later convenience, we introduce the propagators and the vertex functions at the scale $\Lambda^{\prime}$ as
\begin{align}
    [\bmG^{-1} (k ; \Lambda^{\prime}, \calP^{\prime})]_{\alpha \alpha^{\prime}}  \hat{\delta}(k+k^{\prime}) \coloneqq \left. \frac{\delta^2 \calA [\Phi ; \Lambda^{\prime}, \calP^{\prime}]}{\delta \Phi_{\alpha}(k) \delta \Phi_{\alpha^{\prime}}(k^{\prime})} \right|_{\Phi = 0}
\end{align}
and
\begin{align}
    V_{\alpha_1 \alpha_2 \alpha_3} (k_1,k_2 ; \Lambda^{\prime}, \calP^{\prime}) \hat{\delta}(k_1+k_2+k_3) \coloneqq \left. \frac{\delta^3 \calA [\Phi ; \Lambda^{\prime}, \calP^{\prime}]}{\delta \Phi_{\alpha_1}(k_1) \delta \Phi_{\alpha_2}(k_2) \delta \Phi_{\alpha_3}(k_3)} \right|_{\Phi = 0},
\end{align}
respectively, where the action at the scale $\Lambda^{\prime}$ is defined in Eq.~\eqref{eq:defRG}. The propagators and the vertex functions acquire perturbative corrections,
\begin{align*}
    [\bmG^{-1} (k ; \Lambda^{\prime}, \calP^{\prime})]_{\tilt \theta}  &= [\bmG^{-1} (k ; \Lambda, \calP)]_{\tilt \theta}  + \left[ - i \omega \delta \Omega_{\theta} + \nu_{\theta} \delta \nu_{\theta} \bmk^2 \right] \delta l + \cdots,
    \\
    [\bmG^{-1} (k ; \Lambda^{\prime}, \calP^{\prime})]_{\tilt \tilt}  &= [\bmG^{-1} (k ; \Lambda, \calP)]_{\tilt \tilt}  - 2 D_{\theta} \delta D_{\theta} \bmk^2 \cdot \delta l + \cdots,
    \\
    [\bmG^{-1} (k ; \Lambda^{\prime}, \calP^{\prime})]_{\tiln n}  &= [\bmG^{-1} (k ; \Lambda, \calP)]_{\tiln n} + \left[  - i \omega  \delta \Omega_{n} + \nu_{n} \delta \nu_{n} \bmk^2 \right] \delta l + \cdots,
    \\
    [\bmG^{-1} (k ; \Lambda^{\prime}, \calP^{\prime})]_{\tiln \tiln}  &= [\bmG^{-1} (k ; \Lambda, \calP)]_{\tiln \tiln}  - 2 D_n \delta D_n \bmk^2 \cdot \delta l + \cdots,
    \\
    [\bmG^{-1} (k ; \Lambda^{\prime}, \calP^{\prime})]_{\tiln \theta}  &= [\bmG^{-1} (k ; \Lambda, \calP)]_{\tiln \theta}  + d_{\rms} \delta d_{\rms} \bmk^2 \cdot \delta l + \cdots
    \\
    [\bmG^{-1} (k ; \Lambda^{\prime}, \calP^{\prime})]_{\tilt n}  &= [\bmG^{-1} (k ; \Lambda, \calP)]_{\tilt n}  - c_{\rms} \delta c_{\rms} \cdot \delta l + \cdots
    \\
    [\bmG^{-1} (k ; \Lambda^{\prime}, \calP^{\prime})]_{\tile e}  &= [\bmG^{-1} (k ; \Lambda, \calP)]_{\tile e} + \left[ - i \omega \delta \Omega_e + \nu_e \delta \nu_e \bmk^2 \right] \delta l + \cdots
    \\
    [\bmG^{-1} (k ; \Lambda^{\prime}, \calP^{\prime})]_{\tile \tile}  &= [\bmG^{-1} (k ; \Lambda, \calP)]_{\tile \tile}  - 2 D_e \delta D_e \bmk^2 \cdot \delta l + \cdots
\end{align*}
and
\begin{align*}
    V_{\tilt n e} (k_1,k_2 ; \Lambda^{\prime}, \calP^{\prime}) &= V_{\tilt n e} (k_1,k_2 ; \Lambda, \calP) - \lambda_{\theta}  \delta \lambda_{\theta} \cdot \delta l + \cdots,
    \\
    V_{\tiln \theta e} (k_1,k_2 ; \Lambda^{\prime}, \calP^{\prime}) &= V_{\tiln e \theta} (k_1, k_2 ; \Lambda, \calP) - \lambda_n (\bmk_1 \cdot \bmk_2) \delta \lambda_n \cdot \delta l + \cdots,
    \\
    V_{\tile \theta e} (k_1,k_2 ; \Lambda^{\prime}, \calP^{\prime}) &= V_{\tile \theta e} (k_1, k_2 ; \Lambda, \calP) - \lambda_e (\bmk_1 \cdot \bmk_2)  \delta \lambda_e \cdot \delta l + \cdots,
\end{align*}
where the ellipses stand for higher-order terms in $\bmk$ and $\omega$ in the long-wavelength and long-time limit. We calculate these correction terms at one-loop order below.

\hspace{1cm}

\noindent -- Correction to $[\bmG^{-1}]_{\tilt \theta}$

\begin{figure}[h]
    \centering
    \begin{minipage}[b]{0.40\columnwidth}
        \centering
        \includegraphics[width=4cm]{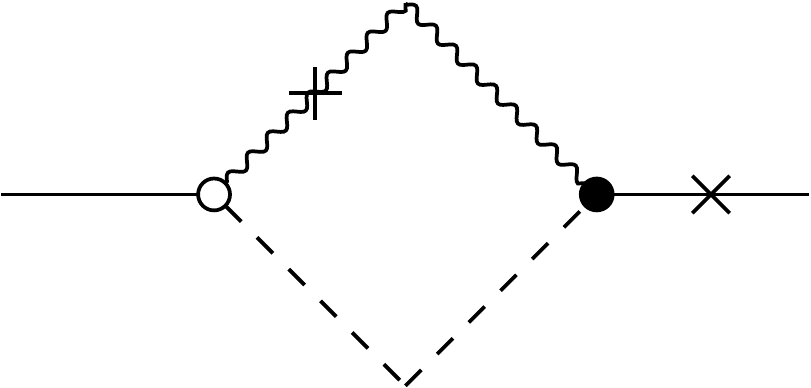}
        \caption*{(a)}
    \end{minipage}
    \begin{minipage}[b]{0.40\columnwidth}
        \centering
        \includegraphics[width=4cm]{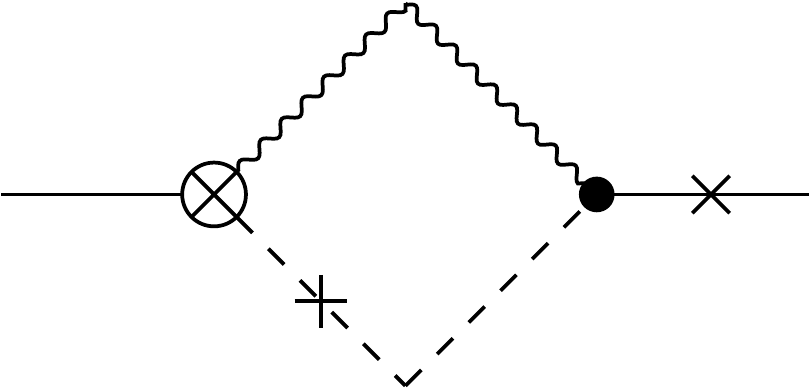}
        \caption*{(b)}
    \end{minipage}
    \caption{One-loop diagrams contributing to $[\bm{G^{-1}}(k)]_{\tilt \theta}$}
    \label{fig:tiltt}
\end{figure}

To keep this paper self-contained, we explain the calculation of the correction to $[\bmG^{-1}]_{\tilt \theta}$ in detail. The corrections to other propagators and vertex functions can be calculated in a similar way.

The one-loop diagrams contributing to $[\bmG^{-1}]_{\tilt \theta}$ are depicted in Fig. \ref{fig:tiltt}. The integral representation corresponding to the diagram in Fig. \ref{fig:tiltt} (a) is
\begin{align}
    \mathrm{(a)} &= \int_{\bbR^d} \frac{d^d \bmq_1}{(2\pi)^d} \int_{\bbR} \frac{d \omega_1}{2 \pi} \int_{\bbR^d} \frac{d^d \bmq_2}{(2\pi)^d} \int_{\bbR} \frac{d \omega_2}{2 \pi} \theta^{+}(|\bmq_1|) \theta^{+}(|\bmq_2|) \hat{\delta}(\bmq_1 + \bmq_2 - \bmk) \hat{\delta}(\omega_1 + \omega_2 - \omega)
    \notag \\
    & \qquad \times (- \lambda_n)(\bmk \cdot \bmq_1) (-\lambda_{\theta}) G_{n \tiln}(\bmq_1,\omega_1) G_{ee}(\bmq_2, \omega - \omega_2)
    \notag \\
    &= \lambda_{\theta} \lambda_n \intq \into ( \bmk \cdot \bmq_{+})  G_{n \tiln}(\bmq_{+}, \Omega) G_{ee}(\bmq_{-}, \omega - \Omega) + \rmO (\bmk^3)
\end{align}
We have used Eq.~\eqref{eq:qpm} in the third line. Using the relationship between the correlation function and the response function in Eq.~\eqref{eq:dbee} and the constraint \eqref{eq:db}, we can write $(\rma)$ as
\begin{align} \label{eq:ttiltdb}
    \mathrm{(a)} &\overset{\mathrm{D.B.}}{=} \lambda_{\theta}^2 \frac{\nu_{\theta}}{D_{\theta}} \frac{D_n}{\nu_n} \frac{D_e}{\nu_e} \intqo (\bmk \cdot \bmq_{+})  G_{n \tiln} (\bmq_{+}, \Omega) G_{e \tile} (\bmq_{-}, \omega - \Omega) + \rmO (\bmk^3).
\end{align}
In this section, we use the notation
\begin{align}
    \intqo \coloneqq \intq \into.
\end{align}
We note that the equality \eqref{eq:ttiltdb} holds under the detailed balance condition because we have used Eqs.~\eqref{eq:dbee} and \eqref{eq:db}. Hereafter, $\overset{\mathrm{D.B.}}{=}$ indicates that this equality follows from the detailed balance condition \eqref{eq:db}. We remark that the term $G_{n \tiln}(\bmq_{+}, \Omega) G_{e \tile}(\bmq_{-}, \Omega - \omega)$ in the above integral \eqref{eq:ttiltdb}, which originates from the second term in Eq.~\eqref{eq:dbee}, vanishes because both propagators have poles only in the lower half-plane. In a similar way, the integral representation corresponding to the diagram in Fig. \ref{fig:tiltt} (b) is
\begin{align*}
    \mathrm{(b)} &=  \lambda_{\theta} \lambda_e \intqo (\bmk \cdot \bmq_{-})  G_{nl}(\bmq_{+},\Omega) G_{e \tile} (\bmq_{-}, \omega - \Omega) + \rmO (\bmk^3)
    \\
    &\overset{\mathrm{D.B.}}{=} \lambda_{\theta}^2 \frac{\nu_{\theta}}{D_{\theta}} \frac{D_n}{\nu_n} \frac{D_e}{\nu_e} \intqo (\bmk \cdot \bmq_{-})  G_{n \tiln} (\bmq_{+}, \Omega) G_{e \tile} (\bmq_{-}, \omega - \Omega) + \rmO (\bmk^3).
\end{align*}
Their sum is
\begin{align*}
    \mathrm{(a)+(b)} &= \lambda_{\theta}^2 \frac{\nu_{\theta}}{D_{\theta}} \frac{D_n}{\nu_n} \frac{D_e}{\nu_e} \bmk^2 \intqo G_{n \tiln}(\bmq_{+}, \Omega) G_{e \tile} (\bmq_{-}, \omega - \Omega) + \rmO (\bmk^3).
\end{align*}
By applying the residue theorem, we can easily perform the integration over the frequency domain as
\begin{align*}
    \mathrm{(a)+(b)} = \lambda_{\theta}^2 \frac{\nu_{\theta}}{D_{\theta}} \frac{D_n}{\nu_n} \frac{D_e}{\nu_e} \bmk^2 \int_{\Lambda^{\prime} \leq |\bmq| \leq \Lambda} \frac{d^d \bmq}{(2\pi)^d} \frac{-i \omega + \nu_e \bmq_{-}^2 + \nu_{\theta} \bmq_{+}^2}{(i \omega - \nu_e \bmq_{-}^2 - i \omega_{+}(\bmq_{+}))(i \omega - \nu_e \bmq_{-}^2 - i \omega_{-}(\bmq_{+}))} + \rmO (\bmk^3)
\end{align*}
To express this perturbative correction in the form given in \ref{subsubsec:potential}, we expand it up to second order in the external momentum $\bmk$ and first order in the external frequency $\omega$. As a result,
\begin{align*}
    \mathrm{(a)+(b)} = \lambda_{\theta}^2 \frac{\nu_{\theta}}{D_{\theta}} \frac{D_n}{\nu_n} \frac{D_e}{\nu_e} \bmk^2 \int_{\Lambda^{\prime} \leq |\bmq| \leq \Lambda} \frac{d^d \bmq}{(2\pi)^d} \frac{\nu_e \bmq^2 + \nu_{\theta} \bmq^2}{(\nu_e \bmq^2 + i \omega_{+}(\bmq))(\nu_e \bmq^2 + i \omega_{-}(\bmq))} + \rmO (\omega \bmk^2, \bmk^3, \cdots).
\end{align*}
It is convenient to rewrite the integration over the loop momentum as
\begin{align*}
    \mathrm{(a)+(b)} = \lambda_{\theta}^2 \frac{\nu_{\theta}}{D_{\theta}} \frac{D_n}{\nu_n} \frac{D_e}{\nu_e} \bmk^2 \int_{\Lambda^{\prime}}^{\Lambda} d|\bmq| |\bmq|^{d-1} \int_{S^{d-1}} \frac{d\hat{\bmq}}{(2\pi)^d} \frac{\nu_e \bmq^2 + \nu_{\theta} \bmq^2}{(\nu_e \bmq^2 + i \omega_{+}(\bmq))(\nu_e \bmq^2 + i \omega_{-}(\bmq))} + \cdots,
\end{align*}
where $\hat{\bmq} = \bmq / |\bmq|$ is the unit vector along the loop momentum $\bmq$. In the limit $\delta l \to 0$, the width of the momentum shell is given by $\Lambda - \Lambda^{\prime} = \Lambda \delta l + \rmO ((\delta l)^2)$. Therefore, the integral over the loop momentum in the shell can be approximated as
\begin{align}
    \mathrm{(a)+(b)} = \lambda_{\theta}^2 \frac{\nu_{\theta}}{D_{\theta}} \frac{D_n}{\nu_n} \frac{D_e}{\nu_e} \bmk^2 \Lambda^d \frac{(\nu_e + \nu_{\theta}) \Lambda^2}{(\nu_e \Lambda^2 + i \omega_{+})(\nu_e \Lambda^2 + i \omega_{-})} \tilde{\Omega}_d \delta l + \cdots,
\end{align}
where
\begin{align}
    \tilde{\Omega}_d \coloneqq \int_{S^{d-1}} \frac{d \hat{\bmq}}{(2\pi)^d}
\end{align}
is the area of the $(d-1)$-dimensional unit sphere divided by $(2 \pi)^d$. Finally, we rewrite the result in terms of the dimensionless parameters introduced in Section \ref{subsubsec:nondim} as
\begin{align}
    \mathrm{(a)+(b)} = \nu_{\theta} \cdot \bart^2 \frac{\barne + \barnt}{\barnt} \frac{1}{(\barnt + \barne)(1+ \barne) + \barc \bard} \bmk^2 \cdot \delta l + \cdots.
\end{align}
The ellipses denote higher-order terms in $\delta l$, $\omega$, and $\bmk$. By comparing this with the expression
\begin{align}
    = \left[ - i \omega  \delta \Omega_{\theta} + \nu_{\theta}  \delta \nu_{\theta} \bmk^2 \right] \delta l + \cdots,
\end{align}
we identity
\begin{align}
    \delta \Omega_{\theta} = 0, \quad \delta \nu_{\theta} &= \bart^2 \frac{\barne + \barnt}{\barnt} \frac{1}{(\barnt + \barne)(1+ \barne) + \barc \bard}.
\end{align}

\hspace{1cm}

\noindent -- Correction to $[\bmG^{-1}]_{\tilt \tilt}$

\begin{figure}[h]
    \centering
    \begin{minipage}[b]{0.40\columnwidth}
        \centering
        \includegraphics[width=4cm]{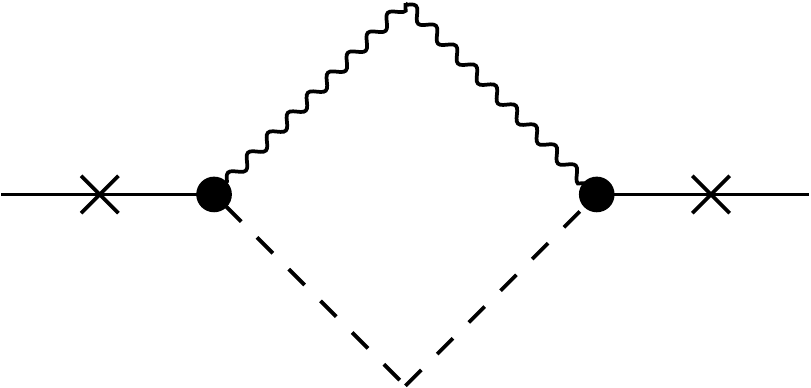}
        \caption*{(a)}
    \end{minipage}
    \caption{One-loop diagram contributing to $[\bm{G}^{-1}(k)]_{\tilt \tilt}$}
    \label{fig:tilttilt}
\end{figure}

The integral representation corresponding to the diagram in Fig. \ref{fig:tilttilt} is
\begin{align*}
    \mathrm{(a)} &= - \frac{\lambda_{\theta}^2}{2} \intqo G_{nl}(\bmq_{+}, \Omega) G_{ee}(\bmq_{-}, \omega - \Omega)
    \\
    & \overset{\mathrm{D.B.}}{=} - \frac{\lambda_{\theta}^2}{2} \frac{D_n}{\nu_n} \frac{D_e}{\nu_e} \intqo (G_{n \tiln}(\bmq_{+}, \Omega) G_{e \tile}(\bmq_{-}, \omega - \Omega) + G_{n \tiln}(\bmq_{+}, - \Omega) G_{e \tile} (\bmq_{-}, - \omega + \Omega))
    \\
    &= - D_{\theta} \cdot \bart^2 \frac{\barne + \barnt}{\barnt} \frac{1}{(\barnt + \barne)(1+ \barne) + \barc \bard} \cdot \delta l + \cdots
\end{align*}
By comparing this with the expression,
\begin{align}
    = - D_{\theta} \delta D_{\theta} \cdot \delta l + \cdots,
\end{align}
we identify
\begin{align}
    \delta D_{\theta} &= \bart^2 \frac{\barne + \barnt}{\barnt} \frac{1}{(\barnt + \barne)(1+ \barne) + \barc \bard} = \delta \nu_{\theta}.
\end{align}

\hspace{1cm}

\noindent -- Correction to $[\bmG^{-1}]_{\tiln n}$

\begin{figure}[h]
    \centering
    \begin{minipage}[b]{0.40\columnwidth}
        \centering
        \includegraphics[width=4cm]{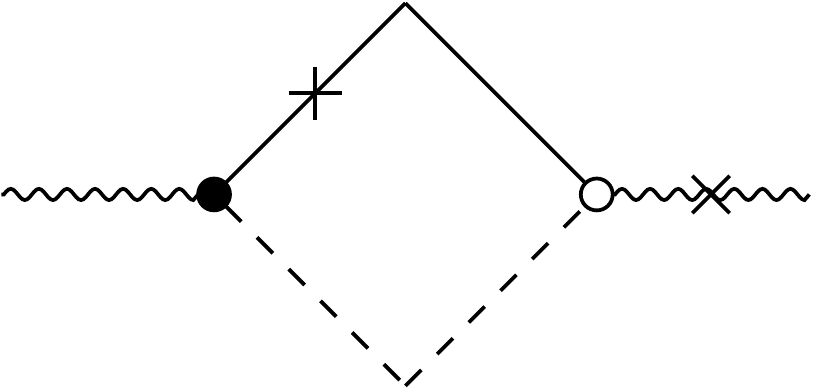}
        \caption*{(a)}
    \end{minipage}
    \begin{minipage}[b]{0.40\columnwidth}
        \centering
        \includegraphics[width=4cm]{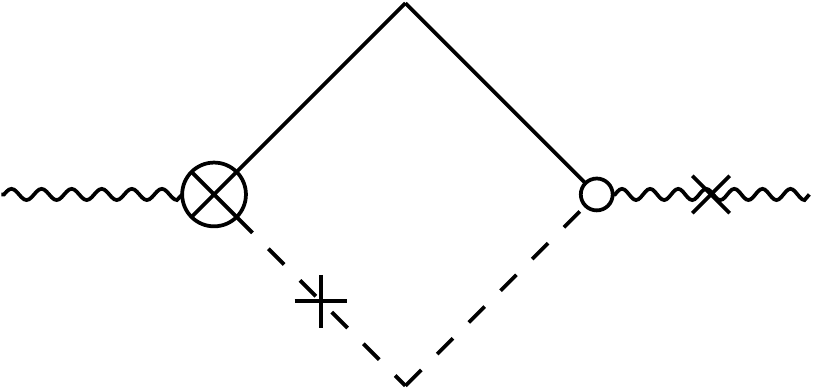}
        \caption*{(b)}
    \end{minipage}
    \caption{One-loop diagrams contributing to $[\bm{G}^{-1}(k)]_{\tiln n}$}
    \label{fig:tilnn}
\end{figure}

The integral representations corresponding to the diagrams in Fig. \ref{fig:tilnn} are
\begin{align*}
    \mathrm{(a)} &= \lambda_{\theta} \lambda_n \intqo  (\bmk \cdot \bmq_{+})  G_{\theta \tilt}(\bmq_{+}, \Omega) G_{ee}(\bmq_{-}, \omega - \Omega)
    \\
    &\overset{\mathrm{D.B.}}{=} \lambda_n^2 \frac{\nu_n}{D_n} \frac{D_{\theta}}{\nu_{\theta}} \frac{D_e}{\nu_e} \intqo (\bmk \cdot \bmq_{+}) G_{\theta \tilt}(\bmq_{+}, \Omega) G_{e \tile} (\bmq_{-}, \omega - \Omega)
\end{align*}
and
\begin{align*}
    \mathrm{(b)} &= \lambda_n \lambda_e \intqo (\bmq_{+} \cdot \bmq_{-})(\bmk \cdot \bmq_{+})  G_{\theta \theta}(\bmq_{+}, \Omega) G_{e \tile} (\bmq_{-}, \omega - \Omega)
    \\
    &\overset{\mathrm{D.B.}}{=} \lambda_n^2 \frac{\nu_n}{D_n} \frac{D_{\theta}}{\nu_{\theta}} \frac{D_e}{\nu_e} \intqo \frac{(\bmq_{+} \cdot \bmq_{-})(\bmk \cdot \bmq_{+})}{\bmq_{+}^2} G_{\theta \tilt}(\bmq_{+}, \Omega) G_{e \tile} (\bmq_{-}, \omega - \Omega).
\end{align*}
Their sum is
\begin{align*}
    \mathrm{(a)+(b)} &= \lambda_n^2 \frac{\nu_n}{D_n} \frac{D_{\theta}}{\nu_{\theta}} \frac{D_e}{\nu_e} \intqo \frac{(\bmk \cdot \bmq_{+})^2}{\bmq_{+}^2} G_{\theta \tilt}(\bmq_{+}, \Omega) G_{e \tile} (\bmq_{-}, \omega - \Omega)
    \\
    &= \nu_n \cdot \frac{1}{d} \barn^2 \frac{\barne+1}{(\barnt + \barne)(1 +\barne) + \barc \bard} \bmk^2 \cdot \delta l + \cdots.
\end{align*}
By comparing this with the expression,
\begin{align}
    = \left[ - i \omega \delta \Omega_n + \nu_n \delta \nu_n \bmk^2 \right] \delta l + \cdots,
\end{align}
we identify
\begin{align}
    \delta \Omega_n = 0, \quad \delta \nu_n = \frac{1}{d} \barn^2 \frac{\barne+1}{(\barnt + \barne)(1 +\barne) + \barc \bard}
\end{align}

\hspace{1cm}

\noindent -- Correction to $[\bmG^{-1}]_{\tiln \tiln}$

\begin{figure}[h]
    \centering
    \begin{minipage}[b]{0.40\columnwidth}
        \centering
        \includegraphics[width=4cm]{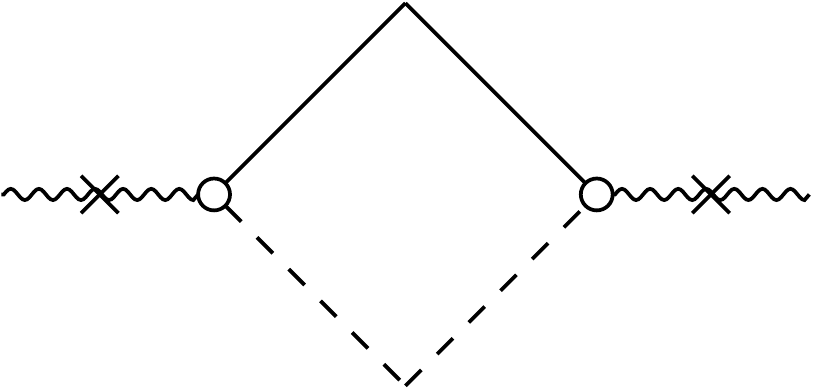}
        \caption*{(a)}
    \end{minipage}
    \caption{One-loop diagram contributing to $[\bm{G}^{-1}(k)]_{\tiln \tiln}$}
    \label{fig:tilntiln}
\end{figure}

The integral representation corresponding to the diagram in Fig. \ref{fig:tilntiln} is
\begin{align*}
    \mathrm{(a)} &= - \frac{\lambda_n^2}{2} \intqo  (\bmk \cdot \bmq_{+})(\bmk \cdot \bmq_{+})  G_{\theta \theta}(\bmq_{+}, \Omega) G_{ee}(\bmq_{-}, \omega - \Omega)
    \\
    &\overset{\mathrm{D.B}}{=} - \frac{1}{2} \lambda_n^2 \frac{D_{\theta}}{\nu_{\theta}} \frac{D_e}{\nu_e} \intqo \frac{(\bmk \cdot \bmq_{+})^2}{\bmq_{+}^2} (G_{\theta \tilt}(\bmq_{+}, \Omega) G_{e \tile} (\bmq_{-}, \omega - \Omega) + G_{\theta \tilt} (\bmq_{+}, - \Omega) G_{e \tile} (\bmq_{-}, - \omega + \Omega))
    \\
    &= - D_n \cdot \frac{1}{d} \barn^2 \frac{\barne+1}{(\barnt + \barne)(1 +\barne) + \barc \bard} \bmk^2 \cdot \delta l + \cdots
\end{align*}
By comparing this with the expression,
\begin{align}
    = - D_n \cdot l \delta D_n \bmk^2 \delta l + \cdots,
\end{align}
we identify
\begin{align}
    \delta D_n &= \frac{1}{d} \barn^2 \frac{\barne+1}{(\barnt + \barne)(1 +\barne) + \barc \bard} = \delta \nu_n.
\end{align}

\hspace{1cm}

\noindent -- Correction to $[\bmG^{-1}]_{\tiln \theta}$

\begin{figure}[h]
    \centering
    \begin{minipage}[b]{0.40\columnwidth}
        \centering
        \includegraphics[width=4cm]{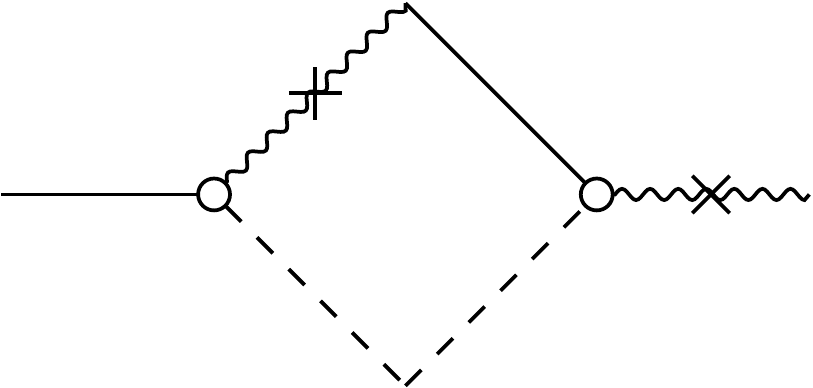}
        \caption*{(a)}
    \end{minipage}
    \begin{minipage}[b]{0.40\columnwidth}
        \centering
        \includegraphics[width=4cm]{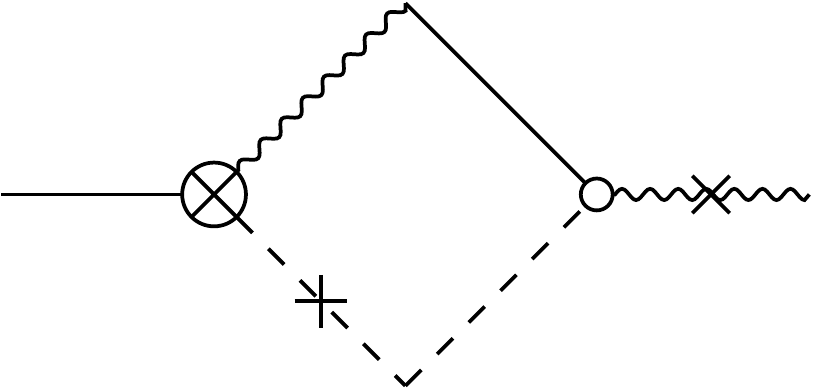}
        \caption*{(b)}
    \end{minipage}
    \caption{One-loop diagrams contributing to $[\bm{G}^{-1}(k)]_{\tiln \theta}$}
    \label{fig:tilnt}
\end{figure}

The integral representations corresponding to the diagrams in Fig. \ref{fig:tilnt}
\begin{align*}
    \mathrm{(a)} &= - \lambda_n^2 \intqo  (\bmk \cdot \bmq_{+})(\bmk \cdot \bmq_{+})  G_{\theta \tiln}(\bmq_{+}, \Omega) G_{ee}(\bmq_{-}, \omega - \Omega)
    \\
    &\overset{\mathrm{D.B.}}{=} - \lambda_n \lambda_e \frac{D_n}{\nu_n} \intqo (\bmk \cdot \bmq_{+})^2 G_{\theta \tiln} (\bmq_{+}, \Omega) G_{e \tile} (\bmq_{-}, \omega - \Omega)
\end{align*}
and
\begin{align*}
    \mathrm{(b)} &= - \lambda_n \lambda_e \intqo (\bmk \cdot \bmq_{-})(\bmk \cdot \bmq_{+})  G_{\theta l}(\bmq_{+}, \Omega) G_{e \tile} (\bmq_{-}, \omega - \Omega)
    \\
    &\overset{\mathrm{D.B.}}{=} - \lambda_n \lambda_e \frac{D_n}{\nu_n} \intqo (\bmk \cdot \bmq_{+})(\bmk \cdot \bmq_{-}) G_{\theta \tiln}(\bmq_{+}, \Omega) G_{e \tile} (\bmq_{-}, \omega - \Omega)
\end{align*}
Their sum is
\begin{align*}
    \mathrm{(a)+(b)} &= - \lambda_n \lambda_e \frac{D_n}{\nu_n} \bmk^2 \intqo (\bmk \cdot \bmq_{+}) G_{\theta \tiln}(\bmq_{+}, \Omega) G_{e \tile} (\bmq_{-}, \omega - \Omega)
    \\
    &= 0 \cdot \bmk^2 + \cdots.
\end{align*}
By comparing this with the expression,
\begin{align}
    = d_{\rms} \delta d_{\rms} \bmk^2 \cdot \delta l + \cdots
\end{align}
we identify
\begin{align}
    \delta d_{\rms} &= 0.
\end{align}

\hspace{1cm}

\noindent -- Correction to $[\bmG^{-1}]_{\tilt n}$

\begin{figure}[h]
    \centering
    \begin{minipage}[b]{0.40\columnwidth}
        \centering
        \includegraphics[width=4cm]{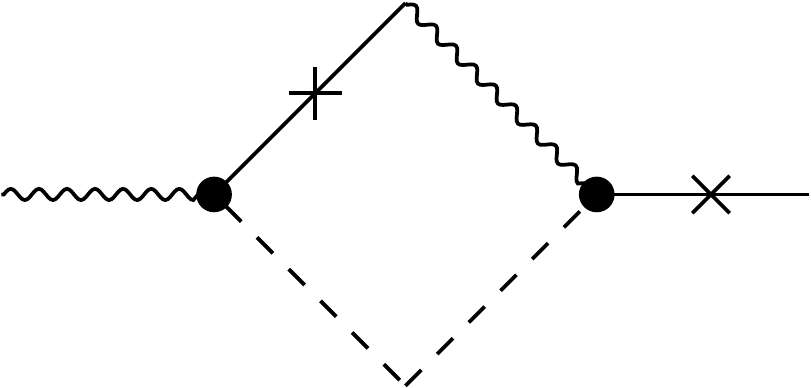}
        \caption*{(a)}
    \end{minipage}
    \begin{minipage}[b]{0.40\columnwidth}
        \centering
        \includegraphics[width=4cm]{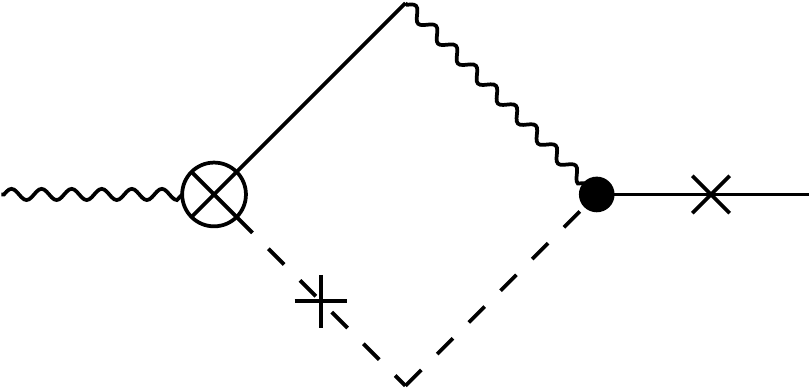}
        \caption*{(b)}
    \end{minipage}
    \caption{One-loop diagrams contributing to $[\bm{G}^{-1}(k)]_{\tilt n}$}
    \label{fig:tiltn}
\end{figure}

The integral representations corresponding to the diagrams in Fig. \ref{fig:tiltn} are
\begin{align*}
    \mathrm{(a)} &= - \lambda_{\theta}^2 \intqo G_{n \tilt}(\bmq_{+}, \Omega) G_{ee}(\bmq_{-}, \omega - \Omega)
    \\
    &\overset{\mathrm{D.B.}}{=} - \lambda_{\theta} \lambda_e \frac{D_{\theta}}{\nu_{\theta}} \intqo G_{n \tilt}(\bmq_{+}, \Omega) G_{e \tile} (\bmq_{-}, \omega - \Omega)
\end{align*}
and
\begin{align*}
    \mathrm{(b)} &= - \lambda_{\theta} \lambda_e \intqo (\bmq_{+} \cdot \bmq_{-})  G_{n \theta}(\bmq_{+}, \Omega) G_{e \tile} (\bmq_{-}, \omega - \Omega)
    \\
    &\overset{\mathrm{D.B.}}{=} - \lambda_{\theta} \lambda_e \frac{D_{\theta}}{\nu_{\theta}} \intqo \frac{(\bmq_{+} \cdot \bmq_{-})}{\bmq_{+}^2} G_{n \tilt}(\bmq_{+}, \Omega) G_{e \tile} (\bmq_{-}, \omega - \Omega).
\end{align*}
Their sum is
\begin{align*}
    \mathrm{(a)+(b)} &= - \lambda_{\theta} \lambda_e \frac{D_{\theta}}{\nu_{\theta}} \intqo \frac{\bmk \cdot \bmq_{+}}{\bmq_{+}^2} G_{n \tilt}(\bmq_{+}, \Omega) G_{e \tile} (\bmq_{-}, \omega - \Omega)
    \\
    &= 0 + \cdots
\end{align*}
By comparing this with the expression,
\begin{align}
     = - c_{\rms} \cdot \delta c_{\rms} \cdot \delta l + \cdots
\end{align}
we identify
\begin{align}
    \delta c_{\rms} = 0.
\end{align}

\hspace{1cm}

\noindent -- Correction to $[\bmG^{-1}]_{\tile e}$

\begin{figure}[h]
    \centering
    \begin{minipage}[b]{0.4\columnwidth}
        \centering
        \includegraphics[width=4cm]{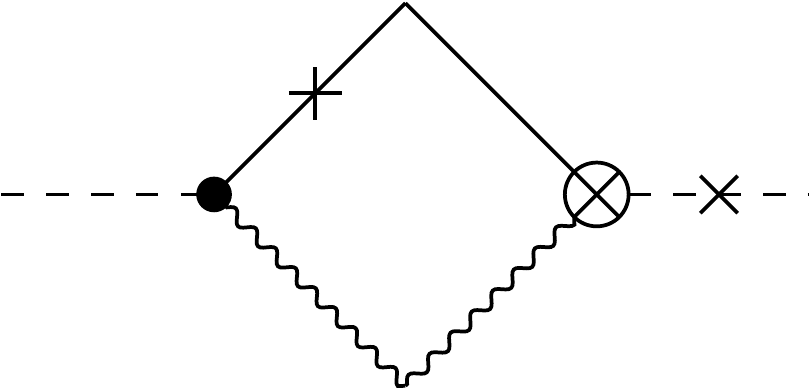}
        \caption*{(a)}
    \end{minipage}
    \begin{minipage}[b]{0.4\columnwidth}
        \centering
        \includegraphics[width=4cm]{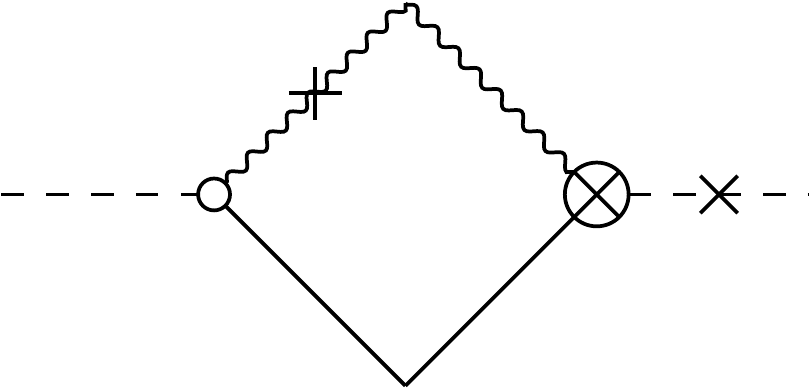}
        \caption*{(b)}
    \end{minipage}

    \hspace{2mm}

    \begin{minipage}[b]{0.4\columnwidth}
        \centering
        \includegraphics[width=4cm]{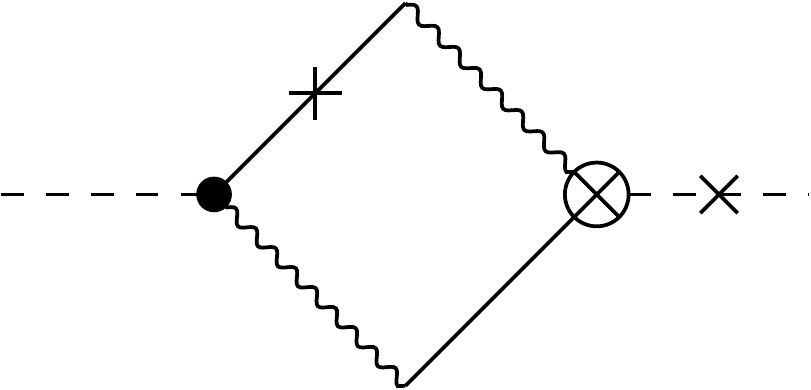}
        \caption*{(c)}
    \end{minipage}
    \begin{minipage}[b]{0.4\columnwidth}
        \centering
        \includegraphics[width=4cm]{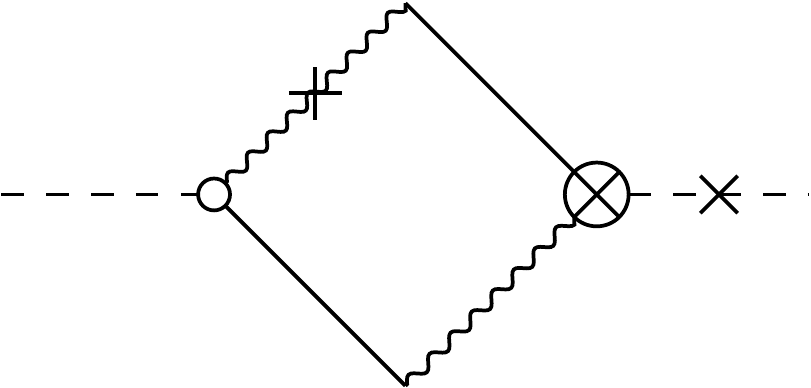}
        \caption*{(d)}
    \end{minipage}
    \caption{One-loop diagrams contributing to $[\bm{G}^{-1}(k)]_{\tile e}$}
    \label{fig:tilee}
\end{figure}

The integral representations corresponding to the diagrams in Fig. \ref{fig:tilee} are
\begin{align*}
    \mathrm{(a)} &= \lambda_{\theta} \lambda_e \intqo (\bmk \cdot \bmq_{+}) G_{\theta \tilt}(\bmq_{+},\Omega) G_{nl}(\bmq_{-},\omega - \Omega)
    \\
    &\dbeq \lambda_e^2 \frac{\nu_e}{D_e} \frac{D_{\theta}}{\nu_{\theta}} \frac{D_n}{\nu_n} \intqo (\bmk \cdot \bmq_{+}) G_{\theta \tilt} (\bmq_{+}, \Omega) G_{n \tiln} (\bmq_{-}, \omega - \Omega),
\end{align*}
\begin{align*}
    \mathrm{(b)} &= \lambda_n \lambda_e \intqo (\bmq_{+}\cdot \bmq_{-})(\bmk \cdot \bmq_{-})  G_{n \tiln}(\bmq_{+},\Omega) G_{\theta \theta}(\bmq_{-}, \omega - \Omega)
    \\
    &\dbeq \lambda_e^2 \frac{\nu_e}{D_e} \frac{D_{\theta}}{\nu_{\theta}} \frac{D_n}{\nu_n} \intqo \frac{(\bmq_{+} \cdot \bmq_{-})(\bmk \cdot \bmq_{+})}{\bmq_{+}^2} G_{\theta \tilt} (\bmq_{+}, \Omega) G_{n \tiln} (\bmq_{-}, \omega - \Omega),
\end{align*}
\begin{align*}
    \mathrm{(c)} &= \lambda_{\theta} \lambda_e \intqo (\bmk \cdot \bmq_{-})  G_{n \tilt}(\bmq_{+}, \Omega) G_{\theta l}(\bmq_{-}, \omega- \Omega)
    \\
    &\dbeq \lambda_e^2 \frac{\nu_e}{D_e} \frac{D_{\theta}}{\nu_{\theta}} \frac{D_n}{\nu_n} \intqo (\bmk \cdot \bmq_{-}) G_{n \tilt}(\bmq_{+}, \Omega) G_{\theta \tiln} (\bmq_{-}, \omega - \Omega),
\end{align*}
and
\begin{align*}
    \mathrm{(d)} &= \lambda_n \lambda_e \intqo (\bmq_{+} \cdot \bmq_{-})(\bmk \cdot \bmq_{+}) G_{\theta \tiln}(\bmq_{+}, \Omega) G_{n \theta}(\bmq_{-}, \omega - \Omega)
    \\
    &\dbeq \lambda_e^2 \frac{\nu_e}{D_e} \frac{D_{\theta}}{\nu_{\theta}} \frac{D_n}{\nu_n} \intqo \frac{(\bmq_{+} \cdot \bmq_{-})(\bmk \cdot \bmq_{-})}{\bmq_{+}^2} G_{n \tilt} (\bmq_{+}, \Omega) G_{\theta \tiln} (\bmq_{-}, \omega - \Omega).
\end{align*}
Their sum is
\begin{align*}
    \mathrm{(a)+(b)+(c)+(d)} &= \nu_e \cdot \frac{1}{d} \bare^2\frac{1}{\barne (\barnt+1)} \bmk^2 \cdot \delta l + \cdots
\end{align*}
By comparing this with the expression
\begin{align}
    = \left[ -i \omega \delta \Omega_e  + \nu_e \delta \nu_e  \bmk^2 \right] \delta l + \cdots,
\end{align}
we identify
\begin{align}
    \delta \Omega_e = 0, \quad \delta \nu_e &= \frac{1}{d} \bare^2 \frac{1}{\barne (\barnt + 1)}.
\end{align}

\hspace{1cm}

\noindent -- Correction to $[\bmG^{-1}]_{\tile \tile}$

\begin{figure}[h]
    \centering
    \begin{minipage}[b]{0.40\columnwidth}
        \centering
        \includegraphics[width=4cm]{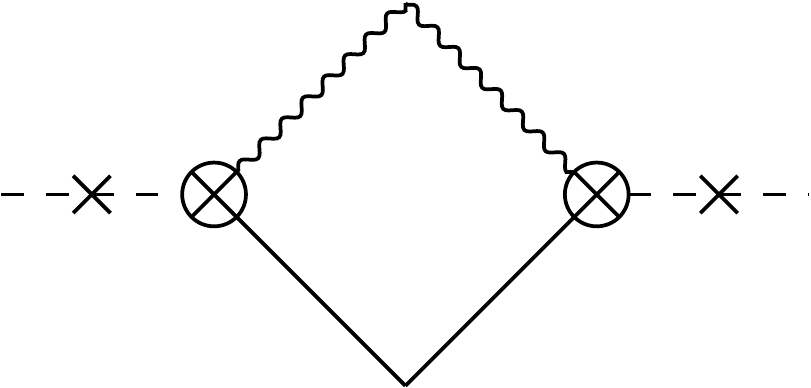}
        \caption*{(a)}
    \end{minipage}
    \begin{minipage}[b]{0.40\columnwidth}
        \centering
        \includegraphics[width=4cm]{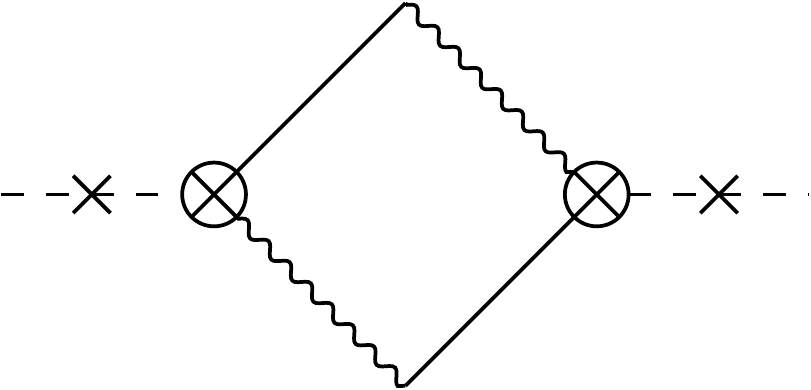}
        \caption*{(b)}
    \end{minipage}
    \caption{One-loop diagrams contributing to $[\bm{G}^{-1}(k)]_{\tile \tile}$}
    \label{fig:tiletile}
\end{figure}

The integral representations corresponding to the diagrams in Fig. \ref{fig:tiletile} are
\begin{align*}
    \mathrm{(a)} &= - \frac{\lambda_e^2}{2} \intqo (\bmk \cdot \bmq_{-})(\bmk \cdot \bmq_{-}) G_{nn}(\bmq_{+},\Omega) G_{\theta \theta} (\bmq_{-}, \omega - \Omega)
    \\
    &\dbeq - \frac{1}{2} \lambda_e^2 \frac{D_{\theta}}{\nu_{\theta}} \frac{D_n}{\nu_n} \intqo \frac{(\bmk \cdot \bmq_{+})^2}{\bmq_{+}^2} (G_{\theta \tilt} (\bmq_{+}, \Omega) G_{n \tiln} (\bmq_{-}, \omega - \Omega)  + G_{\theta \tilt}(\bmq_{+}, - \Omega) G_{n \tiln} (\bmq_{-}, - \omega + \Omega) )
\end{align*}
and
\begin{align*}
    \mathrm{(b)} &= - \frac{\lambda_e^2}{2} \intqo (\bmk \cdot \bmq_{+})(\bmk \cdot \bmq_{-}) G_{n \theta}(\bmq_{+},\Omega) G_{\theta n}(\bmq_{-}, \omega - \Omega)
    \\
    &\dbeq - \frac{1}{2} \lambda_e^2 \frac{D_{\theta}}{\nu_{\theta}} \frac{D_n}{\nu_n} \intqo \frac{(\bmk \cdot \bmq_{+})(\bmk \cdot \bmq_{-})}{\bmq_{+}^2} \left( G_{n \tilt} (\bmq_{+}, \Omega) G_{\theta \tiln} (\bmq_{-}, \omega - \Omega) + G_{n \tilt} (\bmq_{+}, - \Omega) G_{\theta \tiln} (\bmq_{-}, - \omega + \Omega) \right).
\end{align*}
Their sum is
\begin{align*}
    \mathrm{(a)+(b)} &= - D_e \cdot \frac{1}{d} \bare^2\frac{1}{\barne (\barnt+1)} \bmk^2 \cdot \delta l
\end{align*}
By comparing this with the expression,
\begin{align}
    = - D_e  \delta D_e \bmk^2 \cdot \delta l + \cdots
\end{align}
we identify
\begin{align}
    \delta D_e &= \frac{1}{d} \bare^2\frac{1}{\barne (\barnt+1)} = \delta \nu_e.
\end{align}

\hspace{1cm}

\noindent -- Correction to $V_{\tilt n e}$

\begin{figure}[h]
    \centering
    \begin{minipage}[b]{0.3\columnwidth}
        \centering
        \includegraphics[width=3.5cm]{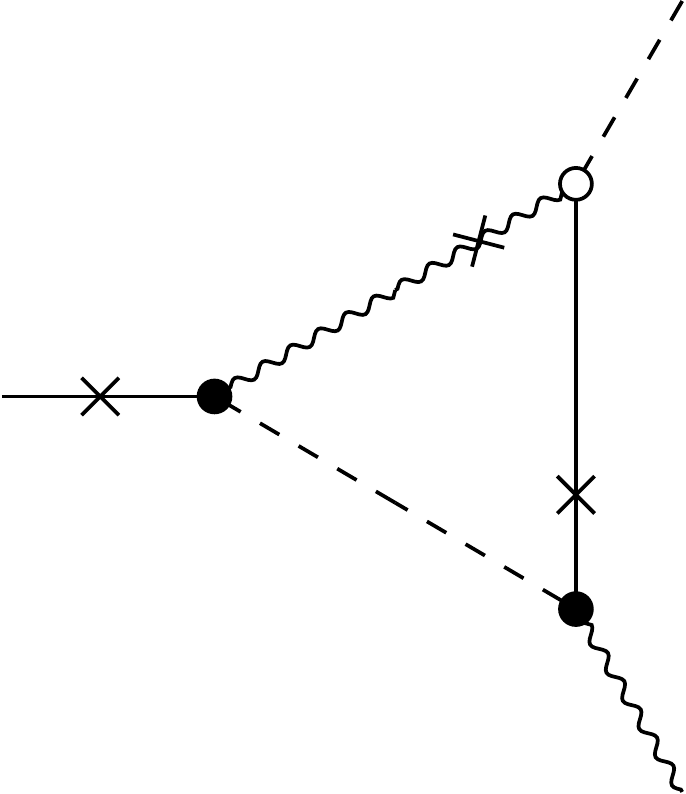}
        \caption*{(a)}
    \end{minipage}
    \begin{minipage}[b]{0.3\columnwidth}
        \centering
        \includegraphics[width=3.5cm]{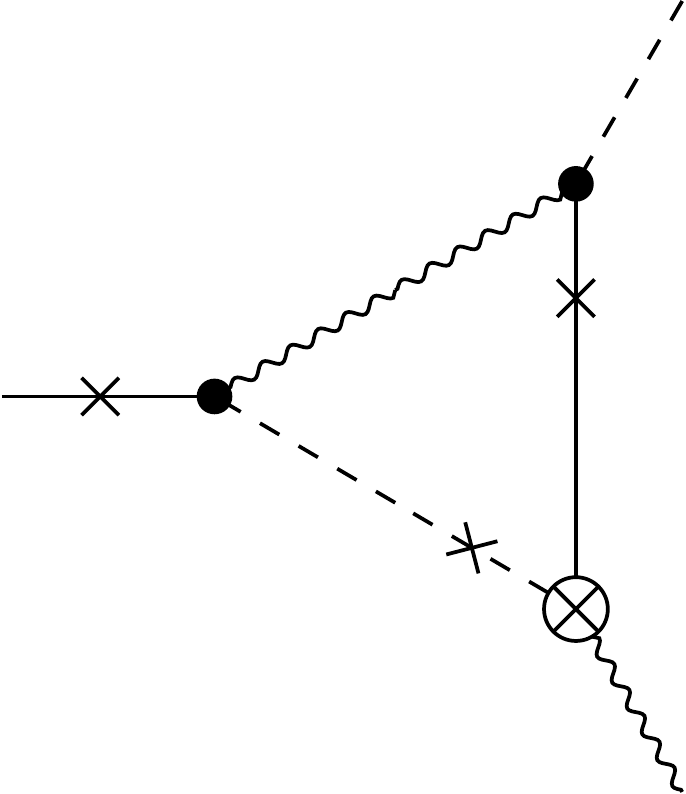}
        \caption*{(b)}
    \end{minipage}
    \begin{minipage}[b]{0.3\columnwidth}
        \centering
        \includegraphics[width=3.5cm]{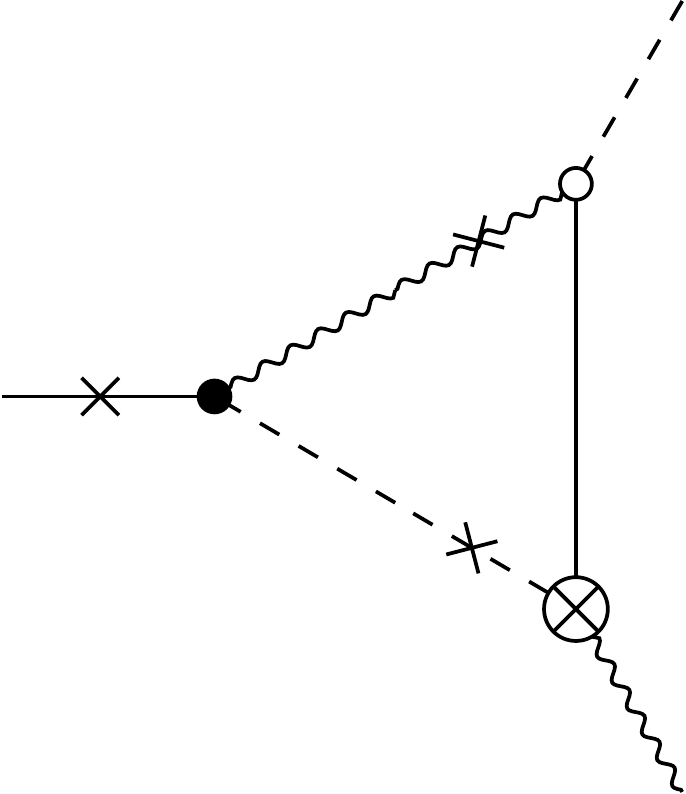}
        \caption*{(c)}
    \end{minipage}

    \hspace{2mm}

    \begin{minipage}[b]{0.3\columnwidth}
        \centering
        \includegraphics[width=3.5cm]{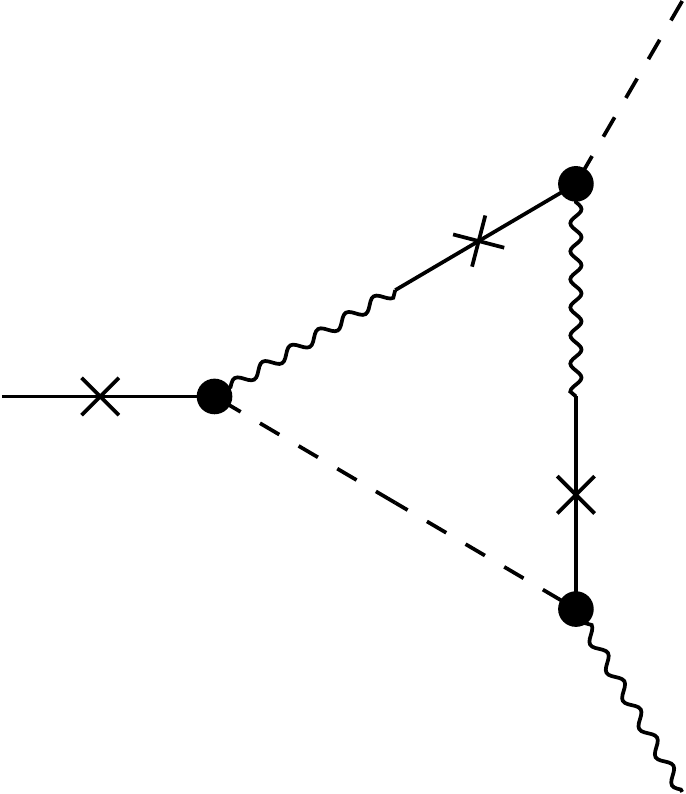}
        \caption*{(d)}
    \end{minipage}
    \begin{minipage}[b]{0.3\columnwidth}
        \centering
        \includegraphics[width=3.5cm]{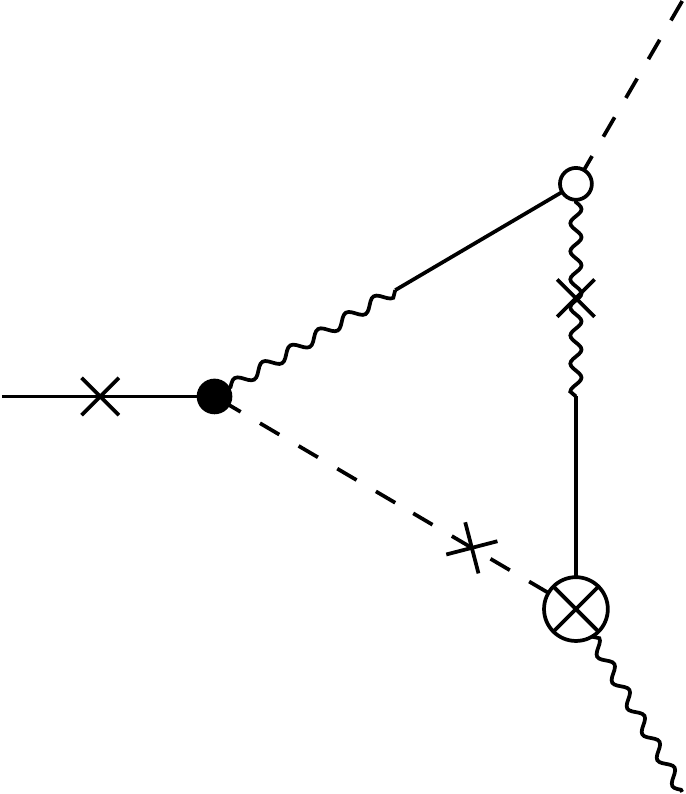}
        \caption*{(e)}
    \end{minipage}
    \begin{minipage}[b]{0.3\columnwidth}
        \centering
        \includegraphics[width=3.5cm]{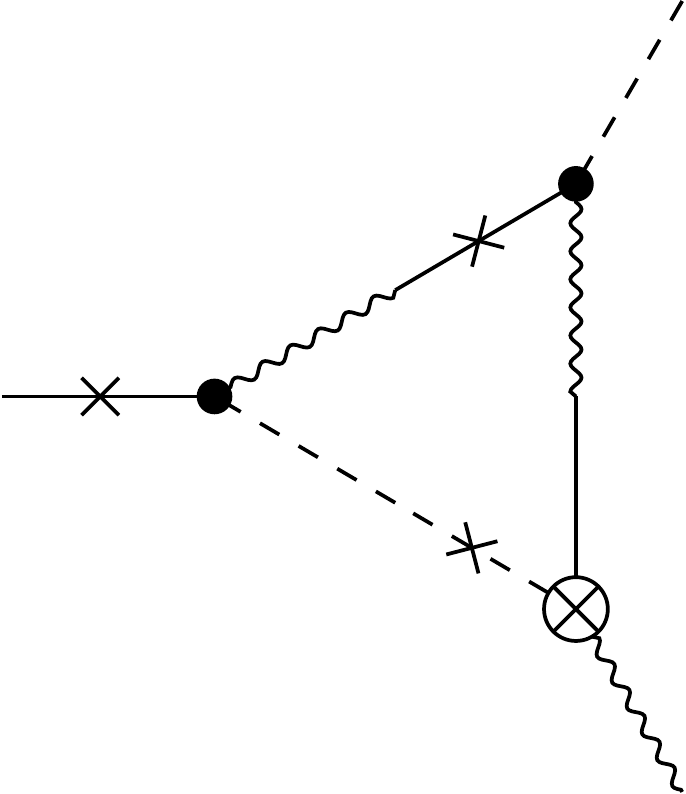}
        \caption*{(f)}
    \end{minipage}
    \caption{One-loop diagrams contributing to $V_{\tilt n e}$}
    \label{fig:tiltne}
\end{figure}

The integral representations corresponding to the diagrams in Fig. \ref{fig:tiltne} are
\begin{align*}
    \mathrm{(a)} &= \lambda_{\theta}^2 \lambda_n \intqo (\bmq + \bmq_1) \cdot (\bmq-\bmq_3) G_{n \tiln} (q+q_1) G_{ee}(-q+q_2) G_{\tilt \theta} (-q+q_3)
    \\
    &\dbeq \lambda_{\theta} \lambda_n \lambda_e \frac{D_{\theta}}{\nu_{\theta}} \intqo (\bmq + \bmq_1) \cdot (\bmq - \bmq_3) G_{n \tiln} (q + q_1) G_{e \tile} (-q+q_2) G_{\theta \tilt} (q-q_3),
\end{align*}
\begin{align*}
    \mathrm{(b)} &= \lambda_{\theta}^2 \lambda_e \intqo (\bmq - \bmq_2) \cdot (\bmq - \bmq_3) G_{nl}(q+q_1) G_{e \tile}(-q+q_2) G_{\theta \tilt} (-q+q_3)
    \\
    &\dbeq \lambda_{\theta} \lambda_n \lambda_e \frac{D_{\theta}}{\nu_{\theta}} \intqo (\bmq - \bmq_2) \cdot (\bmq - \bmq_3) G_{n \tiln} (q+q_1) G_{e \tile} (-q+q_2) G_{\theta \tilt} (-q+q_3),
\end{align*}
\begin{align*}
    \mathrm{(c)} &= - \lambda_{\theta} \lambda_n \lambda_e \intqo [(\bmq + \bmq_1) \cdot (\bmq - \bmq_3)] [(\bmq - \bmq_2) \cdot (\bmq - \bmq_3)] G_{n \tiln}(q+q_1) G_{e \tile} (-q+q_2) G_{\theta \theta} (-q+q_3)
    \\
    &\dbeq - \lambda_{\theta} \lambda_n \lambda_e \frac{D_{\theta}}{\nu_{\theta}} \intqo \frac{[(\bmq + \bmq_1) \cdot (\bmq - \bmq_3)] [(\bmq - \bmq_2) \cdot (\bmq - \bmq_3)]}{|\bmq - \bmq_3|^3} G_{n \tiln} (q + q_1) G_{e \tile} (-q+q_2)
    \\
    &\qquad \times (G_{\theta \tilt}(-q+q_3) + G_{\theta \tilt} (q - q_3)),
\end{align*}
\begin{align*}
    \mathrm{(d)} &= -\lambda_{\theta}^3 \intqo G_{n \tilt} (q+q_1) G_{e e} (-q+q_2) G_{\tilt n} (-q + q_3)
    \\
    &\dbeq \lambda_{\theta} \lambda_n \lambda_e \frac{D_{\theta}}{\nu_{\theta}} \frac{D_e}{\nu_e} \intqo |\bmq - \bmq_3|^2 G_{n \tilt} (q+q_1) G_{e \tile} (-q+q_2) G_{\theta \tiln} (q-q_3),
\end{align*}
\begin{align*}
    \mathrm{(e)} &= - \lambda_{\theta} \lambda_n \lambda_e \intqo [(\bmq + \bmq_1) \cdot (\bmq - \bmq_3)] [(\bmq - \bmq_2) \cdot (\bmq - \bmq_3)] G_{n \theta} (q+q_1) G_{e \tile} (-q+q_2) G_{\theta \tiln} (-q + q_3)
    \\
    &\dbeq - \lambda_{\theta} \lambda_n \lambda_e \frac{D_{\theta}}{\nu_{\theta}} \intqo \frac{[(\bmq + \bmq_1) \cdot (\bmq - \bmq_3)][(\bmq - \bmq_2) \cdot (\bmq - \bmq_3)]}{|\bmq + \bmq_1|^2} G_{n \tilt} (q + q_1) G_{e \tile} (-q + q_2) G_{\theta \tiln} (-q + q_3),
\end{align*}
and
\begin{align*}
    \mathrm{(f)} &= \lambda_{\theta}^2 \lambda_e \cdot \intqo (\bmq - \bmq_2) \cdot (\bmq - \bmq_3)  G_{n \tilt}(q+q_1) G_{e \tile} (-q+q_2) G_{\theta n} (-q+q_3) 
    \\
    &\dbeq \lambda_{\theta} \lambda_n \lambda_e \frac{D_{\theta}}{\nu_{\theta}} \intqo (\bmq - \bmq_2) \cdot (\bmq - \bmq_3) G_{n \tilt} (q+q_1) G_{e \tile} (-q+q_2) (G_{\theta \tiln} (-q+q_3) - G_{\theta \tiln} (q - q_3)).
\end{align*}
Their sum is
\begin{align}
    \mathrm{(a)+(b)+(c)+(d)+(e)+(f)} = 0 + \cdots.
\end{align}
By comparing this with the expression,
\begin{align}
    = - \lambda_{\theta} \delta \lambda_{\theta} \cdot \delta l + \cdots,
\end{align}
we identify
\begin{align}
    \delta \lambda_{\theta} &= 0.
\end{align}

\hspace{1cm}

\noindent -- Correction to $V_{\tiln \theta e}$

\begin{figure}[h]
    \centering
    \begin{minipage}[b]{0.3\columnwidth}
        \centering
        \includegraphics[width=3.5cm]{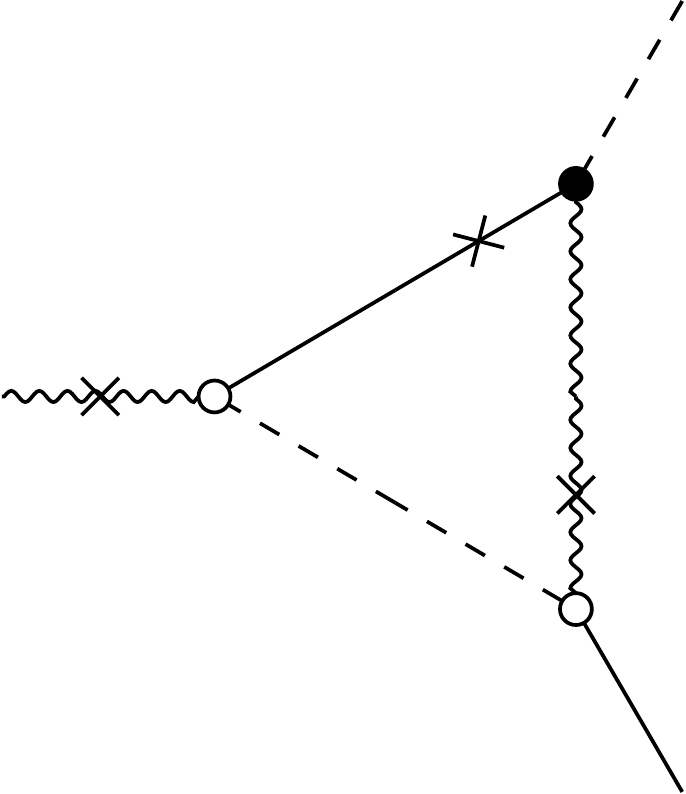}
        \caption*{(a)}
    \end{minipage}
    \begin{minipage}[b]{0.3\columnwidth}
        \centering
        \includegraphics[width=3.5cm]{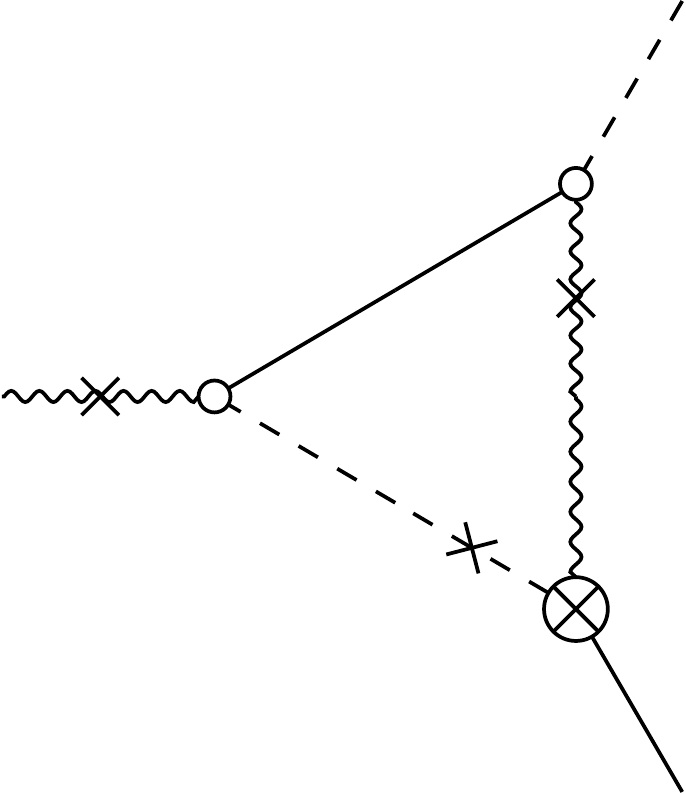}
        \caption*{(b)}
    \end{minipage}
    \begin{minipage}[b]{0.3\columnwidth}
        \centering
        \includegraphics[width=3.5cm]{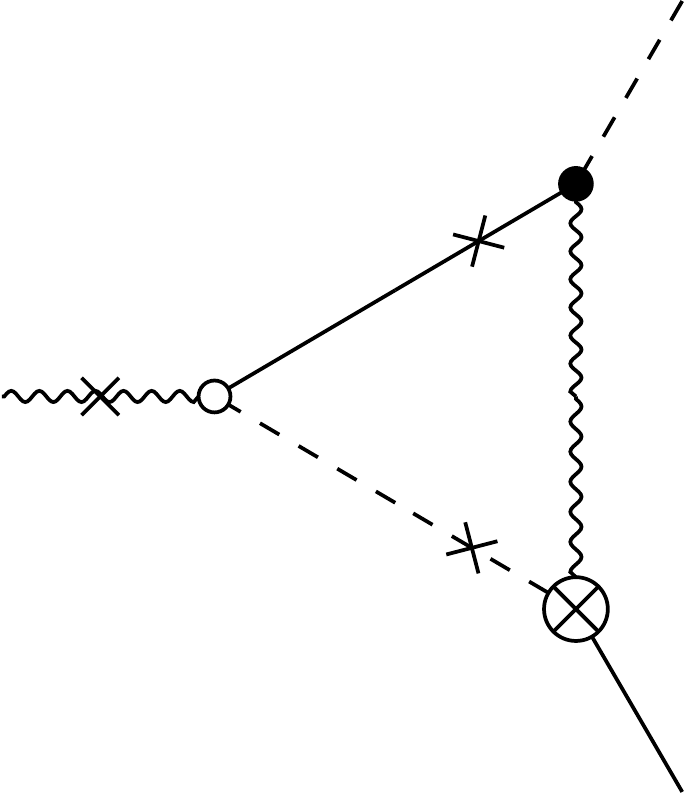}
        \caption*{(c)}
    \end{minipage}

    \hspace{2mm}

    \begin{minipage}[b]{0.3\columnwidth}
        \centering
        \includegraphics[width=3.5cm]{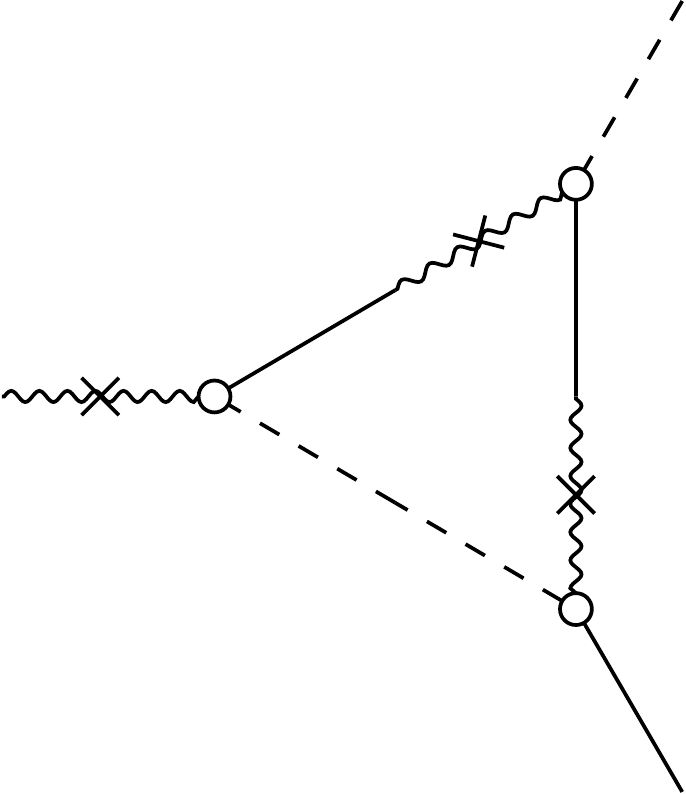}
        \caption*{(d)}
    \end{minipage}
    \begin{minipage}[b]{0.3\columnwidth}
        \centering
        \includegraphics[width=3.5cm]{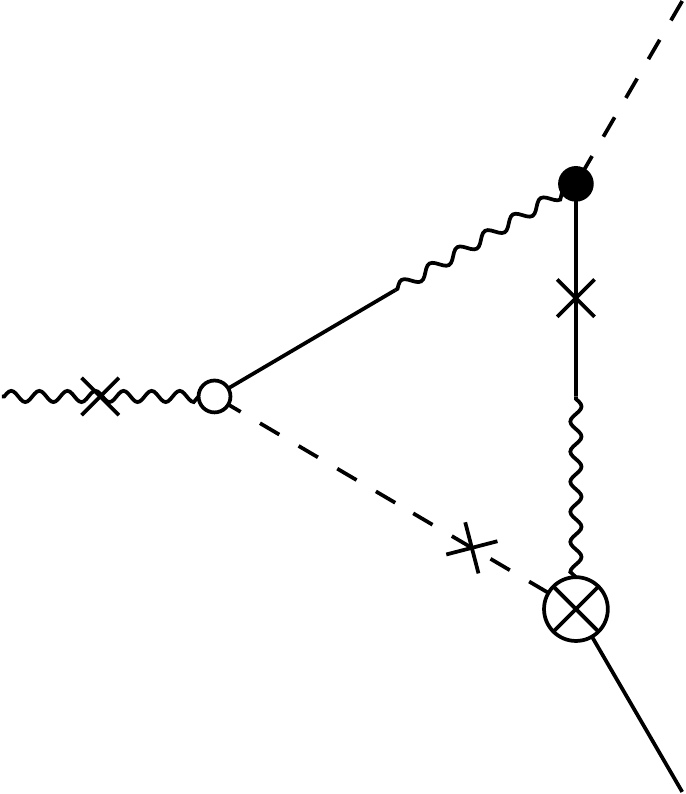}
        \caption*{(e)}
    \end{minipage}
    \begin{minipage}[b]{0.3\columnwidth}
        \centering
        \includegraphics[width=3.5cm]{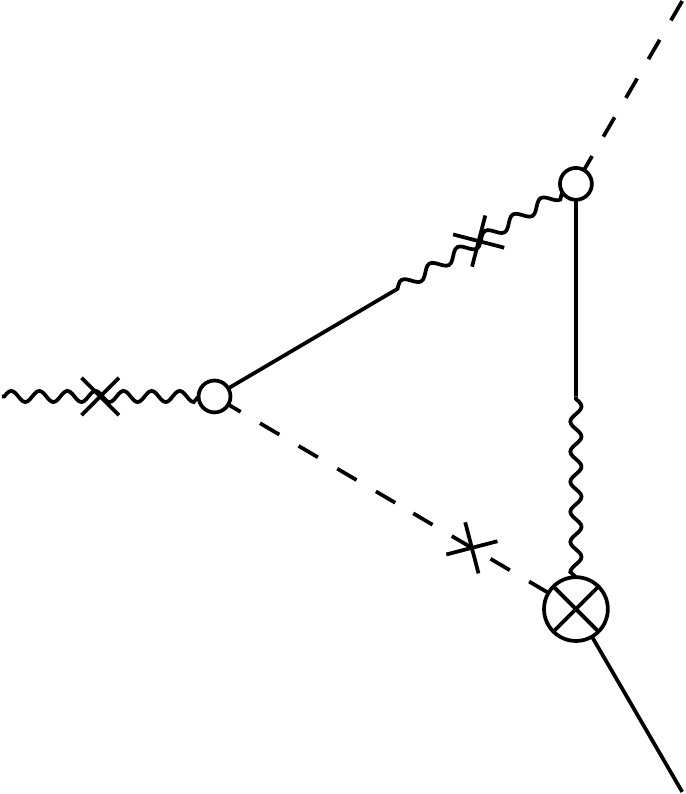}
        \caption*{(f)}
    \end{minipage}
    \caption{One-loop diagrams contributing to $V_{\tiln \theta n}$}
    \label{fig:tilnte}
\end{figure}

The integral representations corresponding to the diagrams in Fig. \ref{fig:tilnte} are
\begin{align*}
    \mathrm{(a)} &= \lambda_{\theta} \lambda_n^2 \intqo [\bmk_1 \cdot (\bmq + \bmq_1)] [\bmk_2 \cdot (\bmq - \bmq_3)] G_{\theta \tilt} (q+q_1) G_{ee} (-q+q_2) G_{\tiln n} (-q+q_3)
    \\
    &\dbeq \lambda_{\theta} \lambda_n \lambda_e \frac{D_n}{\nu_n} \intqo [\bmk_1 \cdot (\bmq + \bmq_1)] [\bmk_2 \cdot (\bmq - \bmq_3)] G_{\theta \tilt} (q+q_1) G_{e \tile} (-q+q_2) G_{n \tiln} (q-q_3),
\end{align*}
\begin{align*}
    \mathrm{(b)} &= \lambda_n^2 \lambda_e \intqo [\bmk_1 \cdot (\bmq + \bmq_1)] [\bmk_2 \cdot (\bmq - \bmq_2)] [(\bmq + \bmq_1) \cdot (\bmq - \bmq_3)]
    \\
    &\qquad \times G_{\theta \theta} (q+q_1) G_{e \tile} (-q+q_2) G_{n \tiln} (-q+q_3)
    \\
    &\dbeq \lambda_{\theta} \lambda_n \lambda_e \frac{D_n}{\nu_n} \intqo [\bmk_1 \cdot (\bmq + \bmq_1)] [\bmk_2 \cdot (\bmq - \bmq_2)] \frac{(\bmq + \bmq_1) \cdot (\bmq - \bmq_3)}{|\bmq + \bmq_1|^2}
    \\
    &\qquad \times G_{\theta \tilt} (q+q_1) G_{e \tile} (-q+q_2) G_{n \tiln} (-q+q_3),
\end{align*}
\begin{align*}
    \mathrm{(c)} &= - \lambda_{\theta} \lambda_n \lambda_e \intqo [\bmk_1 \cdot (\bmq + \bmq_1)] [\bmk_2 \cdot (\bmq - \bmq_2)] G_{\theta \tilt} (q+q_1) G_{e \tile} (-q+q_2) G_{nn}(-q+q_3)
    \\
    &\dbeq - \lambda_{\theta} \lambda_n \lambda_e \frac{D_n}{\nu_n} \intqo [\bmk_1 \cdot (\bmq + \bmq_1)] [ \bmk_2 \cdot (\bmq - \bmq_2)] G_{\theta \tilt} (q+q_1) G_{e \tile} (-q+q_2) 
    \\
    &\qquad \times (G_{n \tiln} (-q+q_3) + G_{n \tiln} (q-q_3)),
\end{align*}
\begin{align*}
    \mathrm{(d)} &= - \lambda_n^3 \intqo [\bmk_1 \cdot (\bmq + \bmq_1)] [\bmk_2 \cdot (\bmq - \bmq_3)] [(\bmq + \bmq_1) \cdot (\bmq - \bmq_3)] 
    \\
    &\qquad \times G_{\theta \tiln} (q+q_1) G_{ee}(-q+q_2) G_{\theta \tiln} (q - q_3)
    \\
    &\dbeq \lambda_{\theta} \lambda_n \lambda_e \frac{D_n}{\nu_n} \intqo [\bmk_1 \cdot (\bmq + \bmq_1)] [\bmk_2 \cdot (\bmq - \bmq_3)] \frac{(\bmq + \bmq_1) \cdot (\bmq - \bmq_3)}{|\bmq - \bmq_3|^2}
    \\
    &\qquad \times G_{\theta \tiln} (q+q_1) G_{e \tile} (-q+q_2) G_{n \tilt} (q-q_3),
\end{align*}
\begin{align*}
    \mathrm{(e)} &= - \lambda_{\theta} \lambda_n \lambda_e \intqo [\bmk_1 \cdot (\bmq + \bmq_1)] [\bmk_2 \cdot (\bmq - \bmq_2)] G_{\theta n}(q+q_1) G_{e \tile} (-q+q_2) G_{n \tilt} (-q+q_3)
    \\
    &\dbeq - \lambda_{\theta} \lambda_n \lambda_e \frac{D_n}{\nu_n} \intqo [\bmk_1 \cdot (\bmq + \bmq_1)] [\bmk_2 \cdot (\bmq - \bmq_2)] G_{\theta \tiln} (q+q_1) G_{e \tile} (-q+q_2) G_{n \tilt} (-q+q_3),
\end{align*}
and
\begin{align*}
    \mathrm{(f)} &= \lambda_n^2 \lambda_e \intqo [\bmk_1 \cdot (\bmq + \bmq_1)] [\bmk_2 \cdot (\bmq - \bmq_3)] [(\bmq + \bmq_1) \cdot (\bmq - \bmq_3)]
    \\
    &\qquad \times G_{\theta \tiln} (q+q_1) G_{e \tile} (-q+q_2) G_{n \theta} (-q+q_3)
    \\
    &\dbeq - \lambda_{\theta} \lambda_n \lambda_e \frac{D_n}{\nu_n} \intqo [\bmk_1 \cdot (\bmq + \bmq_1)] [\bmk_2 \cdot (\bmq - \bmq_2)] \frac{[(\bmq + \bmq_1) \cdot (\bmq - \bmq_3)]}{|\bmq - \bmq_3|^2}
    \\
    &\qquad \times G_{\theta \tiln} (q+q_1) G_{e \tile} (-q+q_2) (G_{n \tilt} (q -q_3) - G_{n \tilt} (-q+q_3)).
\end{align*}
Their sum is 
\begin{align}
    \mathrm{(a)+(b)+(c)+(d)+(e)+(f)} = 0 \cdot (\bmk_1 \cdot \bmk_2) + \cdots.
\end{align}
By comparing this with the expression
\begin{align}
     = - \lambda_n \cdot \delta \lambda_n (\bmk_1 \cdot \bmk_2) l + \cdots,
\end{align}
we identify
\begin{align}
    \delta \lambda_n &= 0.
\end{align}

\hspace{1cm}

\noindent -- Correction to $V_{\tile \theta n}$

\begin{figure}[h]
    \centering
    \begin{minipage}[b]{0.3\columnwidth}
        \centering
        \includegraphics[width=3.5cm]{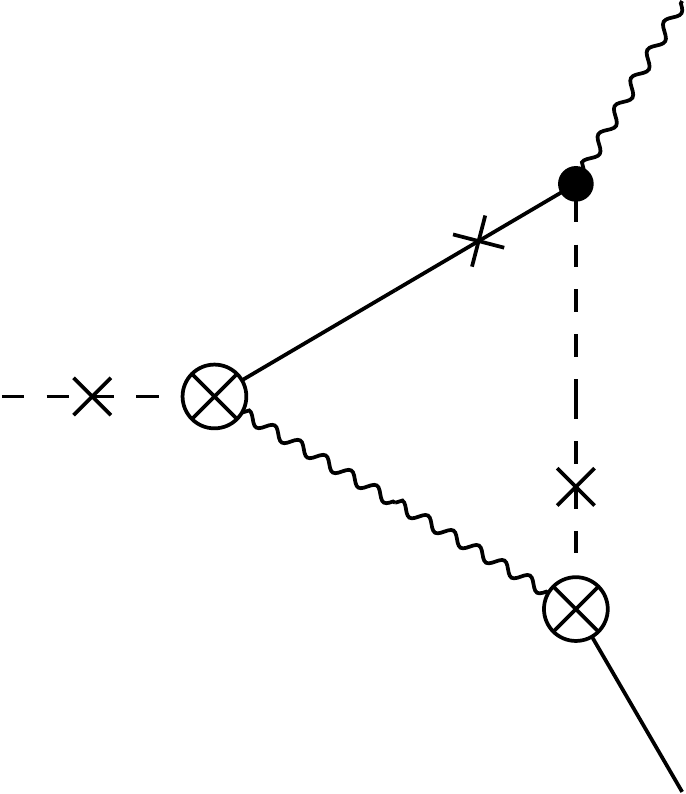}
        \caption*{(a)}
    \end{minipage}
    \begin{minipage}[b]{0.3\columnwidth}
        \centering
        \includegraphics[width=3.5cm]{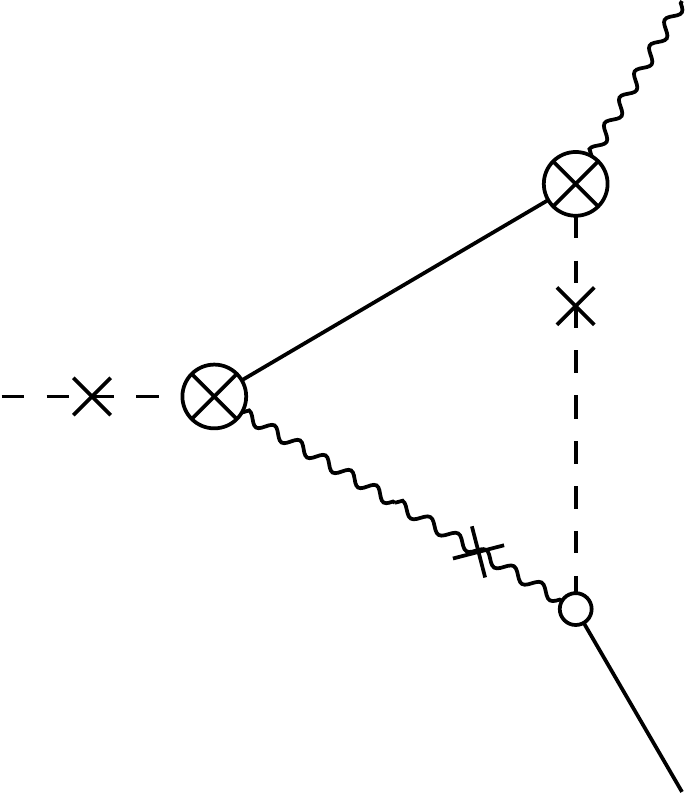}
        \caption*{(b)}
    \end{minipage}
    \begin{minipage}[b]{0.3\columnwidth}
        \centering
        \includegraphics[width=3.5cm]{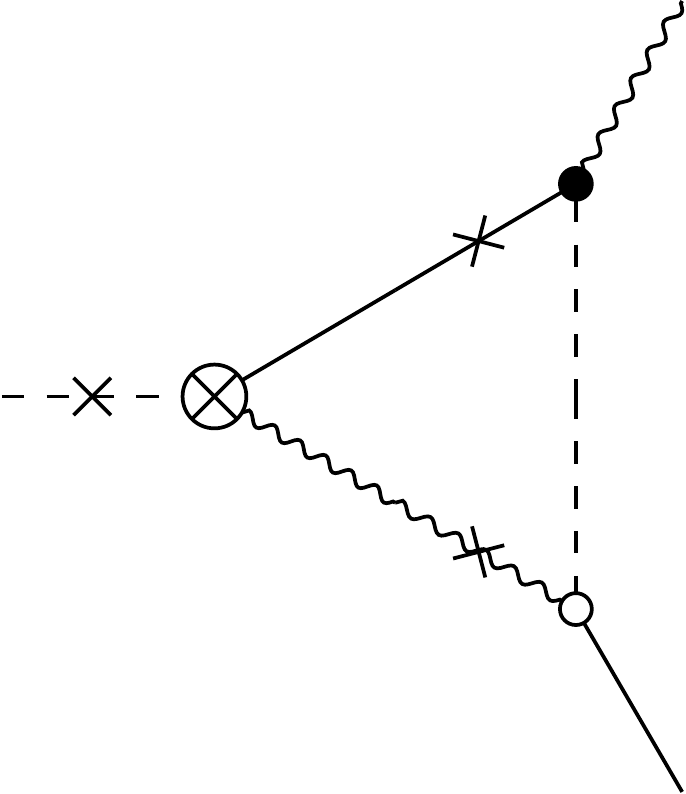}
        \caption*{(c)}
    \end{minipage}

    \hspace{2mm}

    \begin{minipage}[b]{0.3\columnwidth}
        \centering
        \includegraphics[width=3.5cm]{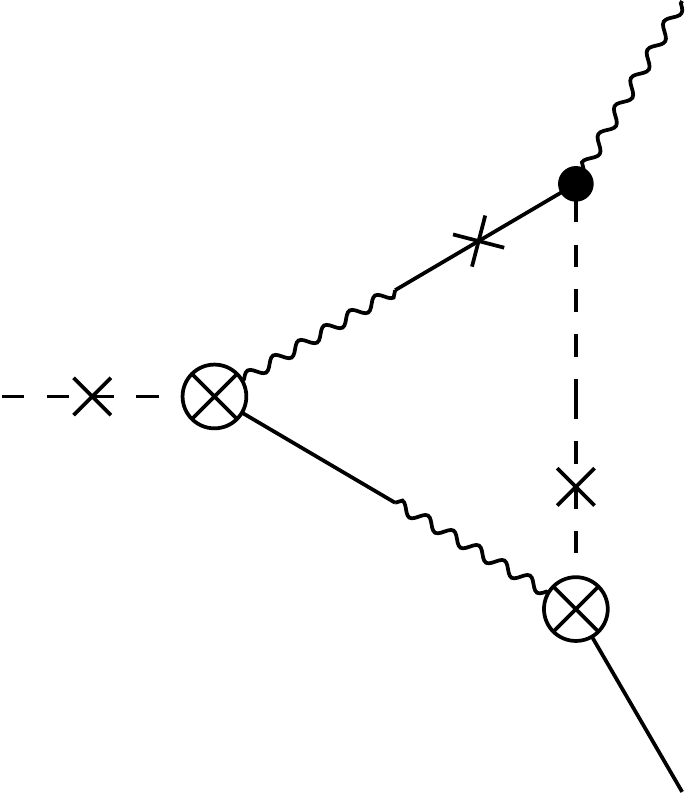}
        \caption*{(d)}
    \end{minipage}
    \begin{minipage}[b]{0.3\columnwidth}
        \centering
        \includegraphics[width=3.5cm]{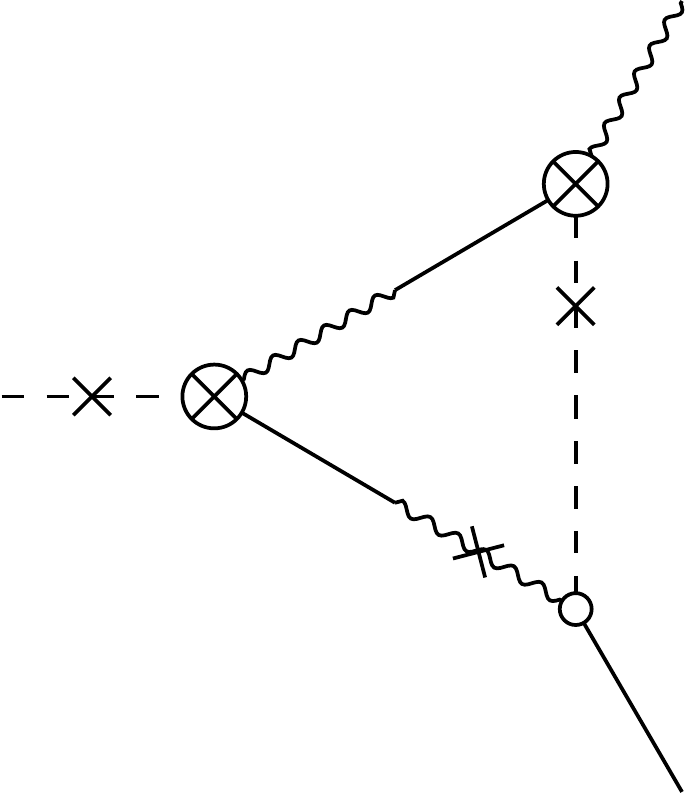}
        \caption*{(e)}
    \end{minipage}
    \begin{minipage}[b]{0.3\columnwidth}
        \centering
        \includegraphics[width=3.5cm]{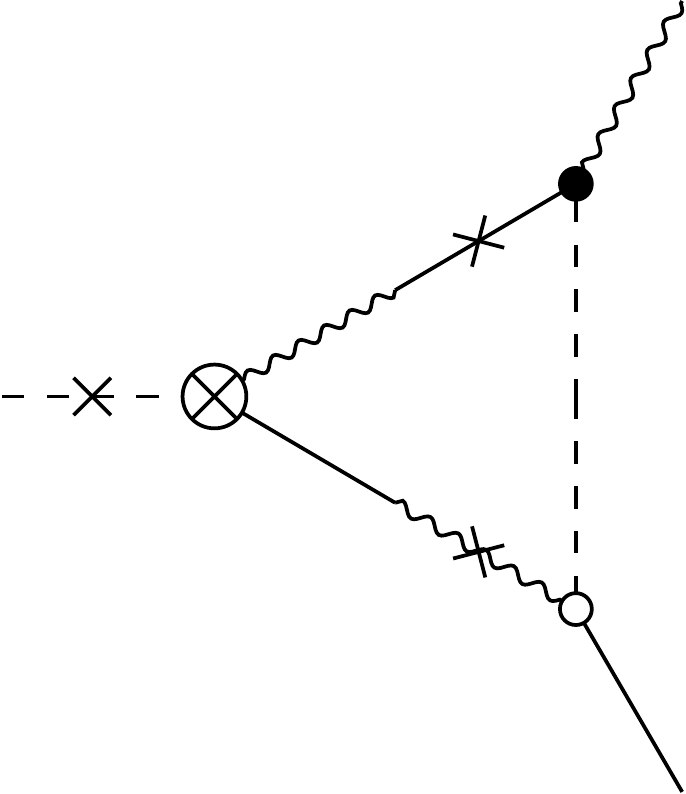}
        \caption*{(f)}
    \end{minipage}
    \caption{One-loop diagrams contributing to $V_{\tile \theta n}$}
    \label{fig:tiletn}
\end{figure}

The integral representations corresponding to the diagrams in Fig. \ref{fig:tiletn} are 
\begin{align*}
    \mathrm{(a)} &= \lambda_{\theta} \lambda_e^2 \intqo [\bmk_1 \cdot (\bmq + \bmq_1)] [\bmk_2 \cdot (\bmq - \bmq_3)] G_{\theta \tilt} (q+q_1) G_{nn} (-q+q_2) G_{\tile e} (-q+q_3)
    \\
    &\dbeq \lambda_{\theta} \lambda_n \lambda_e \frac{D_e}{\nu_e} \intqo [\bmk_1 \cdot (\bmq + \bmq_1)] [\bmk_2 \cdot (\bmq - \bmq_3)] G_{\theta \tilt} (q+q_1) G_{n \tiln} (-q+q_2) G_{e \tile} (q-q_3),
\end{align*}
\begin{align*}
    \mathrm{(b)} &= \lambda_n \lambda_e^2 \intqo [\bmk_1 \cdot (\bmq + \bmq_1)] [\bmk_2 \cdot (\bmq - \bmq_2)] [(\bmq + \bmq_1) \cdot (\bmq - \bmq_3)]
    \\
    &\qquad \times G_{\theta \theta} (q+q_1) G_{n \tiln} (-q+q_2) G_{e \tile} (-q+q_3)
    \\
    &\dbeq \lambda_{\theta} \lambda_n \lambda_e \frac{D_e}{\nu_e} \intqo [\bmk_1 \cdot (\bmq + \bmq_1)] [\bmk_2 \cdot (\bmq - \bmq_2)] \frac{(\bmq + \bmq_1) \cdot (\bmq - \bmq_3)}{|\bmq + \bmq_1|^2}
    \\
    &\qquad \times G_{\theta \tilt} (q+q_1) G_{n \tiln} (-q + q_2) G_{e \tile} (-q+q_3),
\end{align*}
\begin{align*}
    \mathrm{(c)} &= - \lambda_{\theta} \lambda_n \lambda_e \intqo [\bmk_1 \cdot (\bmq + \bmq_1)] [\bmk_2 \cdot (\bmq - \bmq_2)] G_{\theta \tilt} (q+q_1) G_{n \tiln} (-q+q_2) G_{ee} (-q+q_3)
    \\
    &\dbeq - \lambda_{\theta} \lambda_n \lambda_e \intqo [\bmk_1 \cdot (\bmq + \bmq_1)] [\bmk_2 \cdot (\bmq - \bmq_2)] G_{\theta \tilt} (q+q_1) G_{n \tiln} (-q+q_2)
    \\
    &\qquad \times (G_{e \tile} (-q+q_3) + G_{e \tile} (q - q_3)),
\end{align*}
\begin{align*}
    \mathrm{(d)} &= - \lambda_{\theta} \lambda_e^2 \intqo [\bmk_1 \cdot (\bmq - \bmq_2)] [\bmk_2 \cdot (\bmq - \bmq_3)] G_{n \tilt} (q+q_1) G_{\theta n} (-q+q_2) G_{\tile e} (-q+q_3)
    \\
    &\dbeq - \lambda_{\theta} \lambda_n \lambda_e \frac{D_e}{\nu_e} \intqo [\bmk_1 \cdot (\bmq - \bmq_2)] [\bmk_2 \cdot (\bmq - \bmq_3)] G_{n \tilt} (q+q_1) G_{\theta \tiln} (-q+q_2) G_{e \tile} (q-q_3),
\end{align*}
\begin{align*}
    \mathrm{(e)} &= - \lambda_n \lambda_e^2 \intqo [\bmk_1 \cdot (\bmq - \bmq_2)] [\bmk_2 \cdot (\bmq - \bmq_2)] [(\bmq + \bmq_1) \cdot (\bmq - \bmq_3)]
    \\
    &\qquad \times G_{n \theta} (q+q_1) G_{\theta \tiln} (-q+q_2) G_{e \tile} (-q+q_3)
    \\
    &\dbeq - \lambda_{\theta} \lambda_n \lambda_e \frac{D_e}{\nu_e} \intqo [\bmk_1 \cdot (\bmq - \bmq_2)] [\bmk_2 \cdot (\bmq - \bmq_2)] \frac{(\bmq + \bmq_1) \cdot (\bmq - \bmq_3)}{|\bmq + \bmq_1|^2}
    \\
    &\qquad \times G_{n \tilt} (q+q_1) G_{\theta \tiln} (-q+q_2) G_{e \tile} (-q+q_3),
\end{align*}
and
\begin{align*}
    \mathrm{(f)} &= \lambda_{\theta} \lambda_n \lambda_e \intqo [\bmk_1 \cdot (\bmq - \bmq_2)] [\bmk_2 \cdot (\bmq - \bmq_2)] G_{n \theta} (q+q_1) G_{\theta \tiln} (-q+q_2) G_{ee} (-q+q_3)
    \\
    &\dbeq \lambda_{\theta} \lambda_n \lambda_e \frac{D_e}{\nu_e} \intqo [\bmk_1 \cdot (\bmq - \bmq_2)] [\bmk_2 \cdot (\bmq - \bmq_2)] G_{n \tilt} (q+q_1) G_{\theta \tiln} (-q+q_2)
    \\
    &\qquad \times (G_{e \tile} (-q+q_3) + G_{e \tile} (q-q_3))
\end{align*}
Their sum is
\begin{align}
    \mathrm{(a)+(b)+(c)+(d)+(e)+(f)} = 0 \cdot (\bmk_1 \cdot \bmk_2) + \cdots.
\end{align}
By comparing this with the expression,
\begin{align}
    = - \lambda_e \delta \lambda_n (\bmk_1 \cdot \bmk_2) \cdot \delta l + \cdots
\end{align}
we identify
\begin{align}
    \delta \lambda_e &= 0.
\end{align}

\section{Time-reversal symmetry and the Ward-Takahashi identities}
\label{app:db}

We argue that the validity of the detailed balance condition \eqref{eq:db} along the RG flow at one-loop order follows from time-reversal symmetry. Let us define the time-reversal transformation $\sfT$ \cite{AndreanovBiroliLefevre2006} as
\begin{align}
    \mathsf{T}:
    \begin{cases}
    \theta (\bmk, \omega) &\mapsto \theta^{\prime}(\bmk,\omega) = \theta (\bmk, - \omega),
    \\
    n (\bmk, \omega) &\mapsto n^{\prime}(\bmk,\omega) = - n (\bmk, - \omega),
    \\
    e(\bmr, t) &\mapsto e^{\prime}(\bmk, \omega) = e(\bmk, - \omega),
    \\
    \tilt (\bmk, \omega) &\mapsto \tilt^{\prime}(\bmk, \omega) = - \tilt (\bmk, - \omega) + \frac{\nu_{\theta}}{D_{\theta}} \bmk^2 \theta (\bmk, - \omega),
    \\
    \tiln (\bmk, \omega) &\mapsto \tiln^{\prime}(\bmk, \omega) = \tiln (\bmk, -\omega) - \frac{\nu_n}{D_n} n(\bmk, - \omega),
    \\
    \tile(\bmk, \omega) &\mapsto \tile^{\prime}(\bmk, \omega) = - \tile(\bmk, - \omega) + \frac{\nu_e}{D_e} e (\bmk, - \omega).
    \end{cases}
\end{align}
The shift in the response fields can be understood as follows. The time-reversal transformation for $\theta$, $n$, and $e$ reverses the sign of the dissipative terms, which is expected from the meaning of ``dissipative''. However, by simultaneously shifting the response fields $\tilde{\theta}$, $\tilde{n}$, and $e^{\prime}$, this sign inversion can be compensated, restoring the invariance of the action except for the terms associated with the reversible couplings.

Under this transformation, the action changes as
\begin{align}
    \calA [\sfT \Phi ; \Lambda, \calP ] &= \calA [\Phi ; \Lambda, \calP] + \int_k \left[ \left( c_{\rms} \frac{\nu_{\theta}}{D_{\theta}} - d_{\rms} \frac{\nu_n}{D_n} \right) \theta (\bmk, \omega) l(\bmk, - \omega)  \right.
    \notag \\
    &\qquad \left. + \left( \lambda_{\theta} \frac{\nu_{\theta}}{D_{\theta}} \bmk_1 + \lambda_n \frac{\nu_n}{D_n} \bmk_2 + \lambda_e \frac{\nu_e}{D_e} \bmk_3 \right) \cdot \bmk_1 \theta (\bmk_1) l(\bmk_2) e(\bmk_3) \hat{\delta}(k_1+k_2+k_3)  \right].
\end{align}
A sufficient condition for the action to be invariant under the time-reversal transformation is
\begin{align*}
    c_{\rms} \frac{\nu_{\theta}}{D_{\theta}} = d_{\rms} \frac{\nu_n}{D_n}, \quad \lambda_{\theta} \frac{\nu_{\theta}}{D_{\theta}} = \lambda_n \frac{\nu_n}{D_n} = \lambda_e \frac{\nu_e}{D_e},
\end{align*}
which is identical to Eq.~\eqref{eq:db}. Suppose that the detailed balance condition ~\eqref{eq:db} is satisfied at the initial scale $\Lambda$. Since $\sfT \Phi = \sfT \Phi^{<} + (\sfT \Phi)^{>}$, it follows that
\begin{align}
    \calA [\sfT \Phi^{<} ; \Lambda^{\prime}, \calP^{\prime}] &= - \ln \int d[(\sfT \Phi)^{>}] \exp \left[ - \calA [\sfT \Phi ; \Lambda, \calP] \right]
    \notag \\
    &= - \ln \int d[\Phi^{>}] \left| \frac{\delta (\sfT \Phi)^{>}}{\delta \Phi^{>}} \right| \exp \left[ - \calA [\Phi ; \Lambda, \calP] \right]
    \notag \\ \label{eq:Tinv}
    &= \calA [\Phi^{<} ; \Lambda^{\prime}, \calP^{\prime}] - \ln \left| \frac{\delta (\sfT \Phi)^{>}}{\delta \Phi^{>}} \right|.
\end{align}
Because $\sfT$ is a linear transformation, the Jacobian term is constant. By differentiating both sides of Eq.~\eqref{eq:Tinv} twice with respect to the field and the response field, and by taking $\Phi = 0$, we obtain
\begin{align} \label{eq:fdtt}
    [\bmG^{-1}(\bmk, \omega)]_{\tilt \theta} &= - [\bmG^{-1}(\bmk, - \omega)]_{\tilt \theta} - \frac{\nu_{\theta}}{D_{\theta}} \bmk^2 [\bmG^{-1}(\bmk, -\omega)]_{\tilt \tilt},
    \\ \label{eq:fdtn}
    [\bmG^{-1}(\bmk, \omega)]_{\tiln n} &= - [\bmG^{-1}(\bmk, -\omega)]_{\tiln n} - \frac{\nu_n}{D_n} [\bmG^{-1}(\bmk, -\omega)]_{\tiln \tiln},
    \\ \label{eq:fdte}
    [\bmG^{-1}(\bmk, \omega)]_{\tile e} &= - [\bmG^{-1}(\bmk, - \omega)]_{\tile e} - \frac{\nu_e}{D_e} [\bmG^{-1}(\bmk, -\omega)]_{\tile \tile},
    \\
    [\bmG^{-1}(\bmk, \omega)]_{n \theta} &= - [\bmG^{-1}(\bmk, - \omega)]_{n \theta} - \frac{\nu_n}{D_n} [\bmG^{-1}(\bmk, -\omega)]_{\tiln \theta}
    \notag \\ \label{eq:fdtcd}
    &\qquad - \frac{\nu_{\theta}}{D_{\theta}} \bmk^2 [\bmG^{-1}(\bmk, - \omega)]_{n \tilt} - \frac{\nu_n \nu_{\theta}}{D_n D_{\theta}} \bmk^2 [\bmG^{-1}(\bmk, - \omega)]_{\tiln \tilt}.
\end{align}
Here, we have used the following chain rules,
\begin{align*}
    \frac{\delta}{\delta \theta (\bmk, \omega)} &= \frac{\delta}{\delta \theta^{\prime} (\bmk, - \omega)} + \frac{\nu_{\theta}}{D_{\theta}} \bmk^2 \frac{\delta}{\delta \tilt^{\prime}(\bmk, - \omega)},
    \\
    \frac{\delta}{\delta \tilt (\bmk, \omega)} &= - \frac{\delta}{\delta \tilt^{\prime}(\bmk, - \omega)},
    \\
    \frac{\delta}{\delta n (\bmk, \omega)} &= - \frac{\delta}{\delta n^{\prime}(\bmk, \omega)} - \frac{\nu_n}{D_n} \frac{\delta}{\delta \tiln^{\prime}(\bmk, -\omega)}
    \\
    \frac{\delta}{\delta \tiln (\bmk, \omega)} &= \frac{\delta}{\delta \tiln^{\prime}(\bmk, -\omega)},
    \\
    \frac{\delta}{\delta e (\bmk, \omega)} &= \frac{\delta}{\delta e^{\prime}(\bmk, -\omega)} + \frac{\nu_e}{D_e} \frac{\delta}{\delta \tile^{\prime}(\bmk, -\omega)},
    \\
    \frac{\delta}{\delta \tile (\bmk, \omega)} &= - \frac{\delta}{\delta \tile^{\prime}(\bmk, -\omega)}.
\end{align*}
The first three identities, Eqs.~\eqref{eq:fdtt}, \eqref{eq:fdtn}, and \eqref{eq:fdte}, yield the fluctuation-dissipation relations, respectively,
\begin{align} \label{eq:fdt0}
    \frac{\nu_{\theta}^{\prime}}{D_{\theta}^{\prime}} = \frac{\nu_{\theta}}{D_{\theta}}, \quad \frac{\nu_n^{\prime}}{D_n^{\prime}} = \frac{\nu_n}{D_n}, \quad \frac{\nu_{e}^{\prime}}{D_{e}^{\prime}} = \frac{\nu_{e}}{D_{e}},
\end{align}
leading to Eq.~\eqref{eq:fdt} in the main text. At one-loop order, we have $[\bmG^{-1}]_{n \theta} = 0$ and $[\bmG^{-1}]_{\tiln \tilt} \sim \bmk$. Thus, the last equality \eqref{eq:fdtcd} gives
\begin{align}
    c_{\rms}^{\prime} \frac{\nu_{\theta}}{D_{\theta}} &= d_{\rms}^{\prime} \frac{\nu_n}{D_{n}}.
\end{align}
Combining this with Eq.~\eqref{eq:fdt0}, we arrive at
\begin{align}
    c_{\rms}^{\prime} \frac{\nu_{\theta}^{\prime}}{D_{\theta}^{\prime}} &= d_{\rms}^{\prime} \frac{\nu_n^{\prime}}{D_{n}^{\prime}}.
\end{align}
By differentiating both sides of Eq.~\eqref{eq:Tinv} three times with respect to $\theta (k_1)$, $n(k_2)$, and $e(k_3)$, and by taking $\Phi = 0$, we find
\begin{align}
    V_{\theta n e}^{\prime} &= - V_{\theta n e} - \frac{\nu_{\theta}}{D_{\theta}} \bmk_1^2 V_{\tilt n e} - \frac{\nu_n}{D_n} V_{\theta \tiln e} - \frac{\nu_e}{D_e} V_{\theta n \tile} 
    \notag \\ \label{eq:Vtne}
    &\qquad - \frac{\nu_{\theta}}{D_{\theta}} \bmk_1^2 \frac{\nu_n}{D_n} V_{\tilt \tiln e} - \frac{\nu_n}{D_n} \frac{\nu_e}{D_e} V_{\theta \tiln \tile} - \frac{\nu_e}{D_e} \frac{\nu_{\theta}}{D_{\theta}} \bmk_1^2 V_{\tilt n \tile} - \frac{\nu_{\theta}}{D_{\theta}} \bmk_1^2 \frac{\nu_n}{D_n} \frac{\nu_e}{D_e} V_{\tilt \tiln \tile},
\end{align}
where $V^{\prime} = V(\bmk_1, - \omega_1 , \bmk_2, - \omega_2)$. At one-loop order, $V_{\theta n e} = 0$, $V_{\tilt n e} \sim \bmk$, $V_{\theta \tiln \tile} \sim \bmk^3$, $V_{\tilt n \tile} \sim \bmk$, and $V_{\tilt \tiln \tile} \sim \bmk^2$. Thus, Eq.~\eqref{eq:Vtne} yields
\begin{align}
    \left( \lambda_e^{\prime} \frac{\nu_e}{D_e} - \lambda_{\theta}^{\prime} \frac{\nu_{\theta}}{D_{\theta}} \right) \bmk_1^2 + \left( \lambda_e^{\prime} \frac{\nu_e}{D_e} - \lambda_n^{\prime} \frac{\nu_n}{D_n} \right) (\bmk_1 \cdot \bmk_2) = 0.
\end{align}
Combining this with Eq.~\eqref{eq:fdt0}, we have
\begin{align}
    \lambda_{\theta}^{\prime} \frac{\nu_{\theta}^{\prime}}{D_{\theta}^{\prime}} = \lambda_n^{\prime} \frac{\nu_n^{\prime}}{D_n^{\prime}} = \lambda_e^{\prime} \frac{\nu_e^{\prime}}{D_e^{\prime}}.
\end{align}
Thus, the detailed balance condition is preserved along the RG flow within the perturbation theory.

\section*{References}
\bibliographystyle{iopart-num}

\begin{thebibliography}{10}
    \expandafter\ifx\csname url\endcsname\relax
      \def\url#1{{\tt #1}}\fi
    \expandafter\ifx\csname urlprefix\endcsname\relax\def\urlprefix{URL }\fi
    \providecommand{\eprint}[2][]{\url{#2}}
    
    \bibitem{deGrootMazur19962}
    de~Groot S~R and Mazur P 1962 {\em Non-equilibrium thermodynamics\/}
      (Amsterdam, North-Holland)
    
    \bibitem{Chang2008}
    Chang C~W, Okawa D, Garcia H, Majumdar A and Zettl A 2008 {\em Phys. Rev.
      Lett.\/} {\bf 101} 075903
    
    \bibitem{Xu2014}
    Xu X, Pereira L~F~C, Wang Y, Wu J, Zhang K, Zhao X, Bae S, Tinh~Bui C, Xie R,
      Thong J~T~L, Hong B~H, Loh K~P, Donadio D, Li B and {\"O}zyilmaz B 2014 {\em
      Nat. Commun.\/} {\bf 5} 3689
    
    \bibitem{Lee2017}
    Lee V, Wu C~H, Lou Z~X, Lee W~L and Chang C~W 2017 {\em Phys. Rev. Lett.\/}
      {\bf 118} 135901
    
    \bibitem{LepriLiviPoliti2003}
    Lepri S, Livi R and Politi A 2003 {\em Phys. Rep.\/} {\bf 377} 1
    
    \bibitem{Liu2012}
    Liu S, Xu X~F, Xie R~G, Zhang G and Li B~W 2012 {\em Eur. Phys. J. B\/} {\bf
      85} 337
    
    \bibitem{Dhar2008}
    Dhar A 2008 {\em Adv. Phys.\/} {\bf 57} 457
    
    \bibitem{Livi2023}
    Livi R 2023 {\em Physica A\/} {\bf 631} 127779
    
    \bibitem{NarayanRamaswamy2002}
    Narayan O and Ramaswamy S 2002 {\em Phys. Rev. Lett.\/} {\bf 89} 200601
    
    \bibitem{Hohenberg1967}
    Hohenberg P~C 1967 {\em Phys. Rev.\/} {\bf 158} 383
    
    \bibitem{MerminWagner1966}
    Mermin N~D and Wagner H 1966 {\em Phys. Rev. Lett.\/} {\bf 17} 1133
    
    \bibitem{Mermin1968}
    Mermin N~D 1968 {\em Phys. Rev.\/} {\bf 176} 250
    
    \bibitem{AlderWainwright1970}
    Alder B and Wainwright T~E 1970 {\em Phys. Rev. A\/} {\bf 1} 18
    
    \bibitem{PomeauResibois1975}
    Pomeau Y and R{\'e}sibois P 1975 {\em Phys. Rep.\/} {\bf 19} 63
    
    \bibitem{ForsterNelsonStephen1977}
    Forster D, Nelson D~R and Stephen M~J 1977 {\em Phys. Rev. A\/} {\bf 16} 732
    
    \bibitem{LeonciniVergaRuffo1998}
    Leoncini X, Verga A~D and Ruffo S 1998 {\em Phys. Rev. E\/} {\bf 57} 6377
    
    \bibitem{DellagoPosch1997}
    Dellago C and Posch H 1997 {\em Physica A\/} {\bf 237} 95
    
    \bibitem{DelfiniLepriLivi2005}
    Delfini L, Lepri S and Livi R 2005 {\em J. Stat. Mech.\/} {\bf 2005} P05006
    
    \bibitem{Berezinskii1971}
    Berezinskii V~L 1971 {\em Sov. Phys. JETP\/} {\bf 32} 493
    
    \bibitem{Berezinskii1972}
    Berezinskii V~L 1972 {\em Sov. Phys. JETP\/} {\bf 34} 610
    
    \bibitem{KosterlitzThouless1973}
    Kosterlitz J~M and Thouless D~J 1973 {\em J. Phys. Cs\/} {\bf 6} 1181
    
    \bibitem{Kosterlitz1974}
    Kosterlitz J~M 1974 {\em J. Phys. C\/} {\bf 7} 1046
    
    \bibitem{DomanySchickSwendsen1984}
    Domany E, Schick M and Swendsen R~H 1984 {\em Phys. Rev. Lett.\/} {\bf 52} 1535

    \bibitem{ChaikinLubensky1995}
    Chaikin P~M and Lubensky T~C 1995 {\em Principles of Condensed Matter
      Physics\/} (Cambridge University Press)
    
    \bibitem{Hiura2023}
    Hiura K 2023 {\em Phys. Rev. E\/} {\bf 108} 054101
    
    \bibitem{Green1952}
    Green M~S 1952 {\em J. Chem. Phys.\/} {\bf 20} 1281
    
    \bibitem{GrahamHaken1971}
    Graham R and Haken H 1971 {\em Z. Phys. A\/} {\bf 243} 289

    \bibitem{HalperinHohenberg1969}
    Halperin B~I and Hohenberg P~C 1969 {\em Phys. Rev.\/} {\bf 188} 898 

    
    \bibitem{MartinSiggiaRose1973}
    Martin P~C, Siggia E~D and Rose H~A 1973 {\em Phys. Rev. A\/} {\bf 8} 423
    
    \bibitem{Janssen1976}
    Janssen H~K 1976 {\em Z. Phys. B\/} {\bf 23} 377
    
    \bibitem{DeDominicis1976}
    Dominicis C~D 1976 {\em J. Phys. Colloques\/} {\bf 37} C1--247
    
    \bibitem{Sato2020}
    Sato D~S 2020 {\em Phys. Rev. E\/} {\bf 102} 012111
    
    \bibitem{LepriRuffo2001}
    Lepri S and Ruffo S 2001 {\em EPL\/} {\bf 55} 512

    \bibitem{Spohn2014a}
    Spohn H 2014 {\em J. Stat. Phys.\/} {\bf 154} 1191

    \bibitem{Spohn2014b}
    Spohn H 2014 arXiv:1411.3907
    
    \bibitem{GendelmanSavin2000}
    Gendelman O~V and Savin A~V 2000 {\em Phys. Rev. Lett.\/} {\bf 84} 2381
    
    \bibitem{GiardinaLiviPolitiVassalli2000}
    Giardin{\`a} C, Livi R, Politi A and Vassalli M 2000 {\em Phys. Rev. Lett.\/} {\bf 84} 2144
    
    \bibitem{YangHu2005}
    Yang L and Hu B 2005 {\em Phys. Rev. Lett.\/} {\bf 94} 219404
    
    \bibitem{GendelmanSavin2005}
    Gendelman O~V and Savin A~V 2005 {\em Phys. Rev. Lett.\/} {\bf 94} 219405
    
    \bibitem{BereraYoffe2010}
    Berera A and Yoffe S~R 2010 {\em Phys. Rev. E\/} {\bf 82} 066304

    \bibitem{Migdal1975}
    Migdal A~A 1975 {\em Zh. Eksp. Teor. Fiz.\/} {\bf 69} 1457

    \bibitem{Polyakov1975}
    Polyakov A~M 1975 {\em Phys. Lett. B\/} {\bf 59} 79

    \bibitem{BrezinZinnJustin1976}
    Br{\'e}zin E and Zinn-Justin J 1976 {\em Phys. Rev. Lett.\/} {\bf 36} 691

    \bibitem{AndreanovBiroliLefevre2006}
    Andreanov A, Biroli G and Lef{\'e}vre A 2006 {\em J. Stat. Mech.\/} {\bf 2005} P07008
    
    \end{thebibliography}

\end{document}